\documentclass[
 aip, 
 amsmath,amssymb,
 reprint, 
 citeautoscript,  
]{revtex4-1}
\usepackage[utf8]{inputenc}
\usepackage{graphicx}
\usepackage{float}
\usepackage{enumitem}
\usepackage{bm}
\usepackage{bbold}
\usepackage{cases}
\usepackage{upgreek}
\usepackage[usenames, dvipsnames]{color}
\usepackage[colorlinks=true, citecolor=LimeGreen, linkcolor=NavyBlue, urlcolor=NavyBlue]{hyperref}
\usepackage{braket}
\usepackage[normalem]{ulem}
\usepackage{textgreek}

\newcommand \mc[1] { \mathcal{#1} }
\newcommand \dd[1]  { \!\!\textrm d{#1} \,}   
\newcommand \rmm[1]  { \textrm{#1} }

\newcommand \e[1] { \rmm{e}^{#1} }

\makeatletter
\def\@email#1#2{%
 \endgroup
 \patchcmd{\titleblock@produce}
  {\frontmatter@RRAPformat}
  {\frontmatter@RRAPformat{\produce@RRAP{*#1\href{mailto:#2}{#2}}}\frontmatter@RRAPformat}
  {}{}
}%
\makeatother
\begin{document}

\preprint{AIP/123-QED}

\title{Mori generalized master equations offer an efficient route to predict and interpret transport}

\author{Srijan Bhattacharyya}
\affiliation{Department of Chemistry, University of Colorado Boulder, Boulder, CO 80309, USA\looseness=-1}

\author{Thomas Sayer}
\affiliation{Department of Chemistry, University of Colorado Boulder, Boulder, CO 80309, USA\looseness=-1}

\author{Andr\'{e}s Montoya-Castillo}
\homepage{Andres.MontoyaCastillo@colorado.edu}
\affiliation{Department of Chemistry, University of Colorado Boulder, Boulder, CO 80309, USA\looseness=-1} 


\date{\today}

\begin{abstract}
Predicting how a material's microscopic structure and dynamics determine its transport properties remains a fundamental challenge. To alleviate this task's often prohibitive computational expense, we propose a Mori-based generalized quantum master equation (GQME) to predict the frequency-resolved conductivity of small-polaron forming systems described by the dispersive Holstein model. Unlike previous GQME-based approaches to transport that scale with the system size and only give access to the DC conductivity, our method requires only one calculation and yields both the DC and AC mobilities. We further show to easily augment our GQME with numerically accessible derivatives of the current to increase computational efficiency, collectively offering computational cost reductions of up to $90\%$, depending on the transport regime. Finally, we leverage our exact simulations to demonstrate the limited applicability of the celebrated and widely invoked Drude-Smith model in small-polaron forming systems. We instead introduce a cumulant-based analysis of experimentally accessible frequency data to infer the microscopic Hamiltonian parameters. This approach promises to provide valuable insights into material properties and facilitate guided design by linking macroscopic terahertz measurements to the microscopic details of small polaron-forming systems.
\end{abstract}

\maketitle
Prediction of a material's intrinsic charge transport rate is a fundamental goal of theoretical chemistry and materials science, with direct impact on energy and electronics research~\cite{fratini2020charge, nematiaram2020largest}. While an atomistically faithful (twin) model of materials encodes all desired response coefficients, solving the requisite quantum statistical dynamics of such constructions is infeasible in all but the smallest, homogeneous systems. Instead, a minimal but physically motivated model that discards unimportant degrees of freedom, fluctuations, and couplings has become the standard tool for calculating quantities like the macroscopic transport coefficient. This is the source of the celebrated spin-boson~\cite{leggett1987dynamics, 
weiss2012quantum}, Holstein~\cite{holstein1959studies2, holstein1959studies}, Anderson~\cite{anderson1961localized}, and Hubbard~\cite{hubbard1963electron, hubbard1964electron} models, which have been critical in understanding processes ranging from charge transfer reactions in solution~\cite{bader1990role, song1993quantum} to conductivity in polymers~\cite{ghosh2020excitons, fetherolf2020unification} and through nanojunctions~\cite{segal2016vibrational, thoss2018perspective, evers2020advances}, and even magnetism~\cite{ishizuka2021periodic, mielke1993ferromagnetism} and superconductivity~\cite{jiang2022stripe, nosarzewski2021superconductivity}. After this `downfolding' of atomistic complexity, one needs to solve for the dynamics of a sufficiently large model over appropriately long times to enable extraction of macroscopic observables free from finite-size artifacts. Yet, this is generally difficult or even computationally impossible with the available resources. What is more, to unlock general, microscopic insight from the dynamics of the model, one needs a physically transparent interpretation of its defining parameters and how these ultimately determine experimental observables.

Even when one knows how to map an atomistic system to a physically transparent model, its large, potentially infinite-dimensional Hilbert space makes solving its quantum dynamics a significant challenge. An exact, systematically adjustable, and often advantageous solution is to employ projection operator techniques~\cite{grabert2006projection, fick1990quantum} to reduce the dimensionality and predict only the evolution of particular \textit{observables of interest}, $\mc{C}(t)$. While performing a projection sacrifices access to arbitrary observables of the full system, one gains a low-dimensional framework to predict the dynamics of the observables of interest via the Mori-Nakajima-Zwanzig (MNZ) equation~\cite{nakajima1958quantum, zwanzig1960ensemble, mori1965transport}
\begin{equation}\label{eq:MNZ_intro}
    \dot{\mc{C}}(t) = \dot{\mc{C}(0)} \mc{C}(t) - \int_0^t \dd{s}\mc{K}(s)\mc{C}(t-s) + \mc{I}(t).
\end{equation}
In this generalized quantum master equation (GQME), the evolution of the projected variables in $\mc{C}(t)$ requires knowledge of the inhomogeneous term, $\mc{I}(t)$---which can be removed via the proper choice of projection operator and is zero for us---and the memory kernel, $\mc{K}(t<\tau_\mc{K})$, where $\tau_\mc{K}$ is the time after which $\mc{K}(t)=0$. By writing the projected dynamics in terms of this memory kernel, both the complex (non-Markovian) short-time behavior \textit{and} the detailed balance of the long-time populations can be captured using \textit{only short-time data}. A recent example of this principle is the computation of mean first passage times in the folding of large biomolecules, where only 25~ps of reference simulation data contain the information needed to model events over 10s of~$\upmu$s, i.e., three orders of magnitude longer~\cite{dominic2023building}. Yet, the cost savings of this dimensionality reduction procedure rely upon a separation of timescales between the variables of interest and those whose dynamics one does not explicitly track. Indeed, the memory kernel remains as long-lived as the slowest variables excluded from the projected space. It is thus crucial to put all the slowest degrees of freedom in the projected space, even if they are not required for the final calculation of, say, a transport coefficient. 

At the practical level, the choice of the projection operator has significant consequences on computational feasibility. This is because constructing $\mc{K}(t)$, a dynamical $N \times N$ matrix, typically requires at least $N$ distinct simulations. For example, previous work pursued a nonequilibrium strategy of projecting onto the populations of localized electronic states to calculate polaronic transport coefficients along a model one-dimensional chain~\cite{yan2019theoretical}. This GQME formally scales with the number of sites, $N$\footnote{Ref.~\onlinecite{yan2019theoretical} assumed a homogeneous system, which necessitates only a single simulation by symmetry. This is not generally the case.}. Here, we pursue a different strategy via the Kubo formula that relates a material's frequency-resolved conductivity to the equilibrium fluctuations of the current. This relation suggests adopting a Mori-type projection operator~\cite{mori1965transport} with the current operator as the \textit{only} observable of interest. A remarkable consequence of this choice is that \textit{one needs only one equilibrium calculation to construct the GQME, making the method's scaling independent of system size}. Our work shows that this strategy offers a compact and efficient route to encode the current response and frequency-resolved conductivity. 

Why has it taken until now to bridge the Kubo formalism with Mori-Zwanzig theory for electrical conductivity predictions in polaron-forming materials? While path integral simulations on the ground electronic state have become mainstream~\cite{markland2018nuclear, chandler1981exploiting, cao1993new, parrinello1984study, jang1999derivation, craig2004quantum, cao1994formulation}, calculating equilibrium time correlation functions of quantum mechanical systems with nonadiabatic effects remains a fundamental challenge. The challenge can be broken down into two problems. The first centers on obtaining a sufficiently accurate representation of the correlated canonical density of the system, and the second lies in generating the subsequent dynamics. A variety of schemes have emerged to tackle these problems, including path integrals~\cite{shao2002iterative, tanimura2014reduced, song2015calculation, montoya2017path}, semiclassics~\cite{liu2006using, shi2003semiclassical, poulsen2003practical, montoya2017approximate}, and density matrix renormalization group~\cite{barthel2013precise, karrasch2015spin, karrasch2013reducing}. We exploit recent algorithmic advances~\cite{shi2009efficient, song2015time} that have enabled the calculation of these correlation functions using the hierarchical equations of motion (HEOM)~\cite{tanimura1989time}.

The benefits of using the current as the sole projected quantity surpass merely practical considerations. Although non-equilibrium approaches, like population relaxation, are popular and can offer a view of the full relaxation to equilibrium~\cite{bhattacharyya2024anomalous, yan2019theoretical}, they are limited to the zero frequency component of the transport coefficient (i.e., the DC mobility). In contrast, the current autocorrelation function, $C_{JJ}$, encodes the full dynamical conductivity,
\begin{equation}
\begin{split}\label{eq:sigma_omega}
    \mathbb{Re}~\sigma(\omega) & = \frac{1-\e{-\beta \omega}}{2\omega V} \int_{-\infty}^\infty \dd{t} \e{-\rmm{i}\omega t}C_{JJ}(t) \\
    & \equiv \frac{\beta}{V} \int_0^\infty \dd{t} \e{-\rmm{i}\omega t}C^\rmm{Kubo}_{JJ}(t),
\end{split}
\end{equation}
encompassing the system's response to static and alternating fields. Here, $C^\rmm{Kubo}_{JJ}(t)$ is the Kubo-transformed correlation function~\cite{kubo1991statistical}, $\beta = [k_BT]^{-1}$ is the inverse thermal energy, and $V$ is the volume. Furthermore, the structure of the correlation function itself is of fundamental importance as it facilitates the interpretation of the underlying transport mechanism. For example, the phenomenological Drude-Smith model~\cite{smith2001classical} is commonly used to map the experimentally measurable conductivity, $\sigma(\omega)$, onto a mean collision time, $\tau$, and the strength of those collisions, $c$. In this way, the two-parameter Drude-Smith fit is thought to capture much of the behavior observed in experimental terahertz spectra~\cite{  yettapu2016terahertz, pattengale2019metal, kumar2020terahertz, magnanelli2020polarization, li2023charge}. 

To interrogate the advantages of the Mori approach, we employ HEOM to generate numerically exact dynamics of a physically transparent model of polaron formation and transport: the dispersive Holstein Hamiltonian. With our $C_{JJ}$ simulations, which determine $\sigma(\omega)$, we also test the applicability of the Drude-Smith model to small polaron-forming materials and find that it does not offer a satisfying fit. As an alternative, we introduce a frequency-space analysis that reveals a simple relationship between the cumulants of memory function and the parameters of the Hamiltonian that generated it, offering a route to map experimental results directly onto a microscopic Hamiltonian. While we focus on exact quantum dynamics as a means to illustrate our approach to predicting and elucidating polaron transport, our findings are broadly applicable and stand to benefit the calculation of general transport coefficients and systems, whether using quantum or classical dynamics, including \textit{ab initio} molecular dynamics. 

The Holstein Hamiltonian has been extensively used to predict transport in materials spanning organic crystals and polymers~\cite{cheung2008modelling, ortmann2010charge, fetherolf2020unification}, covalent organic frameworks~\cite{ghosh2021topology}, and nanomaterials~\cite{mousavi2010electron}. It describes carriers (excitons, electrons, or holes) that move on a lattice of $N$ sites and interact locally with their nuclear environment to form a small polaron consisting of the original electronic excitation and the material deformation it causes in its immediate environment. While the classic Holstein model assumes localized coupling to a single optical phonon mode~\cite{holstein1959studies, holstein1959studies2, mahan2000many}, we focus on the dispersive Holstein model, which couples to a continuum of phonon modes of varying frequencies and more faithfully describes organic crystals and disordered polymers~\cite{nematiaram2020modeling, yan2018understanding}. To connect with previous studies focusing on transport in organic semiconductors~\cite{song2015new, yan2019theoretical, fetherolf2020unification, bhattacharyya2024anomalous}, we adopt the dilute limit (one electron or hole) in a homogeneous lattice with degenerate lattice sites and only nearest-neighbor hopping, $v$ = 50 cm$^{-1}$, at a temperature of 300~K. We interrogate the rich behavior that the model displays as one varies the strength of carrier-lattice coupling (encoded by the reorganization energy, $\eta$) and the characteristic speed the local lattice relaxes, encoded by frequency $\omega_c$. For additional details, see Appendix~\ref{app:Holstein}. 

Our Mori-type projector focuses on the current operator, $\hat{J}$, yielding a GQME for the current autocorrelation function (see Appendix~\ref{app:gmqe-details}) in the Kubo formula, Eq.~\ref{eq:sigma_omega},
\begin{equation}\label{eq:cjj-def}
    C_{JJ}(t) = \frac{1}{Z}\text{Tr}\Big[ \e{-\beta \hat{H}} \hat{J} \e{\rmm{i}\mathcal{L}t} \hat{J}  \Big],
\end{equation}
where $Z = \mathrm{Tr}[e^{-\beta \hat{H}}]$ is the partition function of the full system. The current operator,
\vspace{-4pt}
\begin{equation}
    \hat{J} = -\rmm{i}d \sum_{\langle mn \rangle} v_{mn} (\hat{a}_m^\dag \hat{a}_n -\hat{a}_n^\dag \hat{a}_m),
    \vspace{-6pt}
\end{equation}
is exclusively an electronic operator and thus accessible from a solver like HEOM. Although $\hat{J}$ sums over hopping terms connecting neighboring sites $\langle mn \rangle$ spaced $d=5$\,\AA~apart, the resulting correlation matrix, $C(t)$, is of size $1 \times 1$. This means that the GQME \textit{requires only a single initial condition for its construction} and hence does not scale with system size. The complicating factor is the correlated initial condition, $\e{-\beta \hat{H}}/Z$. To calculate the equilibrium correlation function~\cite{song2015new}, one performs an `equilibration' HEOM calculation starting from any nonequilibrium condition \footnote{The easiest to construct and fastest initial condition to equilibrate corresponds to a zeroth order approximation to $\e{-\beta \hat{H}}/Z$, i.e., $\e{-\beta \hat{H}_{elec}}/Z_{elec} \times \e{-\beta \hat{H}_{B}}/Z_{B}$.} to converge the auxiliary density matrices to their equilibrium values; one then multiplies the current operator, $\hat{J}$, to generate a new initial condition, $\tilde{\rho}_0 \equiv \e{-\beta \hat{H}}\hat{J}$, that one can then evolve and use to measure $\hat{J}$ at time $t$. See Appendix~\ref{app:HEOM} for computational details. Despite requiring this equilibration step, we show that the protocol offers efficiency gains.

\begin{figure}[!t]
\vspace{-6pt}
\begin{center} 
    \resizebox{.375\textwidth}{!}{\includegraphics[width=0.75\linewidth]{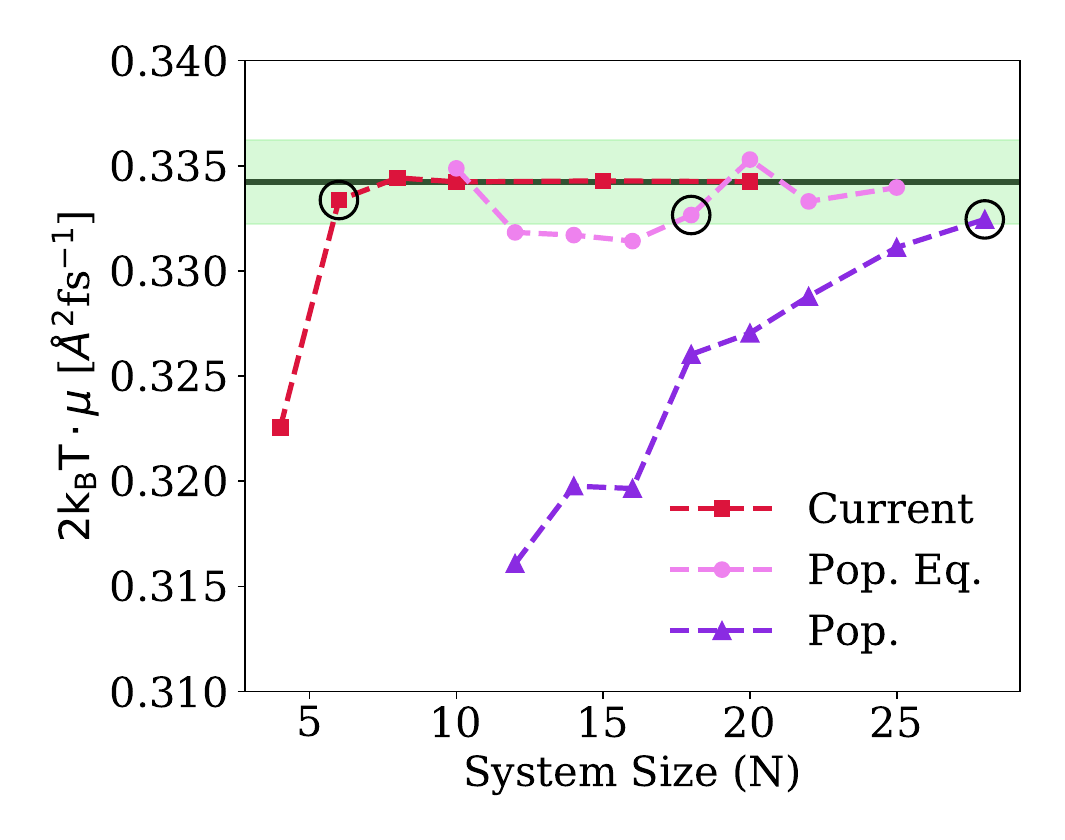}}
\vspace{-12pt}
\caption{\label{fig:mu_convergence_size} Convergence of $\mu$ with system size for the three methods considered and $\eta / v = 6.26$, $\omega_c / v = 0.82$. We take the converged value to be given by the current-based Mori GQME for $N=20$, and allow $\pm 0.002$ precision error (green shaded region). We mark the point where each method enters---and stays within---this region with circles. The population methods have no data for low $N$ since finite size effects preclude a plateau in $\rmm{d}{\rm MSD}(t)/\rmm{d}t$ used to identify diffusive motion (see Ref.~\onlinecite{bhattacharyya2024anomalous} for a full discussion). }
\end{center}
\vspace{-14pt}
\end{figure}

We illustrate the performance of the Mori-type GQME for a dispersive Holstein ring. To compare fairly across methods, we converge each protocol to the macroscopic size limit with the same parameters $\eta/v = 6.26, \omega_c/v = 0.82$, which we choose to align with previous work on organic semiconductors~\cite{song2015new, bhattacharyya2024anomalous}. Figure~\ref{fig:mu_convergence_size} shows the size dependence of the DC mobility given by
\vspace{-1pt}
\begin{equation}\label{eq:mu_def_linear_response}
    \mu = V \lim_{\omega \rightarrow 0} \sigma(\omega).
\end{equation}
For this parameter regime, the Mori GQME requires only $N=6$ sites compared to $N=18$ and $N=28$ for the population-based methods shown below, consistent with our previous findings~\cite{bhattacharyya2024anomalous}.

\begin{figure*}[!th]
\vspace{-0pt}
\begin{center} 
    \resizebox{.91\textwidth}{!}{\includegraphics[trim={15pt 5pt 2pt 0pt},clip]{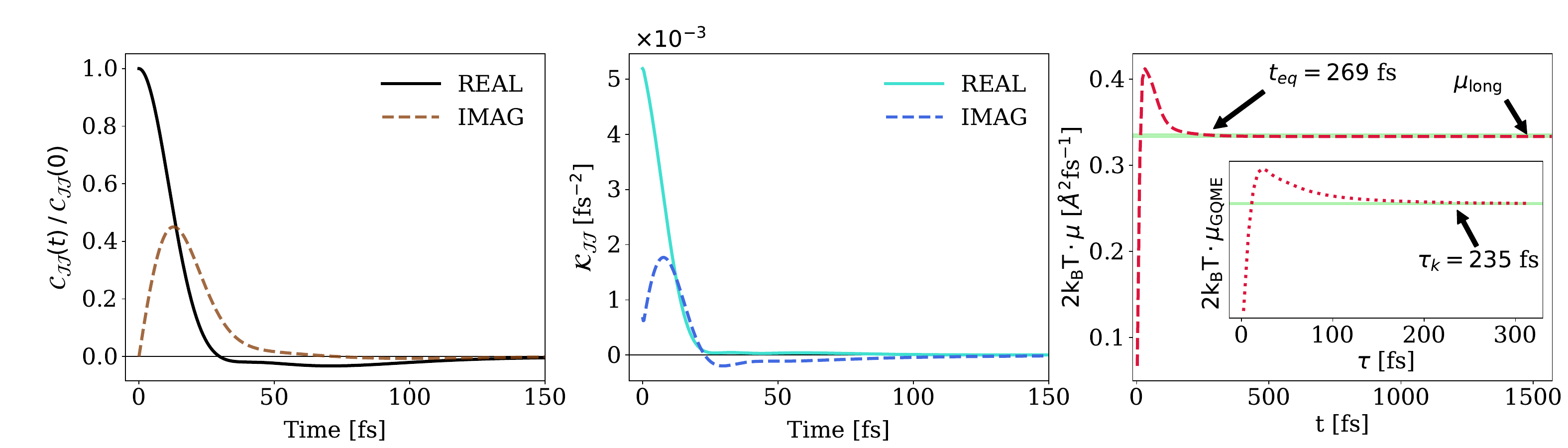}}
\vspace{-0pt}
\caption{\label{fig:current_proj} Dispersive Holstein simulations for $\eta / v = 6.26$, $\omega_c / v = 0.82$. \textbf{Left}: $C_{JJ}(t)$ converges to zero, as expected in dissipative systems. 
\textbf{Middle}: Memory kernel for $C_{JJ}(t)$.
\textbf{Right}: Convergence of $\mu$ as a function of the integral limit in Eq.~\ref{eq:mu_def_linear_response}. $C_{JJ}$ takes $t_\rmm{eq}=269$~fs to reach $\pm 0.002$ of the long-time value. The green shaded region is the same as in Fig.~\ref{fig:mu_convergence_size}.
\textbf{Inset}: Convergence of the GQME estimate of $\mu_\rmm{long}$ with respect to the cutoff time $\tau$. We find $\tau_\mc{K}=235~\rmm{fs}<t_\rmm{eq}$.} 
\end{center}
\vspace{-14pt}
\end{figure*}

Figure~\ref{fig:current_proj}--left displays the real and imaginary parts of the current autocorrelation function. Although, graphically, $C_{JJ}\rightarrow 0$ within 150~fs, Fig.~\ref{fig:current_proj}--right shows that convergence of $\mu$ to three significant figures takes 269~fs, almost twice as long. Figure~\ref{fig:current_proj}--center illustrates the memory kernel associated with these dynamics. The inset of Fig.~\ref{fig:current_proj}--right shows $\mu$ obtained from the dynamics generated from $\mc{K}_{JJ}$ truncated at time $\tau$, demonstrating that $\mu$ can be computed with only $\tau_\mathcal{K} = 235$~fs of data, slightly reducing the cost of the simulation: that is, one needs only $235 / 269 = 87\%$ of the original simulation time. 

\begin{figure}[!t]
\vspace{-4pt}
\begin{center} 
    \resizebox{.50\textwidth}{!}{\includegraphics{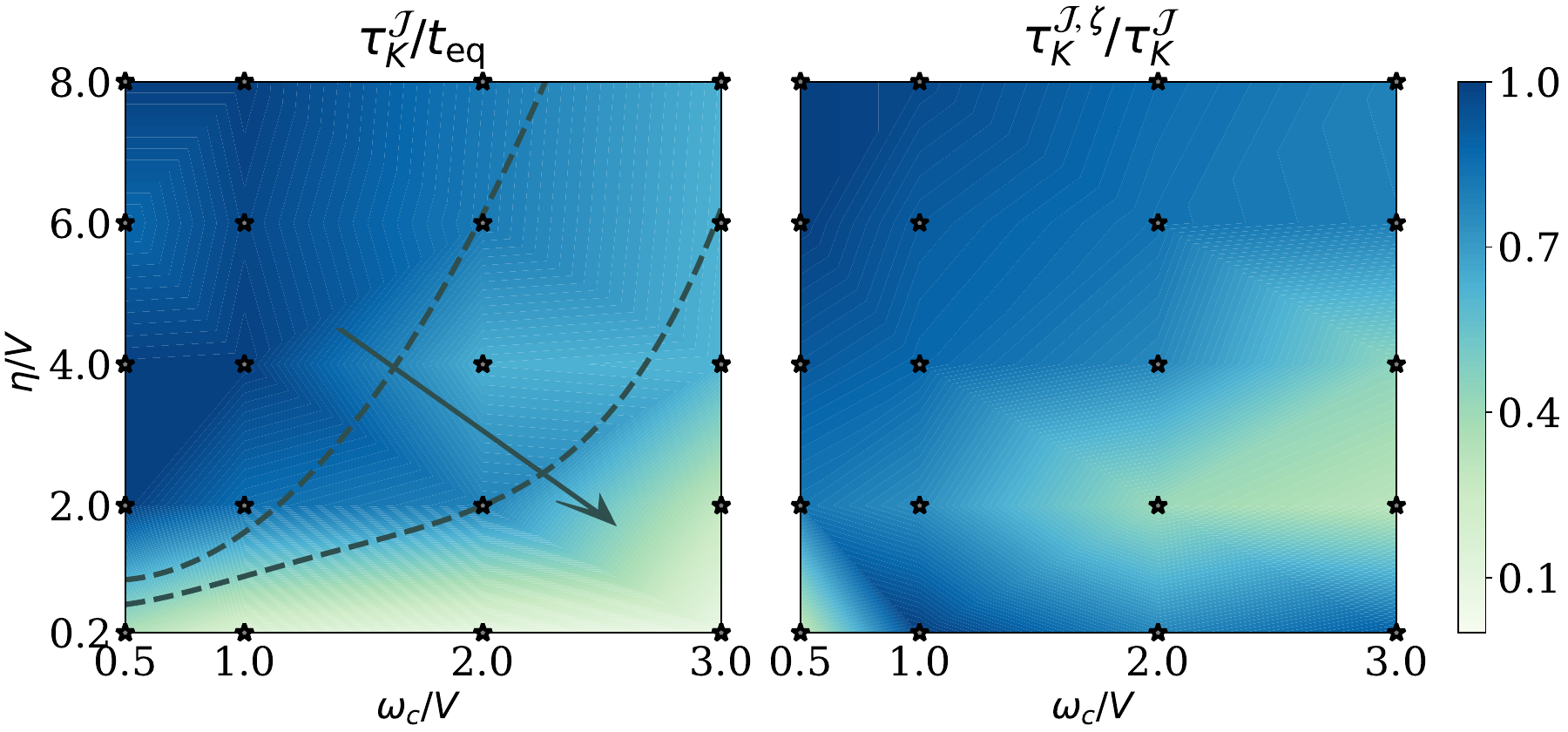}}
    \hspace{-15pt}
\vspace{-14pt}
\caption{\label{fig:theta} Cost reduction from the current and augmented GQMEs for the dispersive Holstein model. \textbf{Left}: Heat map of the ratio $\tau_\mathcal{K}^J/t_\rmm{eq}$, i.e., the MNZ saving factor with the current-based projector, for 20 different parameter regimes of our dispersive Holstein model, each denoted by a star. Dashed lines are guides to the eye showing regions of: no (dark blue), minor (light blue), and large (light green) improvement; the arrow shows the direction of increasing efficacy. \textbf{Right}: Ratio of the cutoff time when the projector is augmented to contain $\zeta \equiv \dot{J}$ to just $\tau_\mc{K}^J$. $t_\rmm{eq}$, $\tau_\mathcal{K}^J$ and $\tau_\mathcal{K}^{J,\zeta}$ are all computed as reaching a $\pm 1 \%$ error of $\mu_\rmm{long}$. }
\vspace{-18pt}
\end{center}
\end{figure}

Since the computational saving is a property of the system parameters, we quantify the effort reductions over a grid of 20~different instances of the model parameters in Fig.~\ref{fig:theta}--left. The dark portion of Fig.~\ref{fig:theta}--left shows the region with $\tau_\mathcal{K}/t_\rmm{eq}> 0.85$, where the brute force calculation and the Mori GQME incur comparable computational cost; the parameters we used in Fig.~\ref{fig:current_proj} lie in this region. The light blue region, where $0.66 < \tau_\mathcal{K}/t_\rmm{eq} \leq 0.85$, offers modest computational savings. The light green region has $\tau_\mathcal{K}/t_\rmm{eq} \sim\! 0.1$ in many places, enabling significant savings with an order-of-magnitude reduction in the required effort. See the Supplementary Material for the current correlation functions, memory kernels, and conductance of the dispersive Holstein model over the parameter space. Hence, the Mori GQME offers maximal efficiency gains when a material displays relatively low charge-lattice coupling (small $\eta$) and fast decorrelating environments (large $\omega_c$).

\onecolumngrid

\pagebreak
\twocolumngrid

To interrogate if one can obtain even greater efficiency gains, we adopt the strategy of including derivatives of the observable in the projector~\cite{mori1965transport, reichman2005mode, janssen2018mode}. While HEOM is unable to sample the current's time derivative explicitly owing to the presence of bath operators, our $C(t)$ has sufficient temporal resolution to use its \textit{numerical} time derivatives to augment the projector (see Appendix~\ref{app:derivatives}). The results, displayed in Fig.~\ref{fig:theta}--right, are surprising. It is indeed possible to decrease the lifetime of the memory kernel required to predict the transport coefficient via $C_{JJ}$ by simply augmenting the projector with derivatives of the motion, but only for particular combinations of the Hamiltonian parameters. Over large regions of the parameter space, such as where $\eta$ is large, there is no advantage. However, when $\omega_c \approx \eta$ we obtain a $\sim$70\% reduction in cost. To our knowledge, this parameter-dependent benefit has not previously been reported and represents a simple and parsimonious way to potentially reduce even further the computational effort required to capture equilibrium time correlation functions.

We are now in a position to contextualize the results from the Mori approach in terms of the advantages offered by the nonequilibrium population-based GQME. Unlike the population-based GQME, which generally requires more simulations to construct its memory kernel, our Mori GQME requires only one, and converges significantly faster with system size, as Fig.~\ref{fig:mu_convergence_size} demonstrates. However, the population-based GQME may still offer a competitive advantage if its memory kernel lifetime, $\tau_\mc{K}$, is sufficiently short. To test this, we consider two different population-based initializations: one corresponding to an instantaneous Franck-Condon excitation and the other to a Marcus-like description of equilibrium charge transfer. Note that to generate the Marcus-like initial condition in HEOM requires a pre-production simulation (see Appendix~\ref{app:HEOM}). Table~\ref{table:mu_compare} summarizes the results.

\begin{table}[!b]
    \centering
     $2k_\rmm{B}T\mu = 0.334 \pm 0.002$~\AA$^2$\,fs$^{-1}$\\
    \vspace{-1.5pt}
    \hspace{-3pt}
    \begin{tabular}{|l||c|ccc|ccc|c|} 
    \hline
    Method & $N^{(1)}$ & \multicolumn{3}{c|}{Sim. time [fs]} & $t_\rmm{job}$ & $n_\rmm{core}$ &  $t_\rmm{tot}^{(2)}$ & Mem.\\ 
                                 \cline{3-5}
     &  &  Pre$^{(3)}$ & Prod.  & ~$\tau_\mathcal{K}$~ & [hours] & & \textbf{[days]} & [GB] \\
     \hline
     \hline
    Pop.      & 28 & --  & 3288 & 981 & 1.85 & 64 & \textbf{137.3} & 4.73 \\
    Pop. eq.  & 18 & 775 & 1081 & 655 & 0.42 & 64 & \textbf{20.14} & 2.14 \\
    Current   &  6 & 725 &  269 & 235 & 0.30 & 64 & \textbf{0.82} & 1.39\\\hline
    \end{tabular}
    
    \caption{\label{table:mu_compare} Computational cost for the three different routes to calculate $\mu$ for $\eta / v = 6.26$, $\omega_c / v = 0.82$. \textbf{(1)}~The number of sites required to converge $\mu$ (see Fig.~\ref{fig:mu_convergence_size}). \textbf{(2)}~The total core time is $n_\rmm{job}\cdot t_\rmm{job}\cdot n_\rmm{core}$, where $n_\rmm{job}=N$ for the first two methods, but only $1$ for `Current'. \textbf{(3)}~Time required for the pre-production step; for the `Pop.~eq.' method this time is independent of $N$, but increases with $N$ for `Current'.}
    \vspace{-4pt}
\end{table}

The computational savings of the Mori GQME are vast: it is more than 20 times cheaper than the pre-equilibrated population-based method, and over two orders of magnitude for the non-equilibrated alternative. What is most impressive is that these are the savings one would obtain for the set of parameters investigated in Fig.~\ref{fig:current_proj}, which lead to the smallest efficiency gains. For this parameter regime, the computational saving arises mainly due to the single entry in the projector, and not from the reduction in cost due to the memory kernel method. Hence, especially in systems with static disorder that require $N$ simulations for the parameterization of the population-based GQME, the Mori GQME stands to yield significant efficiency gains, particularly for systems with small reorganization energies dominated by charge coupling to high-frequency, optical phonon modes. 

Beyond efficiency gains, our approach also yields accurate current autocorrelation functions, which we now employ to analyze the applicability of the frequently invoked phenomenological Drude and Drude-Smith theories. Figure~\ref{fig:current_proj} shows that the real part of $C_{JJ}$ becomes negative. This behavior does not arise in the charge transport of normal metals, where the dynamics is well-described by a decaying exponential~\cite{allen2006electron} with $k_\rmm{Drude} = 1/\tau$ before the onset of interband transitions. Describing this negative region, or `cage effect', necessitates a functional form more complex than a simple exponential. The Smith reformulation of this process assigns these additional terms to Poisson distributed collisions (characterized by constant $\tau$) of the charge carriers with their lattice ions~\cite{smith2001classical}. The coefficients $c_n$ of these terms represent what fraction of the initial velocity is retained after the $n^\rmm{th}$ collision. The Fourier transform of this current autocorrelation function, i.e., the conductivity in Eq.~\ref{eq:sigma_omega}, then takes the form~\cite{smith2001classical}
\begin{equation}\label{eq:drudesmith}
    \sigma(\omega) = \frac{\tau}{1-\rmm{i}\omega \tau} \left[ 1+\sum_{n=1}^\infty \frac{c_n}{(1-\rmm{i}\omega \tau)^n} \right].
\end{equation}
The standard approximation truncates the sum at the first term, with $-1 \leq c \leq 0$, which has been shown to describe experimental data well over the limited frequency range that is normally accessible~\cite{kumar2020terahertz, yettapu2016terahertz, magnanelli2020polarization, pattengale2019metal}. How well does it capture our exact response over the full dynamical range?

To perform this analysis, we consider how to connect the Drude-Smith analysis to our quantum mechanically exact response. The quantum mechanical $C_{JJ}$ is complex and gives the transport coefficient via the linear response relation, Eq.~\ref{eq:sigma_omega}, but the Drude and Drude-Smith expressions consider only classical and real current autocorrelation functions. Thus, to compare to Eq.~\ref{eq:drudesmith}, we replace the classical function with the Kubo-transformed correlation function, $C^{\rm Kubo}_{JJ}$, which displays the same symmetries as classical correlation functions (i.e., it is real and symmetric about $t$)~\cite{craig2004quantum, kubo1991statistical} and is used to derive the harmonic correction factor to approximate quantum correlation functions by their classical counterparts~\cite{egorov1999quantum, bader1994quantum}.

\begin{figure}[!t]
\vspace{-8pt}
\begin{center} 
    \resizebox{.4\textwidth}{!}{\includegraphics{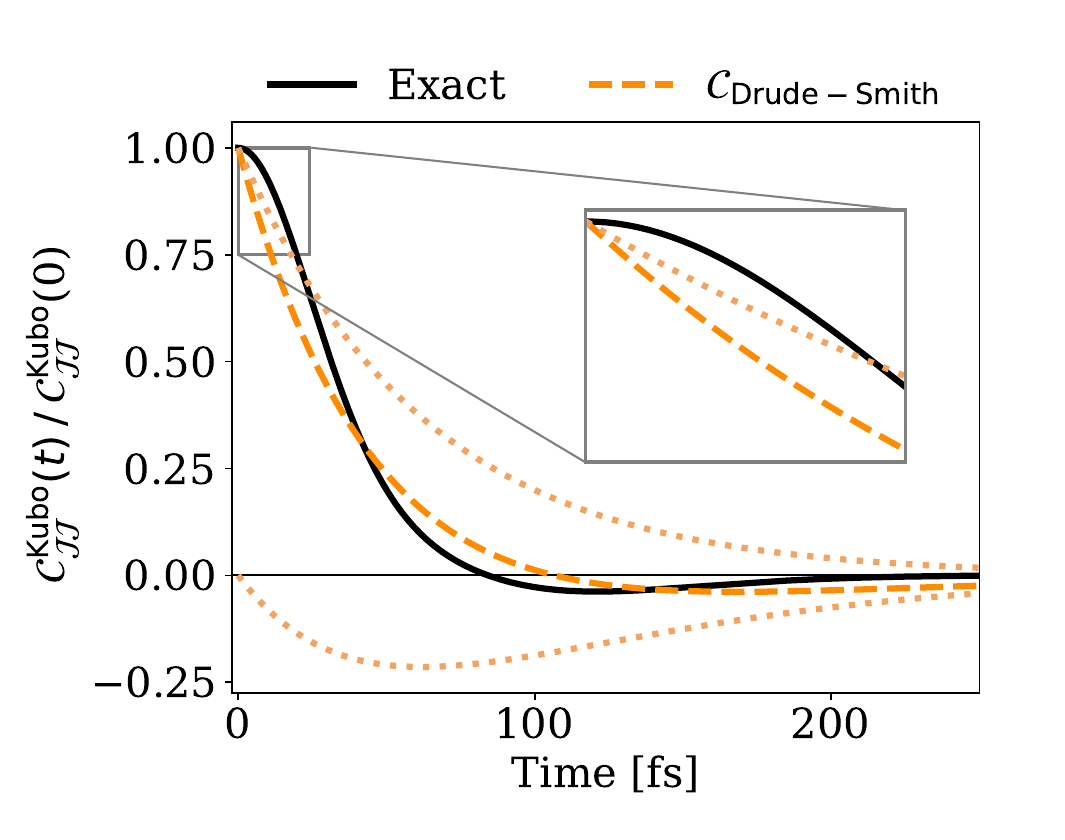}}
\end{center}
\vspace{-22pt}
\caption{\label{fig:drude-fit-example} Drude-smith fit (orange) to the $C_{JJ}^\mathrm{Kubo}$ obtained with HEOM (black) for $\eta / v = 2.0$, $\omega_c / v = 1.0$. The orange dotted lines are the contributions from the pure exponential and exponential-times-linear `c term' respectively. The fit gives $c=-0.58$ and $k_\mathrm{Drude}=0.0161$ fs$^{-1}$. \textbf{Inset}: $C^\mathrm{Kubo}_{JJ}$ at the early time has clear concave curvature which the Drude-Smith form fails to capture.}
\vspace{-8pt}
\end{figure}

We can now assess the applicability of a Drude-Smith decomposition of our $C^{\rm Kubo}_{JJ}(t)$. Performing the inverse transform to Eq.~\ref{eq:drudesmith} truncated at $n=1$ yields~\cite{chen2021drude},
\begin{equation}\label{eq:drudesmith_time}
    C^\rmm{Kubo}_{JJ}(t)/C^\rmm{Kubo}_{JJ}(0) = \left( 1 + ck_\mathrm{Drude}t \right)\e{-k_\mathrm{Drude}t}.
\end{equation}
Focusing first on $\eta/v=2$, $\omega_c/v=1$, Fig.~\ref{fig:drude-fit-example} shows that form of Eq.~\ref{eq:drudesmith_time} cannot capture the qualitative shape of $C^{\rm Kubo}_{JJ}(t)$, missing the curvature near $t=0$. This functional form can correctly capture the long-time decay for some parameter regimes but not for this example, remaining above zero for too long. Finally, although the depth of the negative well is reasonable, the position of its minimum is incorrect. This qualitative description of the mismatch applies across the parameter space, even as the negative region becomes less pronounced.

This is not unexpected since the Drude-Smith form implies the memory kernel decays as a single exponential (see Appendix.~\ref{app:cosine-fit}). Figure~\ref{fig:memeory-fit-example}--(a),(b) shows that the normalized real part $\tilde{\mathcal{K}}(\omega) = \mathbb{Re}\left[ \mathcal{K}^\mathrm{Kubo}(\omega)\right] / \mathbb{Re}\left[ \int \,\dd{\omega} \mathcal{K}^\mathrm{Kubo}(\omega)\right]$ is a complicated function of frequency. Here, the simplest curve shown ($\eta/v=2$, $\omega_c/v=3$) has an approximately Gaussian shape---not Lorentzian, as expected for a single exponential---and as $\eta$ increases the distribution becomes broader and more structured. The complexity of these curves shows that prescribing a simple, few-parameter form for the memory kernel is overly optimistic. What is more, even if one obtained a better fit to the Drude-Smith model in Eq.~\ref{eq:drudesmith} by considering $n>1$, it would be difficult to interpret the coefficients of higher-order collisions in terms of a microscopic picture of polaron formation and transport. 

\begin{figure}[!b]
\vspace{-14pt}
\begin{center} 
    \resizebox{.5\textwidth}{!}{\includegraphics{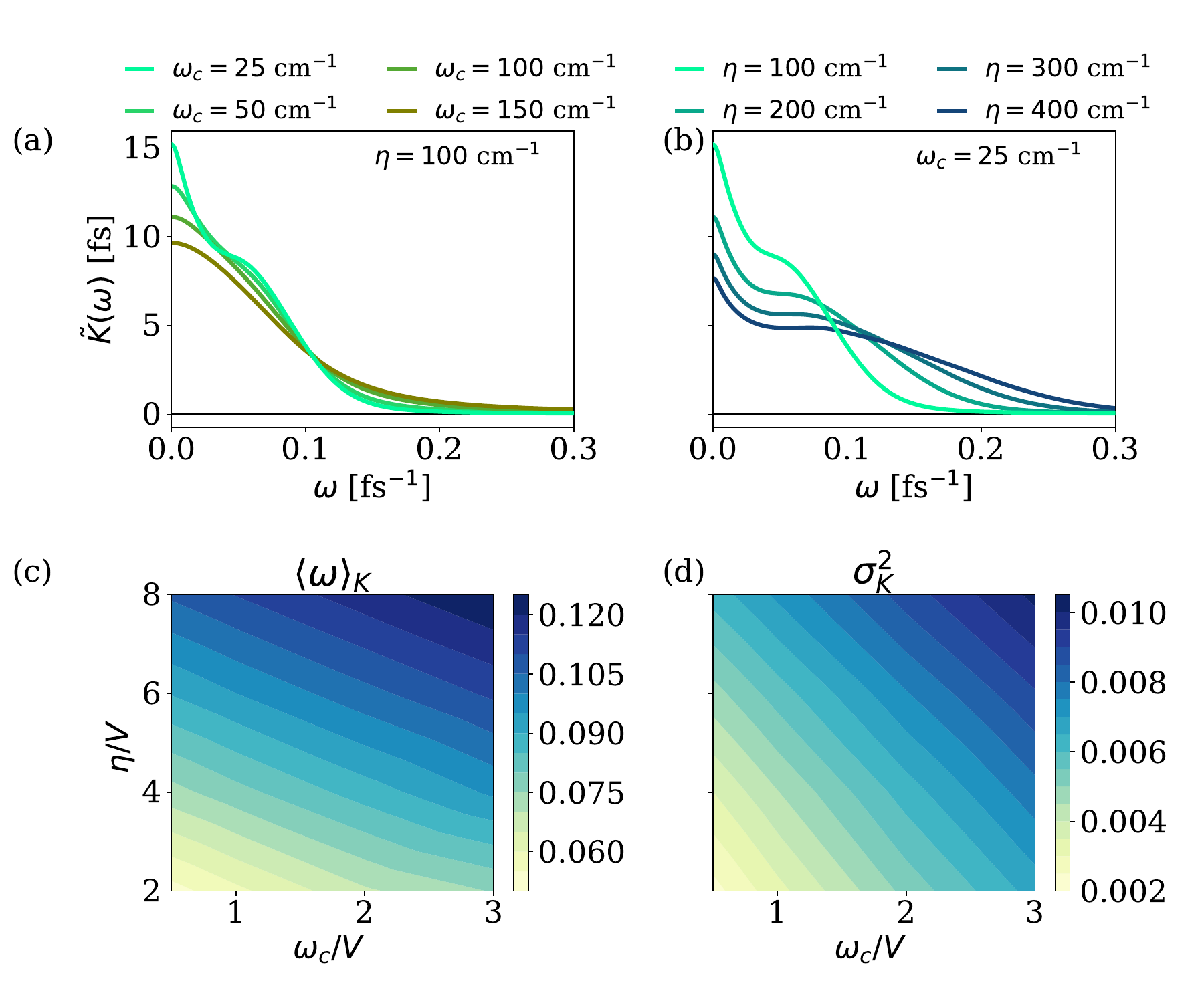}}
\end{center}
\vspace{-22pt}
\caption{\label{fig:memeory-fit-example} Characteristic distributions $\mathcal{K}^{\mathrm{Kubo}} (\omega)$ varying (a) bath correlation time $\omega_c$ while keeping reorganization energy $\eta = 100$ cm$^{-1}$; (b) reorganization energy $\eta$ while keeping bath speed $\omega_c = 25$ cm$^{-1}$. How (c) $\langle \omega_{\mathcal{K}} \rangle$ and (d) $\sigma_{\mathcal{K}}^2$ vary as a function of Hamiltonian parameters $[\eta/v, \omega_c/v]$.}
\end{figure}

\begin{figure}[!t]
\vspace{-6pt}
\begin{center} 
    \resizebox{.4\textwidth}{!}{\includegraphics{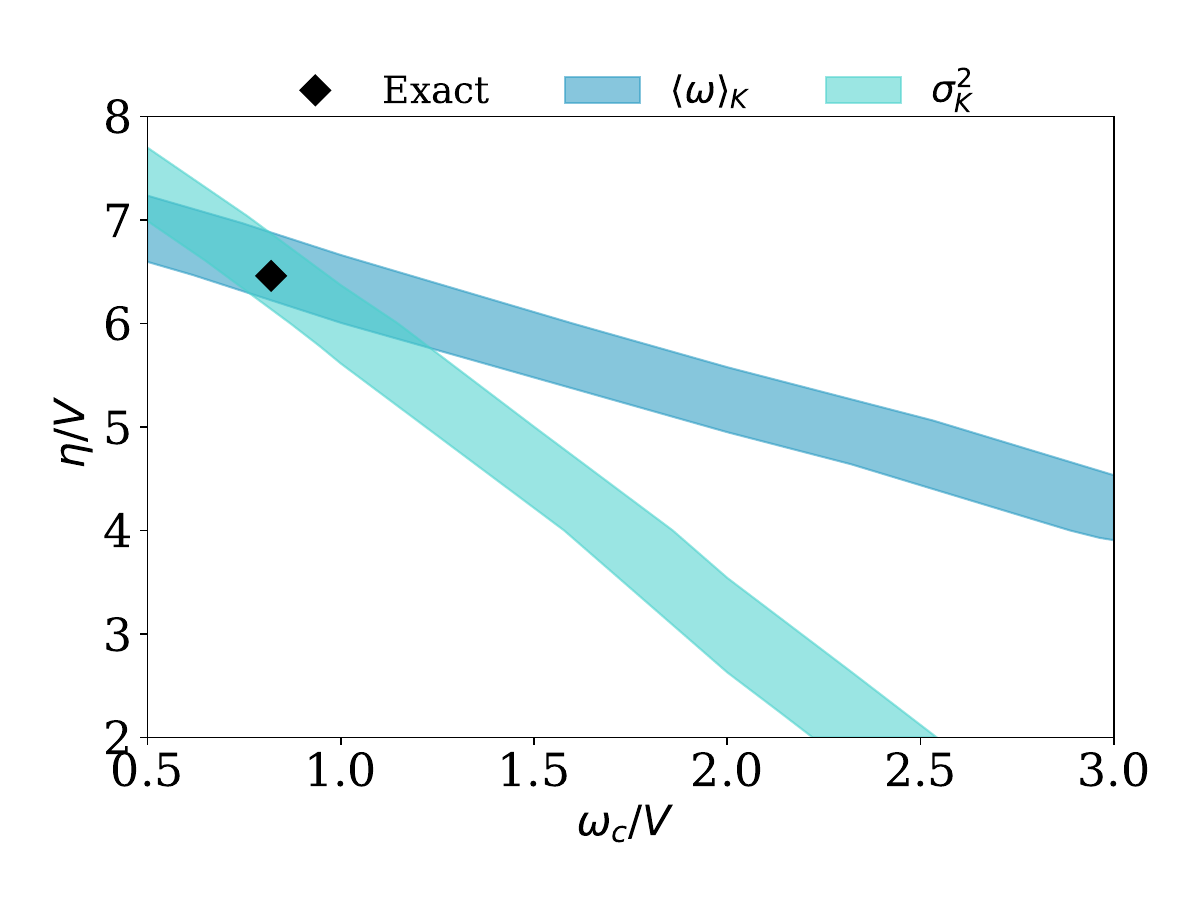}}
\end{center}
\vspace{-22pt}
\caption{\label{fig:backmap} Demonstration of back-mapping from the memory kernel cumulants to the dispersive Holstein parameters. Blue contour line: predicted using $\langle \omega_K \rangle = 0.096$ fs$^{-1}$ from Fig.~~\ref{fig:memeory-fit-example}--(c). Green contour line: predicted using $\sigma_K^2 = 0.0057$  fs$^{-2}$ from Fig.~~\ref{fig:memeory-fit-example}--(d). The intersection of these two contours suggests the location in Hamiltonian parameter space, and the black diamond indicates the exact parameters.}
\end{figure}

To go beyond the limitations of the Drude-Smith model, we show that one can infer the parameters of the microscopic Hamiltonian responsible for the measured signal using \textit{only} the frequency-resolved, complex conductivity, $\sigma(\omega)$. Working with the Kubo memory kernel, which has equivalent information via $\mathcal{K}^{\rm Kubo}(\omega) = 1/\sigma(\omega) - \rmm{i}\omega$, we seek a \textit{statistical} characterization of $\tilde{\mc{K}}(\omega)$ from the measured data. We get this from its moments, noting that all frequency moments of the response function (which is directly related to the memory kernel) \textit{must exist}~\cite{forster2018hydrodynamic}. The $n^\rmm{th}$ moment takes the usual form, 
\begin{equation}\label{nth_moment}
    \langle \omega^n \rangle_K  = \int ~\dd{\omega} \omega^n \tilde{\mathcal{K}}(\omega).
\end{equation}
Since we explore a 2D parameter space, we use the mean $\langle \omega \rangle_K$ and variance $\sigma_K^2 = \langle \omega^2 \rangle_K - \langle \omega \rangle_K^2$ to characterize our set of $[\eta/v, \omega_c/v]$. The contour plots of Fig.~\ref{fig:memeory-fit-example}--(c),(d) show that the first two cumulants have a straightforward, monotonic relationship with the Hamiltonian parameters. Crucially, the dependences of panels (c) and (d) are quantitatively different, which means that together they triangulate the $[\eta/v, \omega_c/v]$ which give rise to a particular combination $[\langle \omega \rangle_K, \sigma_K^2]$. We illustrate this protocol using the data of Fig.~\ref{fig:current_proj}, which are not included in the 16-point grid in Fig.~\ref{fig:memeory-fit-example}--(c),(d). Figure~\ref{fig:backmap} successfully extracts the correct location in parameter space using only the first two moments of $\mc{K}(\omega)$ computed from the measured $\mc{C}(t)$. To characterize a higher dimensional space (for example, as a function of temperature), one would use a simple generalization of this procedure considering higher moments. Therefore, this strategy provides a route to obtain corresponding dispersive Holstein parameters from experimental data, provided sufficient observational range and precision to sample the moments~\cite{ulbricht2011carrier, lloyd2012review, kuvzel2020terahertz}, with recent experiments reaching $\sim\! 75$ THz \cite{cooke2012direct, leitenstorfer20232023}.

In this paper, we employed a Mori-type GQME approach to determine transport coefficients in polaron-forming systems by using current as the primary member of the projection operator. Our approach can offer significant computational advantages of up to 90\% for regimes dominated by weak charge-lattice couplings and fast-decorrelating lattices (i.e., those dominated by coupling to high-frequency optical phonons). In addition, in contrast to previous GQME-based approaches that project onto the nonequilibrium site population dynamics and scale with the size of the system, our approach requires only one calculation for the construction of the GQME, rendering the construction of the memory kernel independent of system size. Furthermore, we introduced a simple protocol to incorporate derivatives of current into the projection operator, leading to additional computational savings that vary across parameter space. 

While we showcased the advantages of this methodology through quantum dynamics simulations of electronic transport in a 1-dimensional periodic Holstein chain, the Mori approach for capturing transport phenomena can be applied to diverse systems and is agnostic to the level of sophistication employed to describe the many-body interactions (physically transparent models, ab initio, semiempirical, empirical) and dynamics (path integrals, semiclassics, and classical dynamics) of the system. What is more, this Mori approach can offer significant computational savings and be easily adapted to a wide range of transport calculations~\cite{lin2017determining, baroni2020heat}.

Beyond computational benefits, our approach yields valuable insights into the experimental measurements of materials. For example, we demonstrated the inadequacy of fitting the Drude-Smith equation to our theoretically unambiguous dynamics of polaron-forming systems, and introduced a cumulant-based method to map experimentally accessible memory kernels (via conductance measurements) to parameters of the dispersive Holstein Hamiltonian. This link enables the interpretation of AC measurements of polaron-forming materials in terms of their underlying Hamiltonian, thereby facilitating material design modifications. While we presented a proof-of-principle example of this backmapping procedure using our theoretical dataset, a comprehensive investigation of backmapping with experimental data from real materials remains a focus for future research.

\section*{Acknowledgements}
\vspace{-4pt}

A.M.C. acknowledges the start-up funds from the University of Colorado Boulder for partial support of this research. Acknowledgment is made to the donors of the American Chemical Society Petroleum Research Fund for partial support of this research (No.~PRF 66836-DNI6). S.B. acknowledges the John Bailar Memorial Endowment for partial support of the research. We thank Prof.~Qiang Shi for sharing his HEOM code with us; Dr.~Matt Beard for helping us navigate the state-of-the-art in terahertz measurements; and Andrew Monaghan for his help in using the Alpine high-performance computing resources at the University of Colorado Boulder. Alpine is jointly funded by the University of Colorado Boulder, the University of Colorado Anschutz, Colorado State University, and the National Science Foundation (award 2201538).


\appendix

\section{Dispersive Holstein Model}\label{app:Holstein}
\vspace{-6pt}

The dispersive Holstein Hamiltonian can be written as
\begin{equation}
\begin{split}
    \hat{H} &= \sum_{i}^{N}\big( \epsilon_i + \hat{V}_{B, i}\big) \hat{a}_i^\dag \hat{a}_i + \sum_{\langle ij \rangle }^{N} v_{ij} \hat{a}_i^\dag \hat{a}_j  +  \sum_{i}^{N} \hat{H}_{B, i} , \label{ham-eq}
\end{split}
\end{equation}
\vspace{-10pt}
\begin{equation}
    \hat{H}_{B,i} = \frac{1}{2}\sum_{\alpha} \big[ \hat{P}_{i,\alpha}^2 +  \omega_{i\alpha}^2 \hat{X}_{i,\alpha}^2 \big], \label{eq:free-bath-ham}
\end{equation}
\vspace{-10pt}
\begin{equation}
    \hat{V}_{B,i} = \sum_{\alpha} c_{i, \alpha} \hat{X}_{i,\alpha} \label{eq:linear-coupling},
\end{equation}
where the fermions (electrons or holes whose creation and annihilation operators are $\hat{a}_i^\dag$ and $\hat{a}_j$) are assumed to be described by a tight binding system Hamiltonian parameterized by site energies $\epsilon_i$ and hopping integrals $v_{ij}$. Local nuclear motions are assumed to cause Gaussian fluctuations of the site energies, enabling one to write the local nuclear environment as a set of harmonic modes localized on site $i$ with momenta $\hat{P}_{\alpha}$, positions $\hat{X}_{\alpha}$, and frequencies $\omega_{\alpha}$. The coupling between the fermions and bosons associated with a given site is linear in the bosonic coordinates, with the coupling constants $c_{i,\alpha}$ given by the site's spectral density
\vspace{-6pt}
\begin{equation}\label{spec_dens_def1}
   \xi_i(\omega)= \dfrac{\pi }{2} \sum_{\alpha} \dfrac{c_{i,\alpha}^2} {\omega_{i\alpha}} \delta(\omega -\omega_{i\alpha}).
\end{equation}
The $\xi_i (\omega)$ for each site are assumed to be equivalent and take the Debye form commonly used to capture the dissipation in the condensed phase
\begin{equation}\label{spec_dens_def2}
    \xi(\omega)= \dfrac{ \eta \omega_c \omega}{\omega^2 + \omega_c^2}.
\end{equation}
Here, $\eta/2$ is the reorganization energy and $1/\omega_c$ is the timescale at which the phonon environment decorrelates. 
Since we focus on a purely homogeneous lattice, all parameters become site-independent, with $\epsilon_i = 0$. Different instances of the model are thus uniquely defined by the set $[\eta, \omega_c, v, \beta]$ which we dimensionalize to $[\eta / v, \omega_c / v]$ with $\beta$ fixed at 300~K throughout. We employ cyclical boundary conditions since our previous work~\cite{bhattacharyya2024anomalous} demonstrated that non-periodic models do not exhibit a well-defined DC mobility, except in the limit of an infinitely large system. 

\vspace{-10pt}
\section{Mori GQME}
\label{app:gmqe-details}
\vspace{-6pt}

To derive a GQME for the current autocorrelation function directly, we employ a Mori-type projector 
\begin{equation}
    \mathcal{P} = |\hat{J}) (\hat{J}|\hat{J})^{-1}(\hat{J}|.
\end{equation}
In contrast to previous uses of the Mori projector for quantum equilibrium time correlation functions~\cite{bosse1995self, reichman2001self, reichman2002self} we define the inner product to yield the direct correlation function rather than the Kubo-transformed counterpart,
\vspace{-10pt}
\begin{equation}
    (A|\mathcal{\hat{O}} |A) = \frac{1}{Z} \mathrm{Tr}[e^{-\beta H} A^{\dagger} \mathcal{\hat{O}} A],
\end{equation}
where $\mathcal{\hat{O}}$ is a superoperator, like the Liouvillian, $\mathcal{L} \equiv [H, \dots]$. This projector satisfies the conditions for its validity: idempotency $\mathcal{P}^2 = \mathcal{P}$ and orthogonality $\mathcal{P}\mathcal{Q} = 0$, where $\mathcal{Q} = 1 - \mathcal{P}$ is the complementary projector. Traditionally, one adopts the Kubo-transformed correlation function because of the natural orthogonality of the cross-correlation function of the dynamical variable in the projector, $\hat{J}$ in this case, and its derivative. When we augment the projector with the derivative of the current in Appendix ~\ref{app:derivatives}, we diagonalize the correlation matrix directly to ensure the idempotency of the projector.

This definition for the projector enables us to construct the current autocorrelation function, $C_{JJ}(t)$, as follows:
\vspace{-4pt}
\begin{equation}\label{mori-proj}
    \mathcal{C}(t) = (A| \e{\rmm{i}\mc{L}t} |A) = \frac{1}{Z} \mathrm{Tr}[e^{-\beta H} \hat{J} \e{\rmm{i}\mc{L}t} \hat{J}].
\end{equation}
We can immediately write down the GQME for this correlation function (for a detailed derivation of the GQME using this notation, see Refs.~\onlinecite{montoya2016approximate, kelly2016generalized, montoya2017approximate}),
\vspace{-6pt}
\begin{equation} \label{eq:MNZ_appendix}
    \dot{\mc{C}}(t) =  \dot{\mc{C}}(0) \mc{C}(t) - \int_0^t \dd{s}\mc{K}(s)\mc{C}(t-s)
\end{equation}
where the memory kernel takes the form
\begin{equation}\label{eq:memK}
    \mathcal{K}(t) = \frac{1}{Z} \text{Tr} \left[ \e{-\beta H} \hat{J} \mathcal{LQ} \e{\rmm{i}\mathcal{QL}t} \mathcal{QL} \hat{J} \right],
\end{equation}
where occurrence of complementary projector in the propagator, $\e{\rmm{i}\mathcal{QL}t}$, 
makes it difficult to obtain. To circumvent this difficulty, we adopt the self-consistent expansion of the memory kernel \cite{shi2003new, zhang2006nonequilibrium, montoya2016approximate, kelly2016generalized, montoya2017approximate} into auxiliary kernels $\mc{K}^{1}$ and $\mc{K}^{3}$\cite{montoya2016approximate}: 
\vspace{-6pt}
\begin{equation}\label{eq:aux-memK}
    \mathcal{K}(t) =  \mathcal{K}^{1}(t) + \int_0^t d\tau \: \mathcal{K}^{3}(t-\tau)\mathcal{K}(\tau).
\end{equation}
One can straightforwardly construct the auxiliary kernels using the derivatives of the correlation matrix $\mathcal{C}(t)$\cite{montoya2016approximate, kelly2016generalized, montoya2017approximate}, \begin{equation}\label{k1-def}
    \mathcal{K}^{1}(t) =  \ddot{\mc{C}}(t) - \{\dot{\mc{C}}(0), \dot{\mc{C}}(t)\} + \dot{\mc{C}}(0) \mc{C}(t) \dot{\mc{C}}(0),
\end{equation}
and 
\vspace{-10pt}
\begin{equation}\label{k3-def}
     \mathcal{K}^{3}(t) = \dot{\mc{C}}(0) \mc{C}(t) -  \dot{\mc{C}}(t).
\end{equation}
Thus, computing numerical derivatives $\dot{\mc{C}}(t)$ and $\ddot{\mc{C}}(t)$, we construct the auxiliary memory kernels. From the auxiliary kernels, one can construct the memory kernel $\mathcal{K}(t)$ using the algorithm presented in Appendix B of Ref.~\onlinecite{pfalzgraff2019efficient} based on the discrete reformulation of the convolution integral in the self-consistent expansion of the memory kernel, Eq.~\ref{eq:aux-memK}. To obtain the current autocorrelation function from knowledge of the memory kernel, we integrate the GQME in Eq.~\ref{eq:MNZ_appendix} employing a Heun integrator\cite{suli2003introduction}.

\vspace{-10pt}
\section{HEOM Details}
\label{app:HEOM}
\vspace{-6pt}

We employ HEOM to obtain the numerically exact dynamics of the dispersive Holstein model in the first excitation subspace, corresponding to the dilute limit of one electron or hole on the lattice. HEOM integrates the bosonic variables and predicts the reduced density matrix,
\vspace{-10pt}
\begin{equation}\label{eq:redfield}
    \mathcal{C}_{kl;ij}(t) = \text{Tr} \left[ \rho_{k,l}(0) \e{\rmm{i}\mathcal{L}t} \hat{a}^\dagger_i \hat{a}_j \right],
\end{equation}
subject to any spectroscopic (nonequilibrium) initial condition, $\rho_{k,l}(0) =  \hat{a}^\dagger_k\hat{a}_l e^{-\beta \hat{H}_b}/Z_b$, where $Z_b = \mathrm{Tr}[e^{-\beta \hat{H}_b}]$ is the partition function of the isolated bath.
 
By noting that the nonequilibrium approach to predicting the transport coefficient only requires the MSD of the polaron, $\mu = \frac{1}{2\mathrm{k_B} T} \lim_{t\rightarrow \infty} \frac{d{\rm MSD(t)}}{dt}$, it suffices to project onto the time-dependent site populations~\cite{sparpaglione1988dielectric, golosov2001reference, yan2019theoretical}. Thus, one needs to perform $\mathcal{O}(N)$ calculations corresponding to cases where $i=j$ and $k=l$ in Eq.~\ref{eq:redfield}. Because our system is fully symmetric, all such population-based starting positions are equivalent, meaning that one only needs to perform one simulation, but this simplification is not generally applicable. Indeed for a disordered system with varying site energies and hopping integrals, all positions are unique. A detailed investigation of the effects of static disorder will form the basis of future work.

To construct the population-based GQMEs in the main text we employ two distinct types of initial condition: Franck-Condon and Marcus-type. HEOM calculations, as described by Eq.~\ref{eq:redfield}, directly access Franck-Condon initial conditions. The Marcus initial condition corresponds to a change in the definition of $\e{-\beta \hat{H}_B} \rightarrow \e{-\beta \hat{H}^{(j)}_B}$, where the superscript denotes that the local bath Hamiltonian for the $j^\rmm{th}$ site is equilibrated with its excited state, $\hat{H}_B^{(j)} = \hat{H}_{B} + V_{B,j}$. To generate this initial condition a pre-equilibration run with all hopping integrals set to zero before a second simulation with finite hopping integrals, as outlined in Refs.~\onlinecite{yan2019theoretical, bhattacharyya2024anomalous}.

\begin{figure}[!t]
\vspace{-6pt}
\begin{center} 
    \resizebox{.4\textwidth}{!}{\includegraphics{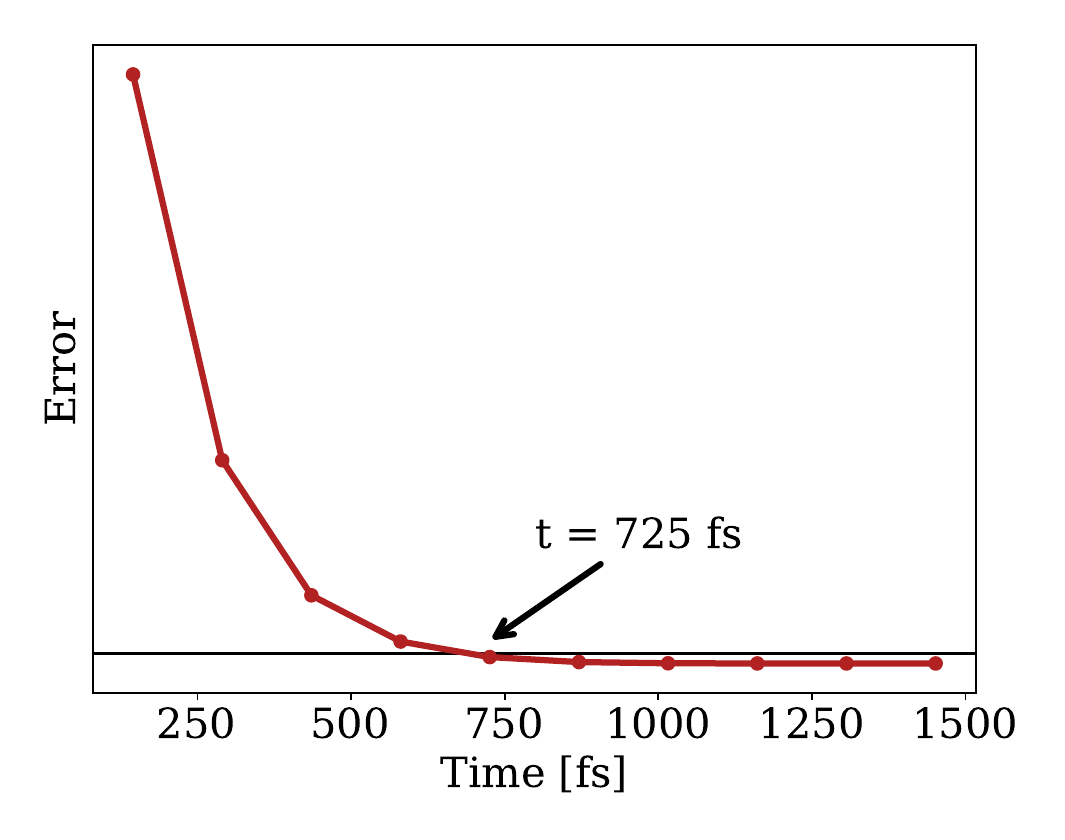}}
\end{center}
\vspace{-22pt}
\caption{\label{fig:mori-error} Error ($||\mathrm{L}||_2$ norm of the difference) between the current autocorrelation function computed with different equilibration times and 4000~fs equilibration time. The cut-off for error is chosen as $10^{-7}$ (black solid line) leading to $725$~fs as sufficient for preparing the initial condition for the equilibrium autocorrelation function. $\eta/v=6.46, \omega_c/v=0.82$, $v=50$ cm$^{-1}$, and ${\rm T} = 300$~K.}
\vspace{-10pt}
\end{figure}

Our methodology for computing equilibrium correlation functions using HEOM follows the protocol outlined in Refs.~\onlinecite{song2015new, bhattacharyya2024anomalous}. It consists of a pre-equilibration step that generates the canonical initial condition $e^{-\beta \hat{H}}/Z$, followed by right-side multiplication by the current operator, $\hat{J}$, followed by real time evolution. We ensured that our pre-equilibration step was successful by varying the time of the pre-equilibration and ensuring that the resulting correlation functions converged as a function of this parameter. By examining the error between the autocorrelation functions obtained with different initialization times and the selected best one, we determine that approximately 725 fs of simulation time are required for the initial canonical density preparation. This analysis is depicted in Fig.~\ref{fig:mori-error}. For our production runs, we used a pre-equilibration time of 4000~fs, which exceeded the minimum required pre-equilibration time in our tests. 

We converged all HEOM calculations with respect to the hierarchical depth $L$, number of Matsubara frequencies $K$, and timestep $\delta t$. For example, we set $L=22$, $K=1$, and $\delta t = 0.25$ fs for  Fig.~\ref{fig:current_proj}. To estimate the computational resource requirements (time and memory) in Table.~\ref{table:mu_compare}, we used the same settings. To compare results across parameter regimes, we employed $L=26$, $K=2$, and $\delta t=0.1$ fs, which was sufficient to converge the most difficult parameter regime: $[\eta/v, \omega_c/v] = [400/50, 150/50]$. We also employed dynamic filtering parameters in our simulations, setting $\delta = 10^{-7}$ atomic units for the nonequilibrium simulations and canonical density preparation. For the second real time propagation in the calculation of equilibrium correlation functions, we set $\delta = 10^{-10}$ atomic units.

To construct the Kubo-transformed correlation function $C^{\rm Kubo}_{JJ}(t)$ from $C_{JJ}(t)$, we leverage their connection in the frequency domain: $C^{\rm Kubo}_{JJ}(\omega) = \frac{1-e^{-\beta \omega}}{\beta \omega} C_{JJ}(\omega)$. The factor $e^{-\beta \omega}$ behaves well for $\omega > 0$ but diverges for $\omega < 0$, complicating its use in this second region. However, since $C^{\rm Kubo}_{JJ}(\omega) = C^{\rm Kubo}_{JJ}(-\omega)$, we mirror this function around $\omega=0$, taking the $\omega \geq 0$ region as the generator of the $\omega \leq 0$ part. Without imposing this symmetry, the inverse transform of $C^{\rm Kubo}_{JJ}(\omega)$ to $C^{\rm Kubo}_{JJ}(t)$ is prone to numerical artifacts. We generate all instances of $C^{\rm Kubo}_{JJ}(t)$ using this approach.

\vspace{-10pt}
\section{Augmenting the equilibrium projector with dynamical derivatives}
\label{app:derivatives}
\vspace{-6pt}

Adding derivatives of the motion to the projector can offer an easy strategy to obtain an even swifter evaluation of the transport coefficient, i.e., a faster-decaying memory kernel. Here we augment our Mori projector of Eq.~\ref{mori-proj} with the time derivative of the current, 
\vspace{-4pt}
\begin{equation}
\vspace{-10pt}
\begin{split} 
    \hat{\zeta} &\equiv i[\hat{H},\hat{J}]\\
    &= d \sum_{\langle mn \rangle} v_{mn} (\hat{a}_m^\dag \hat{a}_n -\hat{a}_n^\dag \hat{a}_m)  \\
    & \qquad \times \Big(\epsilon_m - \epsilon_n + \sum_\alpha c_{m,\alpha}\hat{X}_{m,\alpha} - c_{n,\alpha}\hat{X}_{n,\alpha}\Big),
\end{split}
\end{equation}
which includes the coordinates of the nuclear modes $\hat{X}_{i,\alpha}$. 

\begin{figure}[!t]
\vspace{-6pt}
\begin{center} 
    \resizebox{.4\textwidth}{!}{\includegraphics{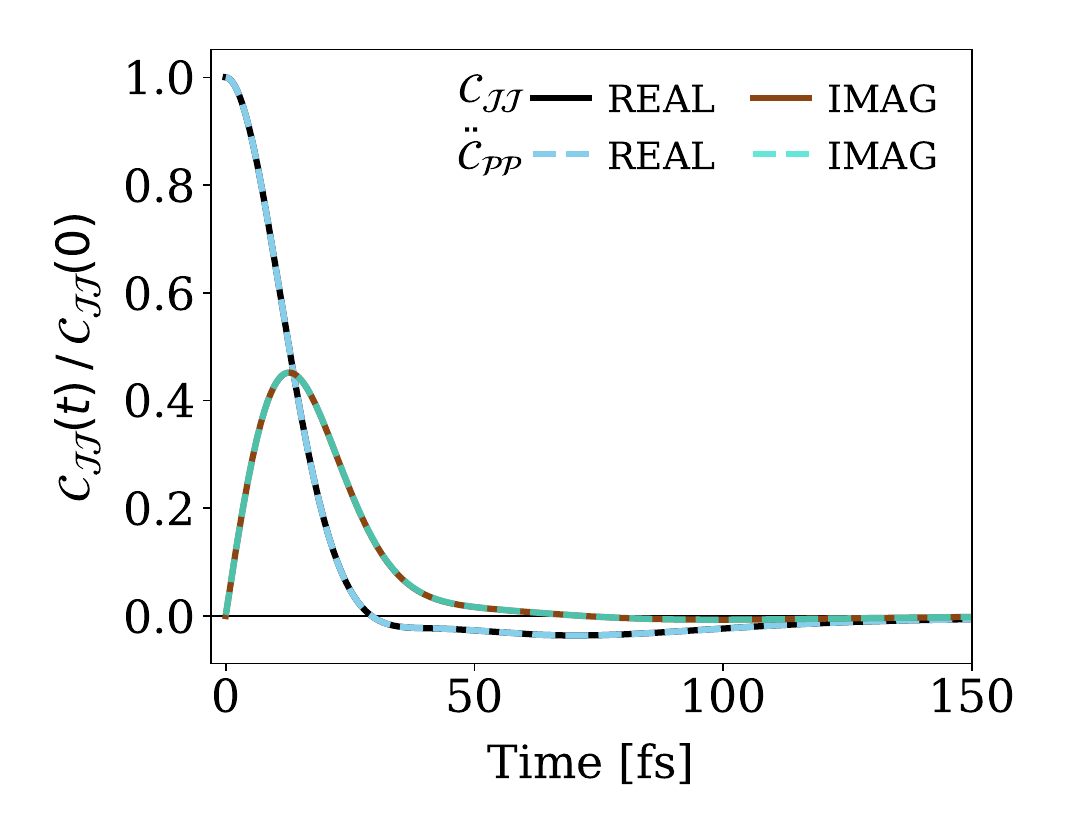}}
\end{center}
\vspace{-24pt}
\caption{\label{fig:derivative-validate} The agreement of real and imaginary parts between the negative of the numerical double derivative of $C_{PP}(t)$ i.e. $-\ddot{C}_{P P} (t)$ and $C_{JJ} (t)$ in the linear topology for $\eta/v=6.46, \omega_c/v=0.81$, $v=50$ cm$^{-1}$, and ${\rm T} = 300$~K.}
\vspace{-10pt}
\end{figure}

Since HEOM integrates over the bosonic environment, one cannot \textit{directly} measure $\zeta$. However, one can employ finite difference on $C_{JJ}$ to obtain a numerical approximation to $\dot{C}_{JJ}(t) = C_{J \zeta}(t) = -C_{\zeta J}(t)$, and, upon applying the finite difference derivative a second time, $\ddot{C}_{JJ}(t) = -C_{\zeta \zeta}(t)$. To confirm that this approximation is sufficiently accurate, we employ HEOM to calculate the polarization autocorrelation function, $C_{P P}(t)$, and compare its numerical time derivatives with the direct HEOM calculation of $C_{PJ}(t)$, $C_{JP}(t)$ and $C_{JJ}(t)$. We employ a linear (open-chain) topology for the Holstein model for this test as it circumvents the ambiguity in the definition of polarization in a periodic topology. As Fig.~\ref{fig:derivative-validate} shows, we obtain numerical agreement between $- \ddot{C}_{P P}(t)$ and $C_{JJ}(t)$. Hence, the time resolution in our HEOM simulations enables us to use numerical time derivatives to augment the projector numerically. 

Augmenting the projection with the derivatives of the current results, naively, in the following $2\times 2$ matrix,
\begin{equation} \label{correlation:derivatives}
    \mathcal{\tilde{C}}(t) = \left( \begin{matrix}
        C_{JJ}(t) & C_{J \zeta}(t) \\
        C_{\zeta J}(t) & C_{\zeta \zeta}(t)
    \end{matrix} \right).
\end{equation}
However, $\mathcal{\tilde{C}}(t = 0) \neq \mathbb{1}$, indicating that the projector is not idempotent. To fix this problem, we multiply the dynamical matrix by its inverse at $t=0$,
\begin{equation}\label{correlation:derivatives-unitary}
      \mathcal{C}(t) =   \mathcal{\tilde{C}}(t)   \mathcal{\tilde{C}}(0)^{-1},
\end{equation}
ensuring $\mathcal{C}(t = 0)= \mathbb{1}$. We can then use Eq.~\ref{eq:MNZ_appendix} to compute the memory kernel. As we show in the main text, augmenting the projector with derivatives of the original dynamical variable can shorten the memory kernel lifetime and result in significant computational savings. 

\vspace{-10pt}
\section{Current and memory kernel fitting}
\label{app:cosine-fit}
\vspace{-6pt}

The widely invoked Drude-Smith model offers a two-parameter analytical approximation for the current autocorrelation function and interprets conductance measurements. The MNZ approach to the current autocorrelation function offers an exact means to obtain its memory kernel. Hence, we employ our Mori GQME to assess the validity of the Drude-Smith approach. 

\begin{table}[!b]
    \centering
    \vspace{-0pt}
    \hspace{-3pt}
    \begin{tabular}{|c|c|c|c|c|c|} 
    \hline
     $i^\mathrm{th}$ term & $a_i$ & $\omega_i$ [fs${}^{-1}$]& $k_i$ [fs${}^{-1}$]\\
     \hline
     1 & 0.7386 & 0.0169 & 0.0143 \\
     \hline
      2 &  0.2614 & 0.0359 & 0.0147 \\ \hline
    \end{tabular} 
    \vspace{-4pt}
    \caption{\label{table:current_fit} Fit parameters for $C_{JJ}^\mathrm{Kubo}(t)$ in Fig.~\ref{fig:cosine-fit} for $\eta/v=2, \omega_c/v=0.5$, $v=50$ cm$^{-1}$, and ${\rm T} = 300$~K. }
    \vspace{-0pt}
\end{table}

\begin{figure}[!t]
\vspace{-6pt}
\begin{center} 
    \resizebox{.4\textwidth}{!}{\includegraphics{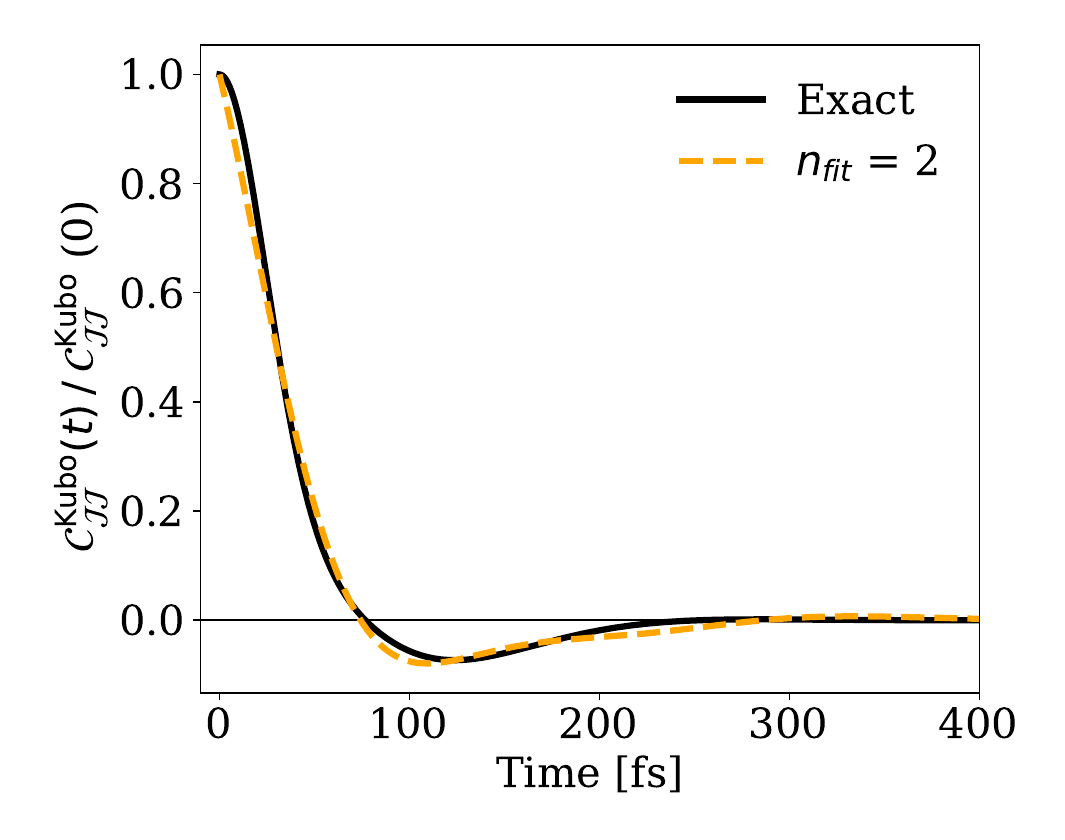}}
\end{center}
\vspace{-24pt}
\caption{\label{fig:cosine-fit} Two damped cosine fit (orange) according to Eq.~\ref{eq:minimalCJJ} to the $C_{JJ}^\mathrm{Kubo}$ obtained with HEOM (black) for $\eta/v=2, \omega_c/v=0.5$, $v=50$ cm$^{-1}$, and ${\rm T} = 300$~K. The 5-parameters fit is not sufficient to capture the  $C_{JJ}^\mathrm{Kubo}$. }
\vspace{-4pt}
\end{figure}

In Laplace space, Eq.~\ref{eq:MNZ_intro} takes the form~\cite{forster2018hydrodynamic},
\vspace{-4pt}
\begin{equation}\label{eq:MoriEq}
    \sigma (\omega) = \frac{1}{\mathcal{K}^\mathrm{Kubo}(\omega)+\rmm{i}\omega}.
\end{equation}
Ref~\onlinecite{chen2021drude} has shown that the Drude-Smith form can arise from setting the memory kernel to ${\mathcal{K}^\mathrm{Kubo}(t)=q\exp{(-rt)}}$, moving to Laplace space, solving for the roots that lead to a singularity in Eq.~\ref{eq:MoriEq}, and considering \textit{only} the case where the two poles are degenerate, i.e., $r=2\sqrt{q}$. A general solution would consider two distinct, complex poles $2\omega_\pm = r \pm \sqrt{r^2 - 4q}$ which give two different decay terms after an inverse Laplace transform, 
\begin{equation}\label{eq:drudeTwoExpo}
    C^\mathrm{Kubo}_{JJ}(t) / C^\mathrm{Kubo}_{JJ}(0) = \frac{(\omega_+ - r) \e{-\omega_+ t} + (r - \omega_-)\e{-\omega_- t}}{(\omega_+ - \omega_-)}.
\end{equation}
This motivates employing a more flexible form to fit our exact current autocorrelation functions,
\begin{equation}\label{eq:minimalCJJ}
    C^\rmm{Kubo}_{JJ}(t)/C^\rmm{Kubo}_{JJ}(0) = \sum_i^{n_\rmm{fit}} a_i \cos (\omega_i t) \e{-k_i t}.
\end{equation}
With 5 parameters ($n_\rmm{fit}=2$ and $a_1+a_2 = 1$ is a constraint) we are guaranteed to obtain a better fit, but one can observe from Fig.~\ref{fig:cosine-fit} that the oscillatory frequencies are qualitatively wrong. By extension, Eq.~\ref{eq:drudeTwoExpo} cannot fit the data. We must conclude that $\mathcal{K}^\mathrm{Kubo}(t)$ is not well-described by a single exponential. Further, a single exponential in time would yield a Lorentzian in frequency, but our simplest $\tilde{\mc{K}}(\omega)$ is instead a Gaussian. Indeed, using just decaying exponentials in time, $\mathcal{K}^\mathrm{Kubo}(t) = \sum_i \alpha_i \exp (-\kappa_i t)$, we find that using even five terms is insufficient to describe the data of Fig.~\ref{fig:drude-fit-example}.

We fit the $C_{JJ}^\mathrm{Kubo} (t)$ using Eq.~\ref{eq:minimalCJJ} for $16$ points in the parameter space with $n_\rmm{fit} = 2$. The fits are closer than the Drude-Smith form, but are clearly qualitatively lacking. Figure~\ref{fig:cosine-fit} shows an example where $[\eta/v, \omega_c/v] = [100/50, 25/50]$. Table.~\ref{table:current_fit} summarizes the corresponding fit parameters. Unlike the Drude-Smith form, neither term has a negative coefficient. Our analysis of this Drude-Smith fitting protocol suggests that decomposing the current autocorrelation function reporting on small polaron transport into decaying and oscillatory exponentials can lead to significant ambiguities, motivating our cumulant-based analysis in the main text. 

\vfill
\pagebreak
\section*{References}
\vspace{-14pt}
\bibliography{ref}

\begin{thebibliography}{91}%
\makeatletter
\providecommand \@ifxundefined [1]{%
 \@ifx{#1\undefined}
}%
\providecommand \@ifnum [1]{%
 \ifnum #1\expandafter \@firstoftwo
 \else \expandafter \@secondoftwo
 \fi
}%
\providecommand \@ifx [1]{%
 \ifx #1\expandafter \@firstoftwo
 \else \expandafter \@secondoftwo
 \fi
}%
\providecommand \natexlab [1]{#1}%
\providecommand \enquote  [1]{``#1''}%
\providecommand \bibnamefont  [1]{#1}%
\providecommand \bibfnamefont [1]{#1}%
\providecommand \citenamefont [1]{#1}%
\providecommand \href@noop [0]{\@secondoftwo}%
\providecommand \href [0]{\begingroup \@sanitize@url \@href}%
\providecommand \@href[1]{\@@startlink{#1}\@@href}%
\providecommand \@@href[1]{\endgroup#1\@@endlink}%
\providecommand \@sanitize@url [0]{\catcode `\\12\catcode `\$12\catcode `\&12\catcode `\#12\catcode `\^12\catcode `\_12\catcode `\%12\relax}%
\providecommand \@@startlink[1]{}%
\providecommand \@@endlink[0]{}%
\providecommand \url  [0]{\begingroup\@sanitize@url \@url }%
\providecommand \@url [1]{\endgroup\@href {#1}{\urlprefix }}%
\providecommand \urlprefix  [0]{URL }%
\providecommand \Eprint [0]{\href }%
\providecommand \doibase [0]{http://dx.doi.org/}%
\providecommand \selectlanguage [0]{\@gobble}%
\providecommand \bibinfo  [0]{\@secondoftwo}%
\providecommand \bibfield  [0]{\@secondoftwo}%
\providecommand \translation [1]{[#1]}%
\providecommand \BibitemOpen [0]{}%
\providecommand \bibitemStop [0]{}%
\providecommand \bibitemNoStop [0]{.\EOS\space}%
\providecommand \EOS [0]{\spacefactor3000\relax}%
\providecommand \BibitemShut  [1]{\csname bibitem#1\endcsname}%
\let\auto@bib@innerbib\@empty
\bibitem [{\citenamefont {Fratini}\ \emph {et~al.}(2020)\citenamefont {Fratini}, \citenamefont {Nikolka}, \citenamefont {Salleo}, \citenamefont {Schweicher},\ and\ \citenamefont {Sirringhaus}}]{fratini2020charge}%
  \BibitemOpen
  \bibfield  {author} {\bibinfo {author} {\bibfnamefont {S.}~\bibnamefont {Fratini}}, \bibinfo {author} {\bibfnamefont {M.}~\bibnamefont {Nikolka}}, \bibinfo {author} {\bibfnamefont {A.}~\bibnamefont {Salleo}}, \bibinfo {author} {\bibfnamefont {G.}~\bibnamefont {Schweicher}}, \ and\ \bibinfo {author} {\bibfnamefont {H.}~\bibnamefont {Sirringhaus}},\ }\bibfield  {title} {\enquote {\bibinfo {title} {Charge transport in high-mobility conjugated polymers and molecular semiconductors},}\ }\href@noop {} {\bibfield  {journal} {\bibinfo  {journal} {Nature Materials}\ }\textbf {\bibinfo {volume} {19}},\ \bibinfo {pages} {491--502} (\bibinfo {year} {2020})}\BibitemShut {NoStop}%
\bibitem [{\citenamefont {Nematiaram}\ \emph {et~al.}(2020)\citenamefont {Nematiaram}, \citenamefont {Padula}, \citenamefont {Landi},\ and\ \citenamefont {Troisi}}]{nematiaram2020largest}%
  \BibitemOpen
  \bibfield  {author} {\bibinfo {author} {\bibfnamefont {T.}~\bibnamefont {Nematiaram}}, \bibinfo {author} {\bibfnamefont {D.}~\bibnamefont {Padula}}, \bibinfo {author} {\bibfnamefont {A.}~\bibnamefont {Landi}}, \ and\ \bibinfo {author} {\bibfnamefont {A.}~\bibnamefont {Troisi}},\ }\bibfield  {title} {\enquote {\bibinfo {title} {On the largest possible mobility of molecular semiconductors and how to achieve it},}\ }\href@noop {} {\bibfield  {journal} {\bibinfo  {journal} {Advanced Functional Materials}\ }\textbf {\bibinfo {volume} {30}},\ \bibinfo {pages} {2001906} (\bibinfo {year} {2020})}\BibitemShut {NoStop}%
\bibitem [{\citenamefont {Leggett}\ \emph {et~al.}(1987)\citenamefont {Leggett}, \citenamefont {Chakravarty}, \citenamefont {Dorsey}, \citenamefont {Fisher}, \citenamefont {Garg},\ and\ \citenamefont {Zwerger}}]{leggett1987dynamics}%
  \BibitemOpen
  \bibfield  {author} {\bibinfo {author} {\bibfnamefont {A.~J.}\ \bibnamefont {Leggett}}, \bibinfo {author} {\bibfnamefont {S.}~\bibnamefont {Chakravarty}}, \bibinfo {author} {\bibfnamefont {A.~T.}\ \bibnamefont {Dorsey}}, \bibinfo {author} {\bibfnamefont {M.~P.}\ \bibnamefont {Fisher}}, \bibinfo {author} {\bibfnamefont {A.}~\bibnamefont {Garg}}, \ and\ \bibinfo {author} {\bibfnamefont {W.}~\bibnamefont {Zwerger}},\ }\bibfield  {title} {\enquote {\bibinfo {title} {Dynamics of the dissipative two-state system},}\ }\href@noop {} {\bibfield  {journal} {\bibinfo  {journal} {Reviews of Modern Physics}\ }\textbf {\bibinfo {volume} {59}},\ \bibinfo {pages} {1} (\bibinfo {year} {1987})}\BibitemShut {NoStop}%
\bibitem [{\citenamefont {Weiss}(2012)}]{weiss2012quantum}%
  \BibitemOpen
  \bibfield  {author} {\bibinfo {author} {\bibfnamefont {U.}~\bibnamefont {Weiss}},\ }\href@noop {} {\emph {\bibinfo {title} {Quantum dissipative systems}}}\ (\bibinfo  {publisher} {World Scientific},\ \bibinfo {year} {2012})\BibitemShut {NoStop}%
\bibitem [{\citenamefont {Holstein}(1959{\natexlab{a}})}]{holstein1959studies2}%
  \BibitemOpen
  \bibfield  {author} {\bibinfo {author} {\bibfnamefont {T.}~\bibnamefont {Holstein}},\ }\bibfield  {title} {\enquote {\bibinfo {title} {Studies of polaron motion: Part i. the molecular-crystal model},}\ }\href@noop {} {\bibfield  {journal} {\bibinfo  {journal} {Annals of Physics}\ }\textbf {\bibinfo {volume} {8}},\ \bibinfo {pages} {325--342} (\bibinfo {year} {1959}{\natexlab{a}})}\BibitemShut {NoStop}%
\bibitem [{\citenamefont {Holstein}(1959{\natexlab{b}})}]{holstein1959studies}%
  \BibitemOpen
  \bibfield  {author} {\bibinfo {author} {\bibfnamefont {T.}~\bibnamefont {Holstein}},\ }\bibfield  {title} {\enquote {\bibinfo {title} {Studies of polaron motion: Part ii. the “small” polaron},}\ }\href@noop {} {\bibfield  {journal} {\bibinfo  {journal} {Annals of Physics}\ }\textbf {\bibinfo {volume} {8}},\ \bibinfo {pages} {343--389} (\bibinfo {year} {1959}{\natexlab{b}})}\BibitemShut {NoStop}%
\bibitem [{\citenamefont {Anderson}(1961)}]{anderson1961localized}%
  \BibitemOpen
  \bibfield  {author} {\bibinfo {author} {\bibfnamefont {P.~W.}\ \bibnamefont {Anderson}},\ }\bibfield  {title} {\enquote {\bibinfo {title} {Localized magnetic states in metals},}\ }\href@noop {} {\bibfield  {journal} {\bibinfo  {journal} {Physical Review}\ }\textbf {\bibinfo {volume} {124}},\ \bibinfo {pages} {41} (\bibinfo {year} {1961})}\BibitemShut {NoStop}%
\bibitem [{\citenamefont {Hubbard}(1963)}]{hubbard1963electron}%
  \BibitemOpen
  \bibfield  {author} {\bibinfo {author} {\bibfnamefont {J.}~\bibnamefont {Hubbard}},\ }\bibfield  {title} {\enquote {\bibinfo {title} {Electron correlations in narrow energy bands},}\ }\href@noop {} {\bibfield  {journal} {\bibinfo  {journal} {Proceedings of the Royal Society of London. Series A. Mathematical and Physical Sciences}\ }\textbf {\bibinfo {volume} {276}},\ \bibinfo {pages} {238--257} (\bibinfo {year} {1963})}\BibitemShut {NoStop}%
\bibitem [{\citenamefont {Hubbard}(1964)}]{hubbard1964electron}%
  \BibitemOpen
  \bibfield  {author} {\bibinfo {author} {\bibfnamefont {J.}~\bibnamefont {Hubbard}},\ }\bibfield  {title} {\enquote {\bibinfo {title} {Electron correlations in narrow energy bands. ii. the degenerate band case},}\ }\href@noop {} {\bibfield  {journal} {\bibinfo  {journal} {Proceedings of the Royal Society of London. Series A. Mathematical and Physical Sciences}\ }\textbf {\bibinfo {volume} {277}},\ \bibinfo {pages} {237--259} (\bibinfo {year} {1964})}\BibitemShut {NoStop}%
\bibitem [{\citenamefont {Bader}, \citenamefont {Kuharski},\ and\ \citenamefont {Chandler}(1990)}]{bader1990role}%
  \BibitemOpen
  \bibfield  {author} {\bibinfo {author} {\bibfnamefont {J.}~\bibnamefont {Bader}}, \bibinfo {author} {\bibfnamefont {R.}~\bibnamefont {Kuharski}}, \ and\ \bibinfo {author} {\bibfnamefont {D.}~\bibnamefont {Chandler}},\ }\bibfield  {title} {\enquote {\bibinfo {title} {Role of nuclear tunneling in aqueous ferrous--ferric electron transfer},}\ }\href@noop {} {\bibfield  {journal} {\bibinfo  {journal} {Journal of Chemical Physics}\ }\textbf {\bibinfo {volume} {93}},\ \bibinfo {pages} {230--236} (\bibinfo {year} {1990})}\BibitemShut {NoStop}%
\bibitem [{\citenamefont {Song}\ and\ \citenamefont {Marcus}(1993)}]{song1993quantum}%
  \BibitemOpen
  \bibfield  {author} {\bibinfo {author} {\bibfnamefont {X.}~\bibnamefont {Song}}\ and\ \bibinfo {author} {\bibfnamefont {R.}~\bibnamefont {Marcus}},\ }\bibfield  {title} {\enquote {\bibinfo {title} {Quantum correction for electron transfer rates. comparison of polarizable versus nonpolarizable descriptions of solvent},}\ }\href@noop {} {\bibfield  {journal} {\bibinfo  {journal} {Journal of Chemical Physics}\ }\textbf {\bibinfo {volume} {99}},\ \bibinfo {pages} {7768--7773} (\bibinfo {year} {1993})}\BibitemShut {NoStop}%
\bibitem [{\citenamefont {Ghosh}\ and\ \citenamefont {Spano}(2020)}]{ghosh2020excitons}%
  \BibitemOpen
  \bibfield  {author} {\bibinfo {author} {\bibfnamefont {R.}~\bibnamefont {Ghosh}}\ and\ \bibinfo {author} {\bibfnamefont {F.~C.}\ \bibnamefont {Spano}},\ }\bibfield  {title} {\enquote {\bibinfo {title} {Excitons and polarons in organic materials},}\ }\href@noop {} {\bibfield  {journal} {\bibinfo  {journal} {Accounts of Chemical Research}\ }\textbf {\bibinfo {volume} {53}},\ \bibinfo {pages} {2201--2211} (\bibinfo {year} {2020})}\BibitemShut {NoStop}%
\bibitem [{\citenamefont {Fetherolf}, \citenamefont {Gole{\v{z}}},\ and\ \citenamefont {Berkelbach}(2020)}]{fetherolf2020unification}%
  \BibitemOpen
  \bibfield  {author} {\bibinfo {author} {\bibfnamefont {J.~H.}\ \bibnamefont {Fetherolf}}, \bibinfo {author} {\bibfnamefont {D.}~\bibnamefont {Gole{\v{z}}}}, \ and\ \bibinfo {author} {\bibfnamefont {T.~C.}\ \bibnamefont {Berkelbach}},\ }\bibfield  {title} {\enquote {\bibinfo {title} {A unification of the holstein polaron and dynamic disorder pictures of charge transport in organic crystals},}\ }\href@noop {} {\bibfield  {journal} {\bibinfo  {journal} {Physical Review X}\ }\textbf {\bibinfo {volume} {10}},\ \bibinfo {pages} {021062} (\bibinfo {year} {2020})}\BibitemShut {NoStop}%
\bibitem [{\citenamefont {Segal}\ and\ \citenamefont {Agarwalla}(2016)}]{segal2016vibrational}%
  \BibitemOpen
  \bibfield  {author} {\bibinfo {author} {\bibfnamefont {D.}~\bibnamefont {Segal}}\ and\ \bibinfo {author} {\bibfnamefont {B.~K.}\ \bibnamefont {Agarwalla}},\ }\bibfield  {title} {\enquote {\bibinfo {title} {Vibrational heat transport in molecular junctions},}\ }\href@noop {} {\bibfield  {journal} {\bibinfo  {journal} {Annual Review of Physical Chemistry}\ }\textbf {\bibinfo {volume} {67}},\ \bibinfo {pages} {185--209} (\bibinfo {year} {2016})}\BibitemShut {NoStop}%
\bibitem [{\citenamefont {Thoss}\ and\ \citenamefont {Evers}(2018)}]{thoss2018perspective}%
  \BibitemOpen
  \bibfield  {author} {\bibinfo {author} {\bibfnamefont {M.}~\bibnamefont {Thoss}}\ and\ \bibinfo {author} {\bibfnamefont {F.}~\bibnamefont {Evers}},\ }\bibfield  {title} {\enquote {\bibinfo {title} {Perspective: Theory of quantum transport in molecular junctions},}\ }\href@noop {} {\bibfield  {journal} {\bibinfo  {journal} {Journal of Chemical Physics}\ }\textbf {\bibinfo {volume} {148}},\ \bibinfo {pages} {030901} (\bibinfo {year} {2018})}\BibitemShut {NoStop}%
\bibitem [{\citenamefont {Evers}\ \emph {et~al.}(2020)\citenamefont {Evers}, \citenamefont {Koryt{\'a}r}, \citenamefont {Tewari},\ and\ \citenamefont {Van~Ruitenbeek}}]{evers2020advances}%
  \BibitemOpen
  \bibfield  {author} {\bibinfo {author} {\bibfnamefont {F.}~\bibnamefont {Evers}}, \bibinfo {author} {\bibfnamefont {R.}~\bibnamefont {Koryt{\'a}r}}, \bibinfo {author} {\bibfnamefont {S.}~\bibnamefont {Tewari}}, \ and\ \bibinfo {author} {\bibfnamefont {J.~M.}\ \bibnamefont {Van~Ruitenbeek}},\ }\bibfield  {title} {\enquote {\bibinfo {title} {Advances and challenges in single-molecule electron transport},}\ }\href@noop {} {\bibfield  {journal} {\bibinfo  {journal} {Reviews of Modern Physics}\ }\textbf {\bibinfo {volume} {92}},\ \bibinfo {pages} {035001} (\bibinfo {year} {2020})}\BibitemShut {NoStop}%
\bibitem [{\citenamefont {Ishizuka}\ and\ \citenamefont {Yanase}(2021)}]{ishizuka2021periodic}%
  \BibitemOpen
  \bibfield  {author} {\bibinfo {author} {\bibfnamefont {J.}~\bibnamefont {Ishizuka}}\ and\ \bibinfo {author} {\bibfnamefont {Y.}~\bibnamefont {Yanase}},\ }\bibfield  {title} {\enquote {\bibinfo {title} {Periodic anderson model for magnetism and superconductivity in ute 2},}\ }\href@noop {} {\bibfield  {journal} {\bibinfo  {journal} {Physical Review B}\ }\textbf {\bibinfo {volume} {103}},\ \bibinfo {pages} {094504} (\bibinfo {year} {2021})}\BibitemShut {NoStop}%
\bibitem [{\citenamefont {Mielke}\ and\ \citenamefont {Tasaki}(1993)}]{mielke1993ferromagnetism}%
  \BibitemOpen
  \bibfield  {author} {\bibinfo {author} {\bibfnamefont {A.}~\bibnamefont {Mielke}}\ and\ \bibinfo {author} {\bibfnamefont {H.}~\bibnamefont {Tasaki}},\ }\bibfield  {title} {\enquote {\bibinfo {title} {Ferromagnetism in the hubbard model: Examples from models with degenerate single-electron ground states},}\ }\href@noop {} {\bibfield  {journal} {\bibinfo  {journal} {Communications in Mathematical Physics}\ }\textbf {\bibinfo {volume} {158}},\ \bibinfo {pages} {341--371} (\bibinfo {year} {1993})}\BibitemShut {NoStop}%
\bibitem [{\citenamefont {Jiang}\ and\ \citenamefont {Kivelson}(2022)}]{jiang2022stripe}%
  \BibitemOpen
  \bibfield  {author} {\bibinfo {author} {\bibfnamefont {H.-C.}\ \bibnamefont {Jiang}}\ and\ \bibinfo {author} {\bibfnamefont {S.~A.}\ \bibnamefont {Kivelson}},\ }\bibfield  {title} {\enquote {\bibinfo {title} {Stripe order enhanced superconductivity in the hubbard model},}\ }\href@noop {} {\bibfield  {journal} {\bibinfo  {journal} {Proceedings of the National Academy of Sciences}\ }\textbf {\bibinfo {volume} {119}},\ \bibinfo {pages} {e2109406119} (\bibinfo {year} {2022})}\BibitemShut {NoStop}%
\bibitem [{\citenamefont {Nosarzewski}\ \emph {et~al.}(2021)\citenamefont {Nosarzewski}, \citenamefont {Huang}, \citenamefont {Dee}, \citenamefont {Esterlis}, \citenamefont {Moritz}, \citenamefont {Kivelson}, \citenamefont {Johnston},\ and\ \citenamefont {Devereaux}}]{nosarzewski2021superconductivity}%
  \BibitemOpen
  \bibfield  {author} {\bibinfo {author} {\bibfnamefont {B.}~\bibnamefont {Nosarzewski}}, \bibinfo {author} {\bibfnamefont {E.}~\bibnamefont {Huang}}, \bibinfo {author} {\bibfnamefont {P.~M.}\ \bibnamefont {Dee}}, \bibinfo {author} {\bibfnamefont {I.}~\bibnamefont {Esterlis}}, \bibinfo {author} {\bibfnamefont {B.}~\bibnamefont {Moritz}}, \bibinfo {author} {\bibfnamefont {S.}~\bibnamefont {Kivelson}}, \bibinfo {author} {\bibfnamefont {S.}~\bibnamefont {Johnston}}, \ and\ \bibinfo {author} {\bibfnamefont {T.}~\bibnamefont {Devereaux}},\ }\bibfield  {title} {\enquote {\bibinfo {title} {Superconductivity, charge density waves, and bipolarons in the holstein model},}\ }\href@noop {} {\bibfield  {journal} {\bibinfo  {journal} {Physical Review B}\ }\textbf {\bibinfo {volume} {103}},\ \bibinfo {pages} {235156} (\bibinfo {year} {2021})}\BibitemShut {NoStop}%
\bibitem [{\citenamefont {Grabert}(2006)}]{grabert2006projection}%
  \BibitemOpen
  \bibfield  {author} {\bibinfo {author} {\bibfnamefont {H.}~\bibnamefont {Grabert}},\ }\href@noop {} {\emph {\bibinfo {title} {Projection operator techniques in nonequilibrium statistical mechanics}}},\ Vol.~\bibinfo {volume} {95}\ (\bibinfo  {publisher} {Springer},\ \bibinfo {year} {2006})\BibitemShut {NoStop}%
\bibitem [{\citenamefont {Fick}, \citenamefont {Sauermann},\ and\ \citenamefont {Brewer}(1990)}]{fick1990quantum}%
  \BibitemOpen
  \bibfield  {author} {\bibinfo {author} {\bibfnamefont {E.}~\bibnamefont {Fick}}, \bibinfo {author} {\bibfnamefont {G.}~\bibnamefont {Sauermann}}, \ and\ \bibinfo {author} {\bibfnamefont {W.~D.}\ \bibnamefont {Brewer}},\ }\href@noop {} {\emph {\bibinfo {title} {The quantum statistics of dynamic processes}}},\ Vol.~\bibinfo {volume} {86}\ (\bibinfo  {publisher} {Springer},\ \bibinfo {year} {1990})\BibitemShut {NoStop}%
\bibitem [{\citenamefont {Nakajima}(1958)}]{nakajima1958quantum}%
  \BibitemOpen
  \bibfield  {author} {\bibinfo {author} {\bibfnamefont {S.}~\bibnamefont {Nakajima}},\ }\bibfield  {title} {\enquote {\bibinfo {title} {On quantum theory of transport phenomena: Steady diffusion},}\ }\href@noop {} {\bibfield  {journal} {\bibinfo  {journal} {Progress of Theoretical Physics}\ }\textbf {\bibinfo {volume} {20}},\ \bibinfo {pages} {948--959} (\bibinfo {year} {1958})}\BibitemShut {NoStop}%
\bibitem [{\citenamefont {Zwanzig}(1960)}]{zwanzig1960ensemble}%
  \BibitemOpen
  \bibfield  {author} {\bibinfo {author} {\bibfnamefont {R.}~\bibnamefont {Zwanzig}},\ }\bibfield  {title} {\enquote {\bibinfo {title} {Ensemble method in the theory of irreversibility},}\ }\href@noop {} {\bibfield  {journal} {\bibinfo  {journal} {Journal of Chemical Physics}\ }\textbf {\bibinfo {volume} {33}},\ \bibinfo {pages} {1338--1341} (\bibinfo {year} {1960})}\BibitemShut {NoStop}%
\bibitem [{\citenamefont {Mori}(1965)}]{mori1965transport}%
  \BibitemOpen
  \bibfield  {author} {\bibinfo {author} {\bibfnamefont {H.}~\bibnamefont {Mori}},\ }\bibfield  {title} {\enquote {\bibinfo {title} {Transport, collective motion, and brownian motion},}\ }\href@noop {} {\bibfield  {journal} {\bibinfo  {journal} {Progress of Theoretical Physics}\ }\textbf {\bibinfo {volume} {33}},\ \bibinfo {pages} {423--455} (\bibinfo {year} {1965})}\BibitemShut {NoStop}%
\bibitem [{\citenamefont {Dominic~III}\ \emph {et~al.}(2023)\citenamefont {Dominic~III}, \citenamefont {Sayer}, \citenamefont {Cao}, \citenamefont {Markland}, \citenamefont {Huang},\ and\ \citenamefont {Montoya-Castillo}}]{dominic2023building}%
  \BibitemOpen
  \bibfield  {author} {\bibinfo {author} {\bibfnamefont {A.~J.}\ \bibnamefont {Dominic~III}}, \bibinfo {author} {\bibfnamefont {T.}~\bibnamefont {Sayer}}, \bibinfo {author} {\bibfnamefont {S.}~\bibnamefont {Cao}}, \bibinfo {author} {\bibfnamefont {T.~E.}\ \bibnamefont {Markland}}, \bibinfo {author} {\bibfnamefont {X.}~\bibnamefont {Huang}}, \ and\ \bibinfo {author} {\bibfnamefont {A.}~\bibnamefont {Montoya-Castillo}},\ }\bibfield  {title} {\enquote {\bibinfo {title} {Building insightful, memory-enriched models to capture long-time biochemical processes from short-time simulations},}\ }\href@noop {} {\bibfield  {journal} {\bibinfo  {journal} {Proceedings of the National Academy of Sciences}\ }\textbf {\bibinfo {volume} {120}},\ \bibinfo {pages} {e2221048120} (\bibinfo {year} {2023})}\BibitemShut {NoStop}%
\bibitem [{\citenamefont {Yan}\ \emph {et~al.}(2019)\citenamefont {Yan}, \citenamefont {Xu}, \citenamefont {Liu},\ and\ \citenamefont {Shi}}]{yan2019theoretical}%
  \BibitemOpen
  \bibfield  {author} {\bibinfo {author} {\bibfnamefont {Y.}~\bibnamefont {Yan}}, \bibinfo {author} {\bibfnamefont {M.}~\bibnamefont {Xu}}, \bibinfo {author} {\bibfnamefont {Y.}~\bibnamefont {Liu}}, \ and\ \bibinfo {author} {\bibfnamefont {Q.}~\bibnamefont {Shi}},\ }\bibfield  {title} {\enquote {\bibinfo {title} {Theoretical study of charge carrier transport in organic molecular crystals using the nakajima-zwanzig-mori generalized master equation},}\ }\href@noop {} {\bibfield  {journal} {\bibinfo  {journal} {Journal of Chemical Physics}\ }\textbf {\bibinfo {volume} {150}},\ \bibinfo {pages} {234101} (\bibinfo {year} {2019})}\BibitemShut {NoStop}%
\bibitem [{Note1()}]{Note1}%
  \BibitemOpen
  \bibinfo {note} {Ref.~\protect \rev@citealpnum {yan2019theoretical} assumed a homogeneous system, which necessitates only a single simulation by symmetry. This is not generally the case.}\BibitemShut {Stop}%
\bibitem [{\citenamefont {Markland}\ and\ \citenamefont {Ceriotti}(2018)}]{markland2018nuclear}%
  \BibitemOpen
  \bibfield  {author} {\bibinfo {author} {\bibfnamefont {T.~E.}\ \bibnamefont {Markland}}\ and\ \bibinfo {author} {\bibfnamefont {M.}~\bibnamefont {Ceriotti}},\ }\bibfield  {title} {\enquote {\bibinfo {title} {Nuclear quantum effects enter the mainstream},}\ }\href@noop {} {\bibfield  {journal} {\bibinfo  {journal} {Nature Reviews Chemistry}\ }\textbf {\bibinfo {volume} {2}},\ \bibinfo {pages} {0109} (\bibinfo {year} {2018})}\BibitemShut {NoStop}%
\bibitem [{\citenamefont {Chandler}\ and\ \citenamefont {Wolynes}(1981)}]{chandler1981exploiting}%
  \BibitemOpen
  \bibfield  {author} {\bibinfo {author} {\bibfnamefont {D.}~\bibnamefont {Chandler}}\ and\ \bibinfo {author} {\bibfnamefont {P.~G.}\ \bibnamefont {Wolynes}},\ }\bibfield  {title} {\enquote {\bibinfo {title} {Exploiting the isomorphism between quantum theory and classical statistical mechanics of polyatomic fluids},}\ }\href@noop {} {\bibfield  {journal} {\bibinfo  {journal} {Journal of Chemical Physics}\ }\textbf {\bibinfo {volume} {74}},\ \bibinfo {pages} {4078--4095} (\bibinfo {year} {1981})}\BibitemShut {NoStop}%
\bibitem [{\citenamefont {Cao}\ and\ \citenamefont {Voth}(1993)}]{cao1993new}%
  \BibitemOpen
  \bibfield  {author} {\bibinfo {author} {\bibfnamefont {J.}~\bibnamefont {Cao}}\ and\ \bibinfo {author} {\bibfnamefont {G.~A.}\ \bibnamefont {Voth}},\ }\bibfield  {title} {\enquote {\bibinfo {title} {A new perspective on quantum time correlation functions},}\ }\href@noop {} {\bibfield  {journal} {\bibinfo  {journal} {Journal of Chemical Physics}\ }\textbf {\bibinfo {volume} {99}},\ \bibinfo {pages} {10070--10073} (\bibinfo {year} {1993})}\BibitemShut {NoStop}%
\bibitem [{\citenamefont {Parrinello}\ and\ \citenamefont {Rahman}(1984)}]{parrinello1984study}%
  \BibitemOpen
  \bibfield  {author} {\bibinfo {author} {\bibfnamefont {M.}~\bibnamefont {Parrinello}}\ and\ \bibinfo {author} {\bibfnamefont {A.}~\bibnamefont {Rahman}},\ }\bibfield  {title} {\enquote {\bibinfo {title} {Study of an f center in molten kcl},}\ }\href@noop {} {\bibfield  {journal} {\bibinfo  {journal} {Journal of Chemical Physics}\ }\textbf {\bibinfo {volume} {80}},\ \bibinfo {pages} {860--867} (\bibinfo {year} {1984})}\BibitemShut {NoStop}%
\bibitem [{\citenamefont {Jang}\ and\ \citenamefont {Voth}(1999)}]{jang1999derivation}%
  \BibitemOpen
  \bibfield  {author} {\bibinfo {author} {\bibfnamefont {S.}~\bibnamefont {Jang}}\ and\ \bibinfo {author} {\bibfnamefont {G.~A.}\ \bibnamefont {Voth}},\ }\bibfield  {title} {\enquote {\bibinfo {title} {A derivation of centroid molecular dynamics and other approximate time evolution methods for path integral centroid variables},}\ }\href@noop {} {\bibfield  {journal} {\bibinfo  {journal} {Journal of Chemical Physics}\ }\textbf {\bibinfo {volume} {111}},\ \bibinfo {pages} {2371--2384} (\bibinfo {year} {1999})}\BibitemShut {NoStop}%
\bibitem [{\citenamefont {Craig}\ and\ \citenamefont {Manolopoulos}(2004)}]{craig2004quantum}%
  \BibitemOpen
  \bibfield  {author} {\bibinfo {author} {\bibfnamefont {I.~R.}\ \bibnamefont {Craig}}\ and\ \bibinfo {author} {\bibfnamefont {D.~E.}\ \bibnamefont {Manolopoulos}},\ }\bibfield  {title} {\enquote {\bibinfo {title} {Quantum statistics and classical mechanics: Real time correlation functions from ring polymer molecular dynamics},}\ }\href@noop {} {\bibfield  {journal} {\bibinfo  {journal} {Journal of Chemical Physics}\ }\textbf {\bibinfo {volume} {121}},\ \bibinfo {pages} {3368--3373} (\bibinfo {year} {2004})}\BibitemShut {NoStop}%
\bibitem [{\citenamefont {Cao}\ and\ \citenamefont {Voth}(1994)}]{cao1994formulation}%
  \BibitemOpen
  \bibfield  {author} {\bibinfo {author} {\bibfnamefont {J.}~\bibnamefont {Cao}}\ and\ \bibinfo {author} {\bibfnamefont {G.~A.}\ \bibnamefont {Voth}},\ }\bibfield  {title} {\enquote {\bibinfo {title} {The formulation of quantum statistical mechanics based on the feynman path centroid density. ii. dynamical properties},}\ }\href@noop {} {\bibfield  {journal} {\bibinfo  {journal} {Journal of Chemical Physics}\ }\textbf {\bibinfo {volume} {100}},\ \bibinfo {pages} {5106--5117} (\bibinfo {year} {1994})}\BibitemShut {NoStop}%
\bibitem [{\citenamefont {Shao}\ and\ \citenamefont {Makri}(2002)}]{shao2002iterative}%
  \BibitemOpen
  \bibfield  {author} {\bibinfo {author} {\bibfnamefont {J.}~\bibnamefont {Shao}}\ and\ \bibinfo {author} {\bibfnamefont {N.}~\bibnamefont {Makri}},\ }\bibfield  {title} {\enquote {\bibinfo {title} {Iterative path integral formulation of equilibrium correlation functions for quantum dissipative systems},}\ }\href@noop {} {\bibfield  {journal} {\bibinfo  {journal} {Journal of Chemical Physics}\ }\textbf {\bibinfo {volume} {116}},\ \bibinfo {pages} {507--514} (\bibinfo {year} {2002})}\BibitemShut {NoStop}%
\bibitem [{\citenamefont {Tanimura}(2014)}]{tanimura2014reduced}%
  \BibitemOpen
  \bibfield  {author} {\bibinfo {author} {\bibfnamefont {Y.}~\bibnamefont {Tanimura}},\ }\bibfield  {title} {\enquote {\bibinfo {title} {Reduced hierarchical equations of motion in real and imaginary time: Correlated initial states and thermodynamic quantities},}\ }\href@noop {} {\bibfield  {journal} {\bibinfo  {journal} {Journal of Chemical Physics}\ }\textbf {\bibinfo {volume} {141}},\ \bibinfo {pages} {044114} (\bibinfo {year} {2014})}\BibitemShut {NoStop}%
\bibitem [{\citenamefont {Song}\ and\ \citenamefont {Shi}(2015{\natexlab{a}})}]{song2015calculation}%
  \BibitemOpen
  \bibfield  {author} {\bibinfo {author} {\bibfnamefont {L.}~\bibnamefont {Song}}\ and\ \bibinfo {author} {\bibfnamefont {Q.}~\bibnamefont {Shi}},\ }\bibfield  {title} {\enquote {\bibinfo {title} {Calculation of correlated initial state in the hierarchical equations of motion method using an imaginary time path integral approach},}\ }\href@noop {} {\bibfield  {journal} {\bibinfo  {journal} {Journal of Chemical Physics}\ }\textbf {\bibinfo {volume} {143}},\ \bibinfo {pages} {194106} (\bibinfo {year} {2015}{\natexlab{a}})}\BibitemShut {NoStop}%
\bibitem [{\citenamefont {Montoya-Castillo}\ and\ \citenamefont {Reichman}(2017{\natexlab{a}})}]{montoya2017path}%
  \BibitemOpen
  \bibfield  {author} {\bibinfo {author} {\bibfnamefont {A.}~\bibnamefont {Montoya-Castillo}}\ and\ \bibinfo {author} {\bibfnamefont {D.~R.}\ \bibnamefont {Reichman}},\ }\bibfield  {title} {\enquote {\bibinfo {title} {Path integral approach to the wigner representation of canonical density operators for discrete systems coupled to harmonic baths},}\ }\href@noop {} {\bibfield  {journal} {\bibinfo  {journal} {Journal of Chemical Physics}\ }\textbf {\bibinfo {volume} {146}},\ \bibinfo {pages} {024107} (\bibinfo {year} {2017}{\natexlab{a}})}\BibitemShut {NoStop}%
\bibitem [{\citenamefont {Liu}\ and\ \citenamefont {Miller}(2006)}]{liu2006using}%
  \BibitemOpen
  \bibfield  {author} {\bibinfo {author} {\bibfnamefont {J.}~\bibnamefont {Liu}}\ and\ \bibinfo {author} {\bibfnamefont {W.~H.}\ \bibnamefont {Miller}},\ }\bibfield  {title} {\enquote {\bibinfo {title} {Using the thermal gaussian approximation for the boltzmann operator in semiclassical initial value time correlation functions},}\ }\href@noop {} {\bibfield  {journal} {\bibinfo  {journal} {Journal of Chemical Physics}\ }\textbf {\bibinfo {volume} {125}},\ \bibinfo {pages} {224104} (\bibinfo {year} {2006})}\BibitemShut {NoStop}%
\bibitem [{\citenamefont {Shi}\ and\ \citenamefont {Geva}(2003{\natexlab{a}})}]{shi2003semiclassical}%
  \BibitemOpen
  \bibfield  {author} {\bibinfo {author} {\bibfnamefont {Q.}~\bibnamefont {Shi}}\ and\ \bibinfo {author} {\bibfnamefont {E.}~\bibnamefont {Geva}},\ }\bibfield  {title} {\enquote {\bibinfo {title} {Semiclassical theory of vibrational energy relaxation in the condensed phase},}\ }\href@noop {} {\bibfield  {journal} {\bibinfo  {journal} {Journal of Physical Chemistry A}\ }\textbf {\bibinfo {volume} {107}},\ \bibinfo {pages} {9059--9069} (\bibinfo {year} {2003}{\natexlab{a}})}\BibitemShut {NoStop}%
\bibitem [{\citenamefont {Poulsen}, \citenamefont {Nyman},\ and\ \citenamefont {Rossky}(2003)}]{poulsen2003practical}%
  \BibitemOpen
  \bibfield  {author} {\bibinfo {author} {\bibfnamefont {J.~A.}\ \bibnamefont {Poulsen}}, \bibinfo {author} {\bibfnamefont {G.}~\bibnamefont {Nyman}}, \ and\ \bibinfo {author} {\bibfnamefont {P.~J.}\ \bibnamefont {Rossky}},\ }\bibfield  {title} {\enquote {\bibinfo {title} {Practical evaluation of condensed phase quantum correlation functions: A feynman--kleinert variational linearized path integral method},}\ }\href@noop {} {\bibfield  {journal} {\bibinfo  {journal} {Journal of Chemical Physics}\ }\textbf {\bibinfo {volume} {119}},\ \bibinfo {pages} {12179--12193} (\bibinfo {year} {2003})}\BibitemShut {NoStop}%
\bibitem [{\citenamefont {Montoya-Castillo}\ and\ \citenamefont {Reichman}(2017{\natexlab{b}})}]{montoya2017approximate}%
  \BibitemOpen
  \bibfield  {author} {\bibinfo {author} {\bibfnamefont {A.}~\bibnamefont {Montoya-Castillo}}\ and\ \bibinfo {author} {\bibfnamefont {D.~R.}\ \bibnamefont {Reichman}},\ }\bibfield  {title} {\enquote {\bibinfo {title} {Approximate but accurate quantum dynamics from the mori formalism. ii. equilibrium time correlation functions},}\ }\href@noop {} {\bibfield  {journal} {\bibinfo  {journal} {Journal of Chemical Physics}\ }\textbf {\bibinfo {volume} {146}},\ \bibinfo {pages} {084110} (\bibinfo {year} {2017}{\natexlab{b}})}\BibitemShut {NoStop}%
\bibitem [{\citenamefont {Barthel}(2013)}]{barthel2013precise}%
  \BibitemOpen
  \bibfield  {author} {\bibinfo {author} {\bibfnamefont {T.}~\bibnamefont {Barthel}},\ }\bibfield  {title} {\enquote {\bibinfo {title} {Precise evaluation of thermal response functions by optimized density matrix renormalization group schemes},}\ }\href@noop {} {\bibfield  {journal} {\bibinfo  {journal} {New Journal of Physics}\ }\textbf {\bibinfo {volume} {15}},\ \bibinfo {pages} {073010} (\bibinfo {year} {2013})}\BibitemShut {NoStop}%
\bibitem [{\citenamefont {Karrasch}, \citenamefont {Kennes},\ and\ \citenamefont {Heidrich-Meisner}(2015)}]{karrasch2015spin}%
  \BibitemOpen
  \bibfield  {author} {\bibinfo {author} {\bibfnamefont {C.}~\bibnamefont {Karrasch}}, \bibinfo {author} {\bibfnamefont {D.}~\bibnamefont {Kennes}}, \ and\ \bibinfo {author} {\bibfnamefont {F.}~\bibnamefont {Heidrich-Meisner}},\ }\bibfield  {title} {\enquote {\bibinfo {title} {Spin and thermal conductivity of quantum spin chains and ladders},}\ }\href@noop {} {\bibfield  {journal} {\bibinfo  {journal} {Physical Review B}\ }\textbf {\bibinfo {volume} {91}},\ \bibinfo {pages} {115130} (\bibinfo {year} {2015})}\BibitemShut {NoStop}%
\bibitem [{\citenamefont {Karrasch}, \citenamefont {Bardarson},\ and\ \citenamefont {Moore}(2013)}]{karrasch2013reducing}%
  \BibitemOpen
  \bibfield  {author} {\bibinfo {author} {\bibfnamefont {C.}~\bibnamefont {Karrasch}}, \bibinfo {author} {\bibfnamefont {J.}~\bibnamefont {Bardarson}}, \ and\ \bibinfo {author} {\bibfnamefont {J.}~\bibnamefont {Moore}},\ }\bibfield  {title} {\enquote {\bibinfo {title} {Reducing the numerical effort of finite-temperature density matrix renormalization group calculations},}\ }\href@noop {} {\bibfield  {journal} {\bibinfo  {journal} {New Journal of Physics}\ }\textbf {\bibinfo {volume} {15}},\ \bibinfo {pages} {083031} (\bibinfo {year} {2013})}\BibitemShut {NoStop}%
\bibitem [{\citenamefont {Shi}\ \emph {et~al.}(2009)\citenamefont {Shi}, \citenamefont {Chen}, \citenamefont {Nan}, \citenamefont {Xu},\ and\ \citenamefont {Yan}}]{shi2009efficient}%
  \BibitemOpen
  \bibfield  {author} {\bibinfo {author} {\bibfnamefont {Q.}~\bibnamefont {Shi}}, \bibinfo {author} {\bibfnamefont {L.}~\bibnamefont {Chen}}, \bibinfo {author} {\bibfnamefont {G.}~\bibnamefont {Nan}}, \bibinfo {author} {\bibfnamefont {R.-X.}\ \bibnamefont {Xu}}, \ and\ \bibinfo {author} {\bibfnamefont {Y.}~\bibnamefont {Yan}},\ }\bibfield  {title} {\enquote {\bibinfo {title} {Efficient hierarchical liouville space propagator to quantum dissipative dynamics},}\ }\href@noop {} {\bibfield  {journal} {\bibinfo  {journal} {Journal of Chemical Physics}\ }\textbf {\bibinfo {volume} {130}},\ \bibinfo {pages} {084105} (\bibinfo {year} {2009})}\BibitemShut {NoStop}%
\bibitem [{\citenamefont {Song}, \citenamefont {Bai},\ and\ \citenamefont {Shi}(2015)}]{song2015time}%
  \BibitemOpen
  \bibfield  {author} {\bibinfo {author} {\bibfnamefont {K.}~\bibnamefont {Song}}, \bibinfo {author} {\bibfnamefont {S.}~\bibnamefont {Bai}}, \ and\ \bibinfo {author} {\bibfnamefont {Q.}~\bibnamefont {Shi}},\ }\bibfield  {title} {\enquote {\bibinfo {title} {A time domain two-particle approximation to calculate the absorption and circular dichroism line shapes of molecular aggregates},}\ }\href@noop {} {\bibfield  {journal} {\bibinfo  {journal} {Journal of Chemical Physics}\ }\textbf {\bibinfo {volume} {143}},\ \bibinfo {pages} {064109} (\bibinfo {year} {2015})}\BibitemShut {NoStop}%
\bibitem [{\citenamefont {Tanimura}\ and\ \citenamefont {Kubo}(1989)}]{tanimura1989time}%
  \BibitemOpen
  \bibfield  {author} {\bibinfo {author} {\bibfnamefont {Y.}~\bibnamefont {Tanimura}}\ and\ \bibinfo {author} {\bibfnamefont {R.}~\bibnamefont {Kubo}},\ }\bibfield  {title} {\enquote {\bibinfo {title} {Time evolution of a quantum system in contact with a nearly gaussian-markoffian noise bath},}\ }\href@noop {} {\bibfield  {journal} {\bibinfo  {journal} {Journal of the Physical Society of Japan}\ }\textbf {\bibinfo {volume} {58}},\ \bibinfo {pages} {101--114} (\bibinfo {year} {1989})}\BibitemShut {NoStop}%
\bibitem [{\citenamefont {Bhattacharyya}, \citenamefont {Sayer},\ and\ \citenamefont {Montoya-Castillo}(2024)}]{bhattacharyya2024anomalous}%
  \BibitemOpen
  \bibfield  {author} {\bibinfo {author} {\bibfnamefont {S.}~\bibnamefont {Bhattacharyya}}, \bibinfo {author} {\bibfnamefont {T.}~\bibnamefont {Sayer}}, \ and\ \bibinfo {author} {\bibfnamefont {A.}~\bibnamefont {Montoya-Castillo}},\ }\bibfield  {title} {\enquote {\bibinfo {title} {Anomalous transport of small polarons arises from transient lattice relaxation or immovable boundaries},}\ }\href@noop {} {\bibfield  {journal} {\bibinfo  {journal} {Journal of Physical Chemistry Letters}\ }\textbf {\bibinfo {volume} {15}},\ \bibinfo {pages} {1382--1389} (\bibinfo {year} {2024})}\BibitemShut {NoStop}%
\bibitem [{\citenamefont {Kubo}, \citenamefont {Toda},\ and\ \citenamefont {Hashitsume}(1991)}]{kubo1991statistical}%
  \BibitemOpen
  \bibfield  {author} {\bibinfo {author} {\bibfnamefont {R.}~\bibnamefont {Kubo}}, \bibinfo {author} {\bibfnamefont {M.}~\bibnamefont {Toda}}, \ and\ \bibinfo {author} {\bibfnamefont {N.}~\bibnamefont {Hashitsume}},\ }\bibfield  {title} {\enquote {\bibinfo {title} {Statistical mechanics of linear response},}\ }\href@noop {} {\bibfield  {journal} {\bibinfo  {journal} {Statistical Physics II: Nonequilibrium Statistical Mechanics}\ ,\ \bibinfo {pages} {146--202}} (\bibinfo {year} {1991})}\BibitemShut {NoStop}%
\bibitem [{\citenamefont {Smith}(2001)}]{smith2001classical}%
  \BibitemOpen
  \bibfield  {author} {\bibinfo {author} {\bibfnamefont {N.}~\bibnamefont {Smith}},\ }\bibfield  {title} {\enquote {\bibinfo {title} {Classical generalization of the drude formula for the optical conductivity},}\ }\href@noop {} {\bibfield  {journal} {\bibinfo  {journal} {Physical Review B}\ }\textbf {\bibinfo {volume} {64}},\ \bibinfo {pages} {155106} (\bibinfo {year} {2001})}\BibitemShut {NoStop}%
\bibitem [{\citenamefont {Yettapu}\ \emph {et~al.}(2016)\citenamefont {Yettapu}, \citenamefont {Talukdar}, \citenamefont {Sarkar}, \citenamefont {Swarnkar}, \citenamefont {Nag}, \citenamefont {Ghosh},\ and\ \citenamefont {Mandal}}]{yettapu2016terahertz}%
  \BibitemOpen
  \bibfield  {author} {\bibinfo {author} {\bibfnamefont {G.~R.}\ \bibnamefont {Yettapu}}, \bibinfo {author} {\bibfnamefont {D.}~\bibnamefont {Talukdar}}, \bibinfo {author} {\bibfnamefont {S.}~\bibnamefont {Sarkar}}, \bibinfo {author} {\bibfnamefont {A.}~\bibnamefont {Swarnkar}}, \bibinfo {author} {\bibfnamefont {A.}~\bibnamefont {Nag}}, \bibinfo {author} {\bibfnamefont {P.}~\bibnamefont {Ghosh}}, \ and\ \bibinfo {author} {\bibfnamefont {P.}~\bibnamefont {Mandal}},\ }\bibfield  {title} {\enquote {\bibinfo {title} {Terahertz conductivity within colloidal cspbbr3 perovskite nanocrystals: remarkably high carrier mobilities and large diffusion lengths},}\ }\href@noop {} {\bibfield  {journal} {\bibinfo  {journal} {Nano Letters}\ }\textbf {\bibinfo {volume} {16}},\ \bibinfo {pages} {4838--4848} (\bibinfo {year} {2016})}\BibitemShut {NoStop}%
\bibitem [{\citenamefont {Pattengale}\ \emph {et~al.}(2019)\citenamefont {Pattengale}, \citenamefont {Neu}, \citenamefont {Ostresh}, \citenamefont {Hu}, \citenamefont {Spies}, \citenamefont {Okabe}, \citenamefont {Brudvig},\ and\ \citenamefont {Schmuttenmaer}}]{pattengale2019metal}%
  \BibitemOpen
  \bibfield  {author} {\bibinfo {author} {\bibfnamefont {B.}~\bibnamefont {Pattengale}}, \bibinfo {author} {\bibfnamefont {J.}~\bibnamefont {Neu}}, \bibinfo {author} {\bibfnamefont {S.}~\bibnamefont {Ostresh}}, \bibinfo {author} {\bibfnamefont {G.}~\bibnamefont {Hu}}, \bibinfo {author} {\bibfnamefont {J.~A.}\ \bibnamefont {Spies}}, \bibinfo {author} {\bibfnamefont {R.}~\bibnamefont {Okabe}}, \bibinfo {author} {\bibfnamefont {G.~W.}\ \bibnamefont {Brudvig}}, \ and\ \bibinfo {author} {\bibfnamefont {C.~A.}\ \bibnamefont {Schmuttenmaer}},\ }\bibfield  {title} {\enquote {\bibinfo {title} {metal--organic framework photoconductivity via time-resolved terahertz spectroscopy},}\ }\href@noop {} {\bibfield  {journal} {\bibinfo  {journal} {Journal of the American Chemical Society}\ }\textbf {\bibinfo {volume} {141}},\ \bibinfo {pages} {9793--9797} (\bibinfo {year} {2019})}\BibitemShut {NoStop}%
\bibitem [{\citenamefont {Kumar}\ \emph {et~al.}(2020)\citenamefont {Kumar}, \citenamefont {Prajapati}, \citenamefont {Dagar}, \citenamefont {Vagadia}, \citenamefont {Rana},\ and\ \citenamefont {Tonouchi}}]{kumar2020terahertz}%
  \BibitemOpen
  \bibfield  {author} {\bibinfo {author} {\bibfnamefont {K.~S.}\ \bibnamefont {Kumar}}, \bibinfo {author} {\bibfnamefont {G.~L.}\ \bibnamefont {Prajapati}}, \bibinfo {author} {\bibfnamefont {R.}~\bibnamefont {Dagar}}, \bibinfo {author} {\bibfnamefont {M.}~\bibnamefont {Vagadia}}, \bibinfo {author} {\bibfnamefont {D.~S.}\ \bibnamefont {Rana}}, \ and\ \bibinfo {author} {\bibfnamefont {M.}~\bibnamefont {Tonouchi}},\ }\bibfield  {title} {\enquote {\bibinfo {title} {Terahertz electrodynamics in transition metal oxides},}\ }\href@noop {} {\bibfield  {journal} {\bibinfo  {journal} {Advanced Optical Materials}\ }\textbf {\bibinfo {volume} {8}},\ \bibinfo {pages} {1900958} (\bibinfo {year} {2020})}\BibitemShut {NoStop}%
\bibitem [{\citenamefont {Magnanelli}\ \emph {et~al.}(2020)\citenamefont {Magnanelli}, \citenamefont {Engmann}, \citenamefont {Wahlstrand}, \citenamefont {Stephenson}, \citenamefont {Richter},\ and\ \citenamefont {Heilweil}}]{magnanelli2020polarization}%
  \BibitemOpen
  \bibfield  {author} {\bibinfo {author} {\bibfnamefont {T.~J.}\ \bibnamefont {Magnanelli}}, \bibinfo {author} {\bibfnamefont {S.}~\bibnamefont {Engmann}}, \bibinfo {author} {\bibfnamefont {J.~K.}\ \bibnamefont {Wahlstrand}}, \bibinfo {author} {\bibfnamefont {J.~C.}\ \bibnamefont {Stephenson}}, \bibinfo {author} {\bibfnamefont {L.~J.}\ \bibnamefont {Richter}}, \ and\ \bibinfo {author} {\bibfnamefont {E.~J.}\ \bibnamefont {Heilweil}},\ }\bibfield  {title} {\enquote {\bibinfo {title} {Polarization dependence of charge conduction in conjugated polymer films investigated with time-resolved terahertz spectroscopy},}\ }\href@noop {} {\bibfield  {journal} {\bibinfo  {journal} {Journal of Physical Chemistry C}\ }\textbf {\bibinfo {volume} {124}},\ \bibinfo {pages} {6993--7006} (\bibinfo {year} {2020})}\BibitemShut {NoStop}%
\bibitem [{\citenamefont {Li}\ \emph {et~al.}(2023)\citenamefont {Li}, \citenamefont {Wei}, \citenamefont {Zhang}, \citenamefont {Wan}, \citenamefont {Cheng}, \citenamefont {Xie}, \citenamefont {Li}, \citenamefont {Zhang}, \citenamefont {Xu}, \citenamefont {Hu} \emph {et~al.}}]{li2023charge}%
  \BibitemOpen
  \bibfield  {author} {\bibinfo {author} {\bibfnamefont {E.}~\bibnamefont {Li}}, \bibinfo {author} {\bibfnamefont {J.}~\bibnamefont {Wei}}, \bibinfo {author} {\bibfnamefont {T.}~\bibnamefont {Zhang}}, \bibinfo {author} {\bibfnamefont {H.}~\bibnamefont {Wan}}, \bibinfo {author} {\bibfnamefont {Y.}~\bibnamefont {Cheng}}, \bibinfo {author} {\bibfnamefont {J.}~\bibnamefont {Xie}}, \bibinfo {author} {\bibfnamefont {H.}~\bibnamefont {Li}}, \bibinfo {author} {\bibfnamefont {K.}~\bibnamefont {Zhang}}, \bibinfo {author} {\bibfnamefont {J.}~\bibnamefont {Xu}}, \bibinfo {author} {\bibfnamefont {J.}~\bibnamefont {Hu}},  \emph {et~al.},\ }\bibfield  {title} {\enquote {\bibinfo {title} {Charge carriers localization effect revealed through terahertz spectroscopy of mxene: Ti3c2tx},}\ }\href@noop {} {\bibfield  {journal} {\bibinfo  {journal} {Small}\ ,\ \bibinfo {pages} {2306200}} (\bibinfo {year} {2023})}\BibitemShut {NoStop}%
\bibitem [{\citenamefont {Cheung}\ and\ \citenamefont {Troisi}(2008)}]{cheung2008modelling}%
  \BibitemOpen
  \bibfield  {author} {\bibinfo {author} {\bibfnamefont {D.~L.}\ \bibnamefont {Cheung}}\ and\ \bibinfo {author} {\bibfnamefont {A.}~\bibnamefont {Troisi}},\ }\bibfield  {title} {\enquote {\bibinfo {title} {Modelling charge transport in organic semiconductors: from quantum dynamics to soft matter},}\ }\href@noop {} {\bibfield  {journal} {\bibinfo  {journal} {Physical Chemistry Chemical Physics}\ }\textbf {\bibinfo {volume} {10}},\ \bibinfo {pages} {5941--5952} (\bibinfo {year} {2008})}\BibitemShut {NoStop}%
\bibitem [{\citenamefont {Ortmann}, \citenamefont {Bechstedt},\ and\ \citenamefont {Hannewald}(2010)}]{ortmann2010charge}%
  \BibitemOpen
  \bibfield  {author} {\bibinfo {author} {\bibfnamefont {F.}~\bibnamefont {Ortmann}}, \bibinfo {author} {\bibfnamefont {F.}~\bibnamefont {Bechstedt}}, \ and\ \bibinfo {author} {\bibfnamefont {K.}~\bibnamefont {Hannewald}},\ }\bibfield  {title} {\enquote {\bibinfo {title} {Charge transport in organic crystals: interplay of band transport, hopping and electron--phonon scattering},}\ }\href@noop {} {\bibfield  {journal} {\bibinfo  {journal} {New Journal of Physics}\ }\textbf {\bibinfo {volume} {12}},\ \bibinfo {pages} {023011} (\bibinfo {year} {2010})}\BibitemShut {NoStop}%
\bibitem [{\citenamefont {Ghosh}\ and\ \citenamefont {Paesani}(2021)}]{ghosh2021topology}%
  \BibitemOpen
  \bibfield  {author} {\bibinfo {author} {\bibfnamefont {R.}~\bibnamefont {Ghosh}}\ and\ \bibinfo {author} {\bibfnamefont {F.}~\bibnamefont {Paesani}},\ }\bibfield  {title} {\enquote {\bibinfo {title} {Topology-mediated enhanced polaron coherence in covalent organic frameworks},}\ }\href@noop {} {\bibfield  {journal} {\bibinfo  {journal} {Journal of Physical Chemistry Letters}\ }\textbf {\bibinfo {volume} {12}},\ \bibinfo {pages} {9442--9448} (\bibinfo {year} {2021})}\BibitemShut {NoStop}%
\bibitem [{\citenamefont {Mousavi}\ and\ \citenamefont {Rezania}(2010)}]{mousavi2010electron}%
  \BibitemOpen
  \bibfield  {author} {\bibinfo {author} {\bibfnamefont {H.}~\bibnamefont {Mousavi}}\ and\ \bibinfo {author} {\bibfnamefont {H.}~\bibnamefont {Rezania}},\ }\bibfield  {title} {\enquote {\bibinfo {title} {Electron--phonon interaction in carbon nanotubes},}\ }\href@noop {} {\bibfield  {journal} {\bibinfo  {journal} {Modern Physics Letters B}\ }\textbf {\bibinfo {volume} {24}},\ \bibinfo {pages} {2947--2954} (\bibinfo {year} {2010})}\BibitemShut {NoStop}%
\bibitem [{\citenamefont {Mahan}(2000)}]{mahan2000many}%
  \BibitemOpen
  \bibfield  {author} {\bibinfo {author} {\bibfnamefont {G.~D.}\ \bibnamefont {Mahan}},\ }\href@noop {} {\emph {\bibinfo {title} {Many-Particle Physics}}}\ (\bibinfo  {publisher} {Springer Science \& Business Media},\ \bibinfo {year} {2000})\BibitemShut {NoStop}%
\bibitem [{\citenamefont {Nematiaram}\ and\ \citenamefont {Troisi}(2020)}]{nematiaram2020modeling}%
  \BibitemOpen
  \bibfield  {author} {\bibinfo {author} {\bibfnamefont {T.}~\bibnamefont {Nematiaram}}\ and\ \bibinfo {author} {\bibfnamefont {A.}~\bibnamefont {Troisi}},\ }\bibfield  {title} {\enquote {\bibinfo {title} {Modeling charge transport in high-mobility molecular semiconductors: Balancing electronic structure and quantum dynamics methods with the help of experiments},}\ }\href@noop {} {\bibfield  {journal} {\bibinfo  {journal} {Journal of Chemical Physics}\ }\textbf {\bibinfo {volume} {152}},\ \bibinfo {pages} {190902} (\bibinfo {year} {2020})}\BibitemShut {NoStop}%
\bibitem [{\citenamefont {Yan}, \citenamefont {Song},\ and\ \citenamefont {Shi}(2018)}]{yan2018understanding}%
  \BibitemOpen
  \bibfield  {author} {\bibinfo {author} {\bibfnamefont {Y.}~\bibnamefont {Yan}}, \bibinfo {author} {\bibfnamefont {L.}~\bibnamefont {Song}}, \ and\ \bibinfo {author} {\bibfnamefont {Q.}~\bibnamefont {Shi}},\ }\bibfield  {title} {\enquote {\bibinfo {title} {Understanding the free energy barrier and multiple timescale dynamics of charge separation in organic photovoltaic cells},}\ }\href@noop {} {\bibfield  {journal} {\bibinfo  {journal} {Journal of Chemical Physics}\ }\textbf {\bibinfo {volume} {148}},\ \bibinfo {pages} {084109} (\bibinfo {year} {2018})}\BibitemShut {NoStop}%
\bibitem [{\citenamefont {Song}\ and\ \citenamefont {Shi}(2015{\natexlab{b}})}]{song2015new}%
  \BibitemOpen
  \bibfield  {author} {\bibinfo {author} {\bibfnamefont {L.}~\bibnamefont {Song}}\ and\ \bibinfo {author} {\bibfnamefont {Q.}~\bibnamefont {Shi}},\ }\bibfield  {title} {\enquote {\bibinfo {title} {A new approach to calculate charge carrier transport mobility in organic molecular crystals from imaginary time path integral simulations},}\ }\href@noop {} {\bibfield  {journal} {\bibinfo  {journal} {Journal of Chemical Physics}\ }\textbf {\bibinfo {volume} {142}},\ \bibinfo {pages} {174103} (\bibinfo {year} {2015}{\natexlab{b}})}\BibitemShut {NoStop}%
\bibitem [{Note2()}]{Note2}%
  \BibitemOpen
  \bibinfo {note} {The easiest to construct and fastest initial condition to equilibrate corresponds to a zeroth order approximation to $ \protect \textrm {e} ^{-\beta \protect \hat {H}} /Z$, i.e., $ \protect \textrm {e} ^{-\beta \protect \hat {H}_{elec}} /Z_{elec} \times \protect \textrm {e} ^{-\beta \protect \hat {H}_{B}} /Z_{B}$.}\BibitemShut {Stop}%
\bibitem [{\citenamefont {Reichman}\ and\ \citenamefont {Charbonneau}(2005)}]{reichman2005mode}%
  \BibitemOpen
  \bibfield  {author} {\bibinfo {author} {\bibfnamefont {D.~R.}\ \bibnamefont {Reichman}}\ and\ \bibinfo {author} {\bibfnamefont {P.}~\bibnamefont {Charbonneau}},\ }\bibfield  {title} {\enquote {\bibinfo {title} {Mode-coupling theory},}\ }\href@noop {} {\bibfield  {journal} {\bibinfo  {journal} {Journal of Statistical Mechanics: Theory and Experiment}\ }\textbf {\bibinfo {volume} {2005}},\ \bibinfo {pages} {P05013} (\bibinfo {year} {2005})}\BibitemShut {NoStop}%
\bibitem [{\citenamefont {Janssen}(2018)}]{janssen2018mode}%
  \BibitemOpen
  \bibfield  {author} {\bibinfo {author} {\bibfnamefont {L.~M.}\ \bibnamefont {Janssen}},\ }\bibfield  {title} {\enquote {\bibinfo {title} {Mode-coupling theory of the glass transition: A primer},}\ }\href@noop {} {\bibfield  {journal} {\bibinfo  {journal} {Frontiers in Physics}\ }\textbf {\bibinfo {volume} {6}},\ \bibinfo {pages} {97} (\bibinfo {year} {2018})}\BibitemShut {NoStop}%
\bibitem [{\citenamefont {Allen}(2006)}]{allen2006electron}%
  \BibitemOpen
  \bibfield  {author} {\bibinfo {author} {\bibfnamefont {P.~B.}\ \bibnamefont {Allen}},\ }\bibfield  {title} {\enquote {\bibinfo {title} {Electron transport},}\ }\href@noop {} {\bibfield  {journal} {\bibinfo  {journal} {Contemporary Concepts of Condensed Matter Science}\ }\textbf {\bibinfo {volume} {2}},\ \bibinfo {pages} {165--218} (\bibinfo {year} {2006})}\BibitemShut {NoStop}%
\bibitem [{\citenamefont {Egorov}, \citenamefont {Everitt},\ and\ \citenamefont {Skinner}(1999)}]{egorov1999quantum}%
  \BibitemOpen
  \bibfield  {author} {\bibinfo {author} {\bibfnamefont {S.}~\bibnamefont {Egorov}}, \bibinfo {author} {\bibfnamefont {K.}~\bibnamefont {Everitt}}, \ and\ \bibinfo {author} {\bibfnamefont {J.}~\bibnamefont {Skinner}},\ }\bibfield  {title} {\enquote {\bibinfo {title} {Quantum dynamics and vibrational relaxation},}\ }\href@noop {} {\bibfield  {journal} {\bibinfo  {journal} {Journal of Physical Chemistry A}\ }\textbf {\bibinfo {volume} {103}},\ \bibinfo {pages} {9494--9499} (\bibinfo {year} {1999})}\BibitemShut {NoStop}%
\bibitem [{\citenamefont {Bader}\ and\ \citenamefont {Berne}(1994)}]{bader1994quantum}%
  \BibitemOpen
  \bibfield  {author} {\bibinfo {author} {\bibfnamefont {J.~S.}\ \bibnamefont {Bader}}\ and\ \bibinfo {author} {\bibfnamefont {B.}~\bibnamefont {Berne}},\ }\bibfield  {title} {\enquote {\bibinfo {title} {Quantum and classical relaxation rates from classical simulations},}\ }\href@noop {} {\bibfield  {journal} {\bibinfo  {journal} {Journal of Chemical Physics}\ }\textbf {\bibinfo {volume} {100}},\ \bibinfo {pages} {8359--8366} (\bibinfo {year} {1994})}\BibitemShut {NoStop}%
\bibitem [{\citenamefont {Chen}\ and\ \citenamefont {Marcus}(2021)}]{chen2021drude}%
  \BibitemOpen
  \bibfield  {author} {\bibinfo {author} {\bibfnamefont {W.-C.}\ \bibnamefont {Chen}}\ and\ \bibinfo {author} {\bibfnamefont {R.~A.}\ \bibnamefont {Marcus}},\ }\bibfield  {title} {\enquote {\bibinfo {title} {The drude-smith equation and related equations for the frequency-dependent electrical conductivity of materials: Insight from a memory function formalism},}\ }\href@noop {} {\bibfield  {journal} {\bibinfo  {journal} {ChemPhysChem}\ }\textbf {\bibinfo {volume} {22}},\ \bibinfo {pages} {1667--1674} (\bibinfo {year} {2021})}\BibitemShut {NoStop}%
\bibitem [{\citenamefont {Forster}(2018)}]{forster2018hydrodynamic}%
  \BibitemOpen
  \bibfield  {author} {\bibinfo {author} {\bibfnamefont {D.}~\bibnamefont {Forster}},\ }\href@noop {} {\emph {\bibinfo {title} {Hydrodynamic fluctuations, broken symmetry, and correlation functions}}}\ (\bibinfo  {publisher} {CRC Press},\ \bibinfo {year} {2018})\BibitemShut {NoStop}%
\bibitem [{\citenamefont {Ulbricht}\ \emph {et~al.}(2011)\citenamefont {Ulbricht}, \citenamefont {Hendry}, \citenamefont {Shan}, \citenamefont {Heinz},\ and\ \citenamefont {Bonn}}]{ulbricht2011carrier}%
  \BibitemOpen
  \bibfield  {author} {\bibinfo {author} {\bibfnamefont {R.}~\bibnamefont {Ulbricht}}, \bibinfo {author} {\bibfnamefont {E.}~\bibnamefont {Hendry}}, \bibinfo {author} {\bibfnamefont {J.}~\bibnamefont {Shan}}, \bibinfo {author} {\bibfnamefont {T.~F.}\ \bibnamefont {Heinz}}, \ and\ \bibinfo {author} {\bibfnamefont {M.}~\bibnamefont {Bonn}},\ }\bibfield  {title} {\enquote {\bibinfo {title} {Carrier dynamics in semiconductors studied with time-resolved terahertz spectroscopy},}\ }\href@noop {} {\bibfield  {journal} {\bibinfo  {journal} {Reviews of Modern Physics}\ }\textbf {\bibinfo {volume} {83}},\ \bibinfo {pages} {543} (\bibinfo {year} {2011})}\BibitemShut {NoStop}%
\bibitem [{\citenamefont {Lloyd-Hughes}\ and\ \citenamefont {Jeon}(2012)}]{lloyd2012review}%
  \BibitemOpen
  \bibfield  {author} {\bibinfo {author} {\bibfnamefont {J.}~\bibnamefont {Lloyd-Hughes}}\ and\ \bibinfo {author} {\bibfnamefont {T.-I.}\ \bibnamefont {Jeon}},\ }\bibfield  {title} {\enquote {\bibinfo {title} {A review of the terahertz conductivity of bulk and nano-materials},}\ }\href@noop {} {\bibfield  {journal} {\bibinfo  {journal} {Journal of Infrared, Millimeter, and Terahertz Waves}\ }\textbf {\bibinfo {volume} {33}},\ \bibinfo {pages} {871--925} (\bibinfo {year} {2012})}\BibitemShut {NoStop}%
\bibitem [{\citenamefont {Ku{\v{z}}el}\ and\ \citenamefont {N{\v{e}}mec}(2020)}]{kuvzel2020terahertz}%
  \BibitemOpen
  \bibfield  {author} {\bibinfo {author} {\bibfnamefont {P.}~\bibnamefont {Ku{\v{z}}el}}\ and\ \bibinfo {author} {\bibfnamefont {H.}~\bibnamefont {N{\v{e}}mec}},\ }\bibfield  {title} {\enquote {\bibinfo {title} {Terahertz spectroscopy of nanomaterials: a close look at charge-carrier transport},}\ }\href@noop {} {\bibfield  {journal} {\bibinfo  {journal} {Advanced Optical Materials}\ }\textbf {\bibinfo {volume} {8}},\ \bibinfo {pages} {1900623} (\bibinfo {year} {2020})}\BibitemShut {NoStop}%
\bibitem [{\citenamefont {Cooke}, \citenamefont {Krebs},\ and\ \citenamefont {Jepsen}(2012)}]{cooke2012direct}%
  \BibitemOpen
  \bibfield  {author} {\bibinfo {author} {\bibfnamefont {D.}~\bibnamefont {Cooke}}, \bibinfo {author} {\bibfnamefont {F.~C.}\ \bibnamefont {Krebs}}, \ and\ \bibinfo {author} {\bibfnamefont {P.~U.}\ \bibnamefont {Jepsen}},\ }\bibfield  {title} {\enquote {\bibinfo {title} {Direct observation of sub-100 fs mobile charge generation in a polymer-fullerene film},}\ }\href@noop {} {\bibfield  {journal} {\bibinfo  {journal} {Physical Review Letters}\ }\textbf {\bibinfo {volume} {108}},\ \bibinfo {pages} {056603} (\bibinfo {year} {2012})}\BibitemShut {NoStop}%
\bibitem [{\citenamefont {Leitenstorfer}\ \emph {et~al.}(2023)\citenamefont {Leitenstorfer}, \citenamefont {Moskalenko}, \citenamefont {Kampfrath}, \citenamefont {Kono}, \citenamefont {Castro-Camus}, \citenamefont {Peng}, \citenamefont {Qureshi}, \citenamefont {Turchinovich}, \citenamefont {Tanaka}, \citenamefont {Markelz} \emph {et~al.}}]{leitenstorfer20232023}%
  \BibitemOpen
  \bibfield  {author} {\bibinfo {author} {\bibfnamefont {A.}~\bibnamefont {Leitenstorfer}}, \bibinfo {author} {\bibfnamefont {A.~S.}\ \bibnamefont {Moskalenko}}, \bibinfo {author} {\bibfnamefont {T.}~\bibnamefont {Kampfrath}}, \bibinfo {author} {\bibfnamefont {J.}~\bibnamefont {Kono}}, \bibinfo {author} {\bibfnamefont {E.}~\bibnamefont {Castro-Camus}}, \bibinfo {author} {\bibfnamefont {K.}~\bibnamefont {Peng}}, \bibinfo {author} {\bibfnamefont {N.}~\bibnamefont {Qureshi}}, \bibinfo {author} {\bibfnamefont {D.}~\bibnamefont {Turchinovich}}, \bibinfo {author} {\bibfnamefont {K.}~\bibnamefont {Tanaka}}, \bibinfo {author} {\bibfnamefont {A.~G.}\ \bibnamefont {Markelz}},  \emph {et~al.},\ }\bibfield  {title} {\enquote {\bibinfo {title} {The 2023 terahertz science and technology roadmap},}\ }\href@noop {} {\bibfield  {journal} {\bibinfo  {journal} {Journal of Physics D: Applied Physics}\ }\textbf {\bibinfo {volume} {56}},\ \bibinfo {pages} {223001} (\bibinfo {year} {2023})}\BibitemShut {NoStop}%
\bibitem [{\citenamefont {Lin}, \citenamefont {Bierbaum},\ and\ \citenamefont {Cohen}(2017)}]{lin2017determining}%
  \BibitemOpen
  \bibfield  {author} {\bibinfo {author} {\bibfnamefont {N.~Y.}\ \bibnamefont {Lin}}, \bibinfo {author} {\bibfnamefont {M.}~\bibnamefont {Bierbaum}}, \ and\ \bibinfo {author} {\bibfnamefont {I.}~\bibnamefont {Cohen}},\ }\bibfield  {title} {\enquote {\bibinfo {title} {Determining quiescent colloidal suspension viscosities using the green-kubo relation and image-based stress measurements},}\ }\href@noop {} {\bibfield  {journal} {\bibinfo  {journal} {Physical Review Letters}\ }\textbf {\bibinfo {volume} {119}},\ \bibinfo {pages} {138001} (\bibinfo {year} {2017})}\BibitemShut {NoStop}%
\bibitem [{\citenamefont {Baroni}\ \emph {et~al.}(2020)\citenamefont {Baroni}, \citenamefont {Bertossa}, \citenamefont {Ercole}, \citenamefont {Grasselli},\ and\ \citenamefont {Marcolongo}}]{baroni2020heat}%
  \BibitemOpen
  \bibfield  {author} {\bibinfo {author} {\bibfnamefont {S.}~\bibnamefont {Baroni}}, \bibinfo {author} {\bibfnamefont {R.}~\bibnamefont {Bertossa}}, \bibinfo {author} {\bibfnamefont {L.}~\bibnamefont {Ercole}}, \bibinfo {author} {\bibfnamefont {F.}~\bibnamefont {Grasselli}}, \ and\ \bibinfo {author} {\bibfnamefont {A.}~\bibnamefont {Marcolongo}},\ }\bibfield  {title} {\enquote {\bibinfo {title} {Heat transport in insulators from ab initio green-kubo theory},}\ }\href@noop {} {\bibfield  {journal} {\bibinfo  {journal} {Handbook of Materials Modeling: Applications: Current and Emerging Materials}\ ,\ \bibinfo {pages} {809--844}} (\bibinfo {year} {2020})}\BibitemShut {NoStop}%
\bibitem [{\citenamefont {Bosse}\ and\ \citenamefont {Kaneko}(1995)}]{bosse1995self}%
  \BibitemOpen
  \bibfield  {author} {\bibinfo {author} {\bibfnamefont {J.}~\bibnamefont {Bosse}}\ and\ \bibinfo {author} {\bibfnamefont {Y.}~\bibnamefont {Kaneko}},\ }\bibfield  {title} {\enquote {\bibinfo {title} {Self-diffusion in supercooled binary liquids},}\ }\href@noop {} {\bibfield  {journal} {\bibinfo  {journal} {Physical Review Letters}\ }\textbf {\bibinfo {volume} {74}},\ \bibinfo {pages} {4023} (\bibinfo {year} {1995})}\BibitemShut {NoStop}%
\bibitem [{\citenamefont {Reichman}\ and\ \citenamefont {Rabani}(2001)}]{reichman2001self}%
  \BibitemOpen
  \bibfield  {author} {\bibinfo {author} {\bibfnamefont {D.~R.}\ \bibnamefont {Reichman}}\ and\ \bibinfo {author} {\bibfnamefont {E.}~\bibnamefont {Rabani}},\ }\bibfield  {title} {\enquote {\bibinfo {title} {Self-consistent mode-coupling theory for self-diffusion in quantum liquids},}\ }\href@noop {} {\bibfield  {journal} {\bibinfo  {journal} {Physical Review Letters}\ }\textbf {\bibinfo {volume} {87}},\ \bibinfo {pages} {265702} (\bibinfo {year} {2001})}\BibitemShut {NoStop}%
\bibitem [{\citenamefont {Reichman}\ and\ \citenamefont {Rabani}(2002)}]{reichman2002self}%
  \BibitemOpen
  \bibfield  {author} {\bibinfo {author} {\bibfnamefont {D.~R.}\ \bibnamefont {Reichman}}\ and\ \bibinfo {author} {\bibfnamefont {E.}~\bibnamefont {Rabani}},\ }\bibfield  {title} {\enquote {\bibinfo {title} {A self-consistent mode-coupling theory for dynamical correlations in quantum liquids: Application to liquid para-hydrogen},}\ }\href@noop {} {\bibfield  {journal} {\bibinfo  {journal} {Journal of Chemical Physics}\ }\textbf {\bibinfo {volume} {116}},\ \bibinfo {pages} {6279--6285} (\bibinfo {year} {2002})}\BibitemShut {NoStop}%
\bibitem [{\citenamefont {Montoya-Castillo}\ and\ \citenamefont {Reichman}(2016)}]{montoya2016approximate}%
  \BibitemOpen
  \bibfield  {author} {\bibinfo {author} {\bibfnamefont {A.}~\bibnamefont {Montoya-Castillo}}\ and\ \bibinfo {author} {\bibfnamefont {D.~R.}\ \bibnamefont {Reichman}},\ }\bibfield  {title} {\enquote {\bibinfo {title} {Approximate but accurate quantum dynamics from the mori formalism: I. nonequilibrium dynamics},}\ }\href@noop {} {\bibfield  {journal} {\bibinfo  {journal} {Journal of Chemical Physics}\ }\textbf {\bibinfo {volume} {144}},\ \bibinfo {pages} {184104} (\bibinfo {year} {2016})}\BibitemShut {NoStop}%
\bibitem [{\citenamefont {Kelly}\ \emph {et~al.}(2016)\citenamefont {Kelly}, \citenamefont {Montoya-Castillo}, \citenamefont {Wang},\ and\ \citenamefont {Markland}}]{kelly2016generalized}%
  \BibitemOpen
  \bibfield  {author} {\bibinfo {author} {\bibfnamefont {A.}~\bibnamefont {Kelly}}, \bibinfo {author} {\bibfnamefont {A.}~\bibnamefont {Montoya-Castillo}}, \bibinfo {author} {\bibfnamefont {L.}~\bibnamefont {Wang}}, \ and\ \bibinfo {author} {\bibfnamefont {T.~E.}\ \bibnamefont {Markland}},\ }\bibfield  {title} {\enquote {\bibinfo {title} {Generalized quantum master equations in and out of equilibrium: When can one win?}}\ }\href@noop {} {\bibfield  {journal} {\bibinfo  {journal} {Journal of Chemical Physics}\ }\textbf {\bibinfo {volume} {144}},\ \bibinfo {pages} {184105} (\bibinfo {year} {2016})}\BibitemShut {NoStop}%
\bibitem [{\citenamefont {Shi}\ and\ \citenamefont {Geva}(2003{\natexlab{b}})}]{shi2003new}%
  \BibitemOpen
  \bibfield  {author} {\bibinfo {author} {\bibfnamefont {Q.}~\bibnamefont {Shi}}\ and\ \bibinfo {author} {\bibfnamefont {E.}~\bibnamefont {Geva}},\ }\bibfield  {title} {\enquote {\bibinfo {title} {A new approach to calculating the memory kernel of the generalized quantum master equation for an arbitrary system--bath coupling},}\ }\href@noop {} {\bibfield  {journal} {\bibinfo  {journal} {Journal of Chemical Physics}\ }\textbf {\bibinfo {volume} {119}},\ \bibinfo {pages} {12063--12076} (\bibinfo {year} {2003}{\natexlab{b}})}\BibitemShut {NoStop}%
\bibitem [{\citenamefont {Zhang}, \citenamefont {Ka},\ and\ \citenamefont {Geva}(2006)}]{zhang2006nonequilibrium}%
  \BibitemOpen
  \bibfield  {author} {\bibinfo {author} {\bibfnamefont {M.-L.}\ \bibnamefont {Zhang}}, \bibinfo {author} {\bibfnamefont {B.~J.}\ \bibnamefont {Ka}}, \ and\ \bibinfo {author} {\bibfnamefont {E.}~\bibnamefont {Geva}},\ }\bibfield  {title} {\enquote {\bibinfo {title} {Nonequilibrium quantum dynamics in the condensed phase via the generalized quantum master equation},}\ }\href@noop {} {\bibfield  {journal} {\bibinfo  {journal} {Journal of Chemical Physics}\ }\textbf {\bibinfo {volume} {125}},\ \bibinfo {pages} {044106} (\bibinfo {year} {2006})}\BibitemShut {NoStop}%
\bibitem [{\citenamefont {Pfalzgraff}\ \emph {et~al.}(2019)\citenamefont {Pfalzgraff}, \citenamefont {Montoya-Castillo}, \citenamefont {Kelly},\ and\ \citenamefont {Markland}}]{pfalzgraff2019efficient}%
  \BibitemOpen
  \bibfield  {author} {\bibinfo {author} {\bibfnamefont {W.~C.}\ \bibnamefont {Pfalzgraff}}, \bibinfo {author} {\bibfnamefont {A.}~\bibnamefont {Montoya-Castillo}}, \bibinfo {author} {\bibfnamefont {A.}~\bibnamefont {Kelly}}, \ and\ \bibinfo {author} {\bibfnamefont {T.~E.}\ \bibnamefont {Markland}},\ }\bibfield  {title} {\enquote {\bibinfo {title} {Efficient construction of generalized master equation memory kernels for multi-state systems from nonadiabatic quantum-classical dynamics},}\ }\href@noop {} {\bibfield  {journal} {\bibinfo  {journal} {Journal of Chemical Physics}\ }\textbf {\bibinfo {volume} {150}},\ \bibinfo {pages} {244109} (\bibinfo {year} {2019})}\BibitemShut {NoStop}%
\bibitem [{\citenamefont {S{\"u}li}\ and\ \citenamefont {Mayers}(2003)}]{suli2003introduction}%
  \BibitemOpen
  \bibfield  {author} {\bibinfo {author} {\bibfnamefont {E.}~\bibnamefont {S{\"u}li}}\ and\ \bibinfo {author} {\bibfnamefont {D.~F.}\ \bibnamefont {Mayers}},\ }\href@noop {} {\emph {\bibinfo {title} {An introduction to numerical analysis}}}\ (\bibinfo  {publisher} {Cambridge university press},\ \bibinfo {year} {2003})\BibitemShut {NoStop}%
\bibitem [{\citenamefont {Sparpaglione}\ and\ \citenamefont {Mukamel}(1988)}]{sparpaglione1988dielectric}%
  \BibitemOpen
  \bibfield  {author} {\bibinfo {author} {\bibfnamefont {M.}~\bibnamefont {Sparpaglione}}\ and\ \bibinfo {author} {\bibfnamefont {S.}~\bibnamefont {Mukamel}},\ }\bibfield  {title} {\enquote {\bibinfo {title} {Dielectric friction and the transition from adiabatic to nonadiabatic electron transfer. i. solvation dynamics in liouville space},}\ }\href@noop {} {\bibfield  {journal} {\bibinfo  {journal} {Journal of Chemical Physics}\ }\textbf {\bibinfo {volume} {88}},\ \bibinfo {pages} {3263--3280} (\bibinfo {year} {1988})}\BibitemShut {NoStop}%
\bibitem [{\citenamefont {Golosov}\ and\ \citenamefont {Reichman}(2001)}]{golosov2001reference}%
  \BibitemOpen
  \bibfield  {author} {\bibinfo {author} {\bibfnamefont {A.~A.}\ \bibnamefont {Golosov}}\ and\ \bibinfo {author} {\bibfnamefont {D.~R.}\ \bibnamefont {Reichman}},\ }\bibfield  {title} {\enquote {\bibinfo {title} {Reference system master equation approaches to condensed phase charge transfer processes. i. general formulation},}\ }\href@noop {} {\bibfield  {journal} {\bibinfo  {journal} {Journal of Chemical Physics}\ }\textbf {\bibinfo {volume} {115}},\ \bibinfo {pages} {9848--9861} (\bibinfo {year} {2001})}\BibitemShut {NoStop}%
\end{thebibliography}%

\setcounter{section}{0}
\setcounter{equation}{0}
\setcounter{figure}{0}
\setcounter{table}{0}

\renewcommand{\theequation}{S\arabic{equation}}
\renewcommand{\thefigure}{S\arabic{figure}}


\onecolumngrid

\vfill\pagebreak

\section*{Supplementary Information: Current autocorrelation functions and conductivity responses for the dispersive Holstein model}
\label{SI-section:transport-figures}
\vspace{-12pt}

Here, we display the current autocorrelation functions and conductivity for the homogeneous, 1-dimensional dispersive Holstein model in the thermodynamic limit as one varies the charge-lattice coupling, $\eta$, and characteristic frequency of the local lattice phonons, $\omega_c$, for $v=50$ cm$^{-1}$ and ${\rm T}=300$~K.

\begin{figure*}[!th]
\vspace{-12pt}
\begin{center} 
    \resizebox{.91\textwidth}{!}{\includegraphics[trim={15pt 5pt 2pt 0pt},clip]{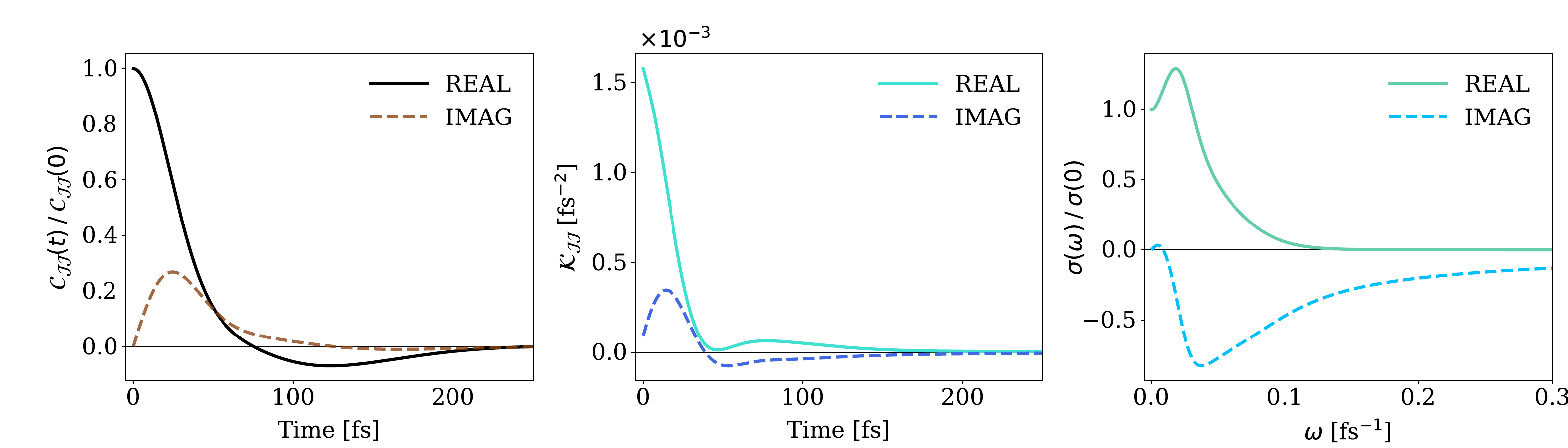}}
\vspace{-8pt}
\caption{\label{fig:set1} Parameters: $\eta / v = 2.0$, $\omega_c / v = 0.5$. \textbf{Left}: Current autocorrelation function, $C_{JJ}(t)$. \textbf{Middle}: Memory kernel, $\mathcal{K}(t)$ with lifetime $\tau_K = 317$~fs. \textbf{Right}: Real and imaginary part of the conductance, $\sigma(\omega)$.} 
\end{center}
\vspace{-14pt}
\end{figure*}


\begin{figure*}[!th]
\vspace{-16pt}
\begin{center} 
    \resizebox{.91\textwidth}{!}{\includegraphics[trim={15pt 5pt 2pt 0pt},clip]{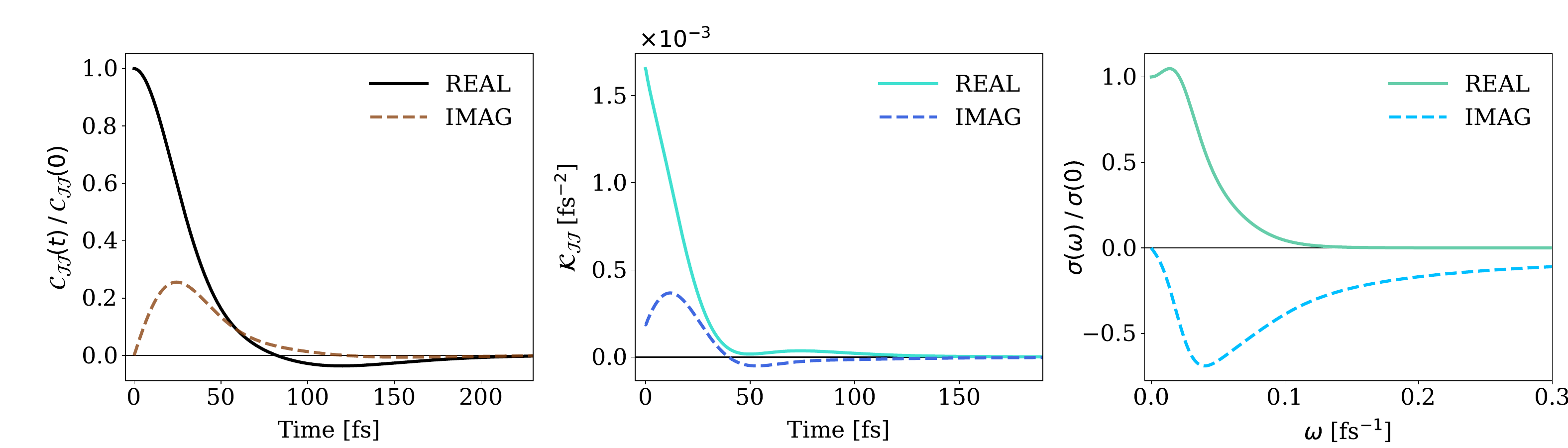}}
\vspace{-8pt}
\caption{\label{fig:set2} Parameters: $\eta / v = 2.0$, $\omega_c / v = 1.0$. \textbf{Left}: Current autocorrelation function, $C_{JJ}(t)$. \textbf{Middle}: Memory kernel, $\mathcal{K}(t)$ with lifetime $\tau_K = 164$~fs. \textbf{Right}: Real and imaginary part of the conductance, $\sigma(\omega)$.} 
\end{center}
\vspace{-14pt}
\end{figure*}


\begin{figure*}[!th]
\vspace{-16pt}
\begin{center} 
    \resizebox{.91\textwidth}{!}{\includegraphics[trim={15pt 5pt 2pt 0pt},clip]{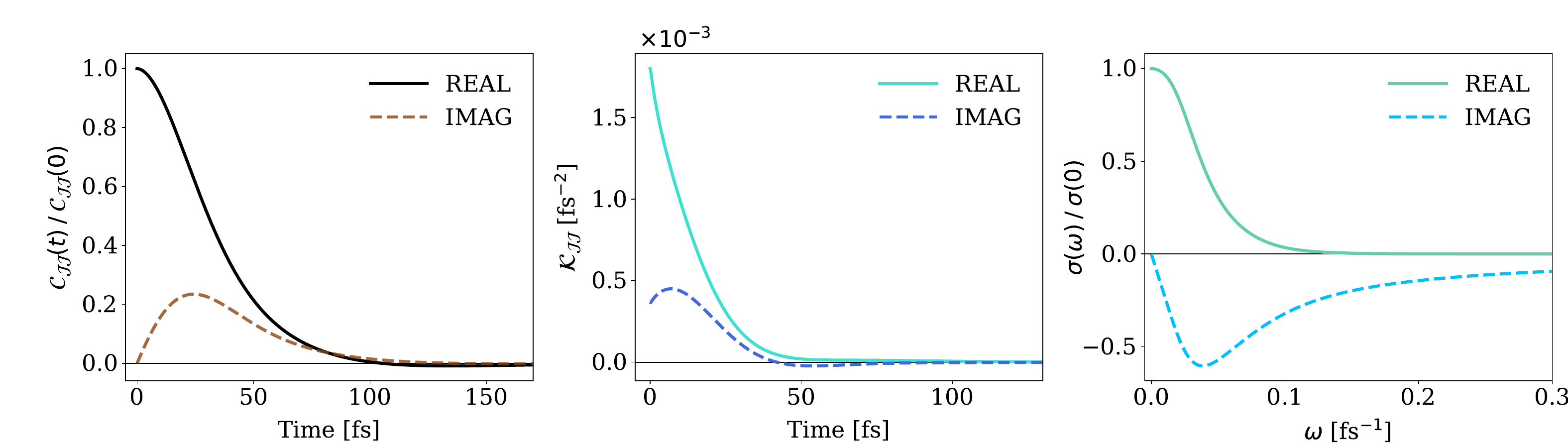}}
\vspace{-8pt}
\caption{\label{fig:set3} Parameters: $\eta / v = 2.0$, $\omega_c / v = 2.0$. \textbf{Left}: Current autocorrelation function, $C_{JJ}(t)$. \textbf{Middle}: Memory kernel, $\mathcal{K}(t)$ with lifetime $\tau_K = 167$~fs. \textbf{Right}: Real and imaginary part of the conductance, $\sigma(\omega)$.} 
\end{center}
\vspace{-14pt}
\end{figure*}


\begin{figure*}[!th]
\vspace{-16pt}
\begin{center} 
    \resizebox{.91\textwidth}{!}{\includegraphics[trim={15pt 5pt 2pt 0pt},clip]{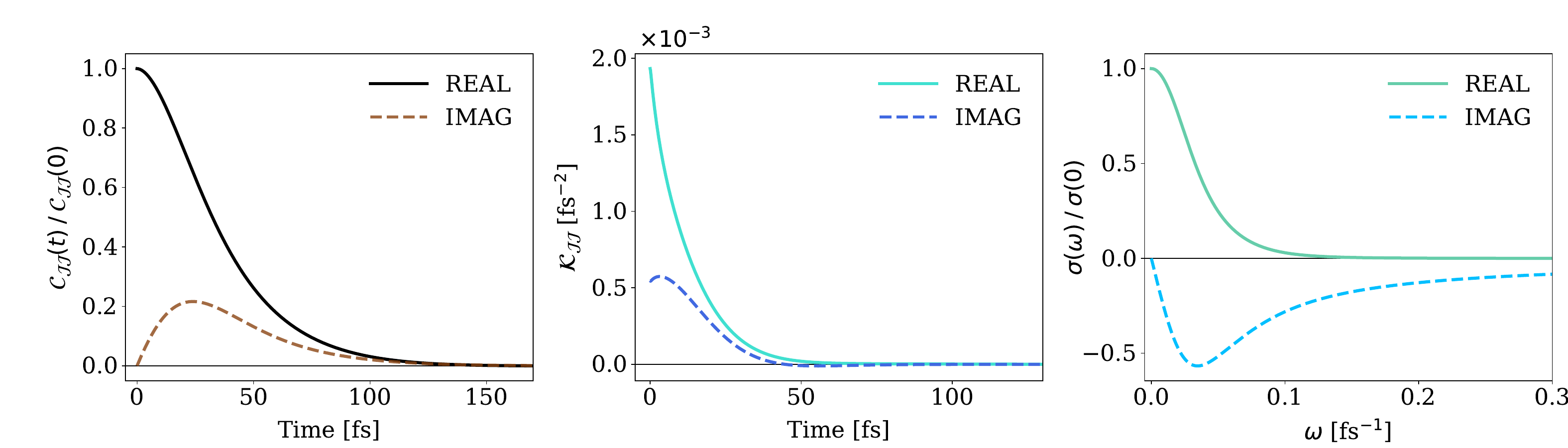}}
\vspace{-8pt}
\caption{\label{fig:set4} Parameters: $\eta / v = 2.0$, $\omega_c / v = 3.0$. \textbf{Left}: Current autocorrelation function, $C_{JJ}(t)$. \textbf{Middle}: Memory kernel, $\mathcal{K}(t)$ with lifetime $\tau_K = 65$~fs. \textbf{Right}: Real and imaginary part of the conductance, $\sigma(\omega)$.} 
\end{center}
\vspace{-84pt}
\end{figure*}


\begin{figure*}[!th]
\vspace{-8pt}
\begin{center} 
    \resizebox{.91\textwidth}{!}{\includegraphics[trim={15pt 5pt 2pt 0pt},clip]{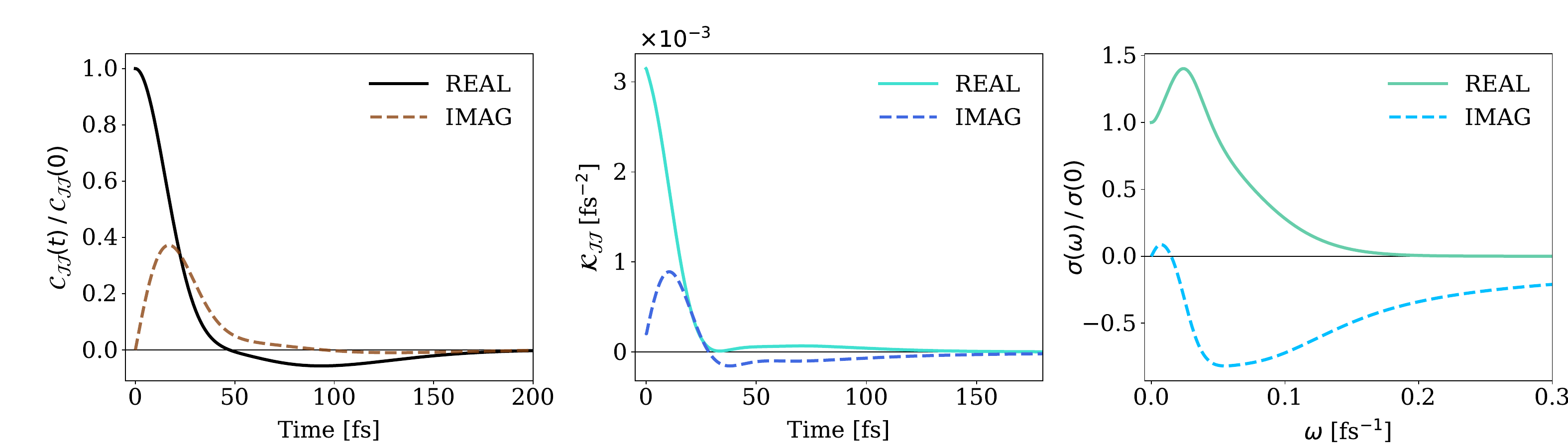}}
\vspace{-0pt}
\caption{\label{fig:set5} Parameters: $\eta / v = 4.0$, $\omega_c / v = 0.5$. \textbf{Left}: Current autocorrelation function, $C_{JJ}(t)$. \textbf{Middle}: Memory kernel, $\mathcal{K}(t)$ with lifetime $\tau_K = 337$~fs. \textbf{Right}: Real and imaginary part of the conductance, $\sigma(\omega)$.} 
\end{center}
\vspace{-14pt}
\end{figure*}


\begin{figure*}[!th]
\vspace{-8pt}
\begin{center} 
    \resizebox{.91\textwidth}{!}{\includegraphics[trim={15pt 5pt 2pt 0pt},clip]{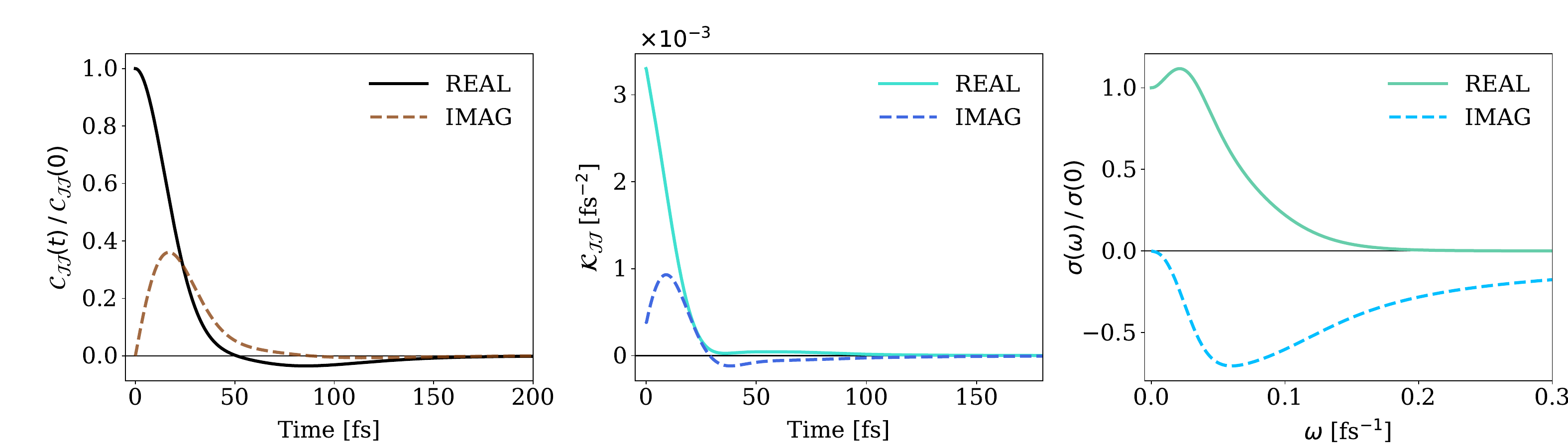}}
\vspace{-0pt}
\caption{\label{fig:set6} Parameters: $\eta / v = 4.0$, $\omega_c / v = 1.0$. \textbf{Left}: Current autocorrelation function, $C_{JJ}(t)$. \textbf{Middle}: Memory kernel, $\mathcal{K}(t)$ with lifetime $\tau_K = 178$~fs. \textbf{Right}: Real and imaginary part of the conductance, $\sigma(\omega)$.} 
\end{center}
\vspace{-14pt}
\end{figure*}


\begin{figure*}[!th]
\vspace{-8pt}
\begin{center} 
    \resizebox{.91\textwidth}{!}{\includegraphics[trim={15pt 5pt 2pt 0pt},clip]{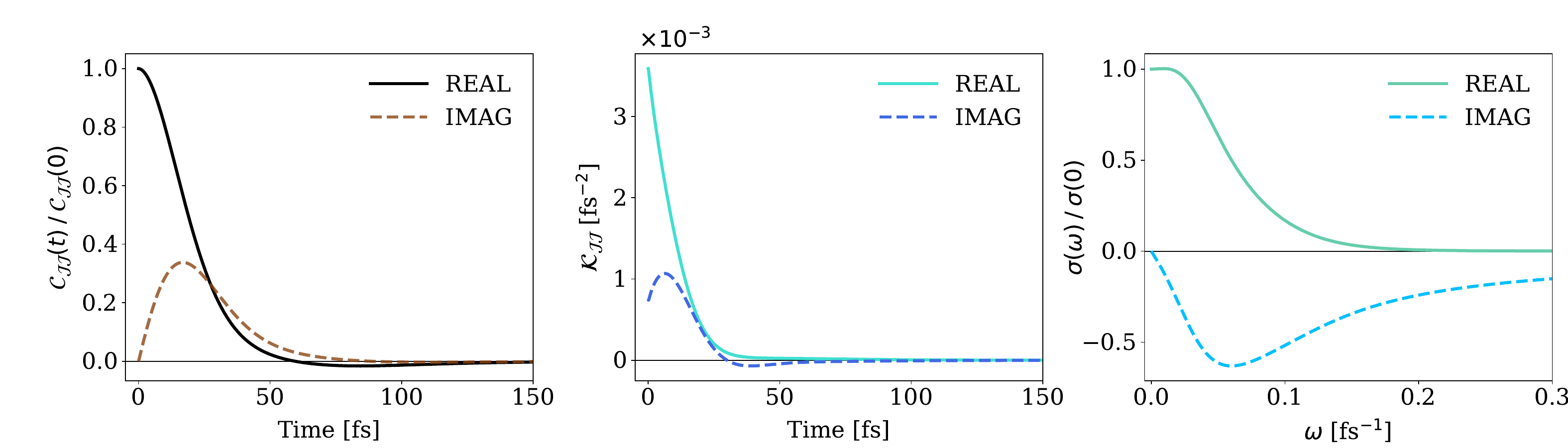}}
\vspace{-0pt}
\caption{\label{fig:set7} Parameters: $\eta / v = 4.0$, $\omega_c / v = 2.0$. \textbf{Left}: Current autocorrelation function, $C_{JJ}(t)$. \textbf{Middle}: Memory kernel, $\mathcal{K}(t)$ with lifetime $\tau_K = 93$~fs. \textbf{Right}: Real and imaginary part of the conductance, $\sigma(\omega)$.} 
\end{center}
\vspace{-14pt}
\end{figure*}


\begin{figure*}[!th]
\vspace{-8pt}
\begin{center} 
    \resizebox{.91\textwidth}{!}{\includegraphics[trim={15pt 5pt 2pt 0pt},clip]{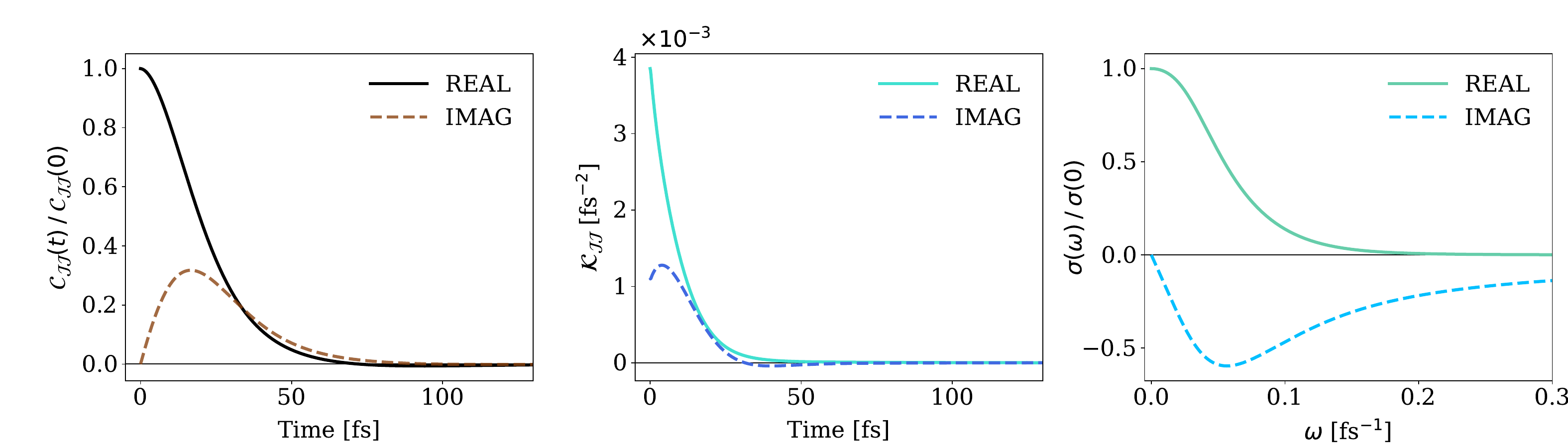}}
\vspace{-0pt}
\caption{\label{fig:set8} Parameters: $\eta / v = 4.0$, $\omega_c / v = 3.0$. \textbf{Left}: Current autocorrelation function, $C_{JJ}(t)$. \textbf{Middle}: Memory kernel, $\mathcal{K}(t)$ with lifetime $\tau_K = 60$~fs. \textbf{Right}: Real and imaginary part of the conductance, $\sigma(\omega)$.} 
\end{center}
\vspace{-14pt}
\end{figure*}


\begin{figure*}[!th]
\vspace{-8pt}
\begin{center} 
    \resizebox{.91\textwidth}{!}{\includegraphics[trim={15pt 5pt 2pt 0pt},clip]{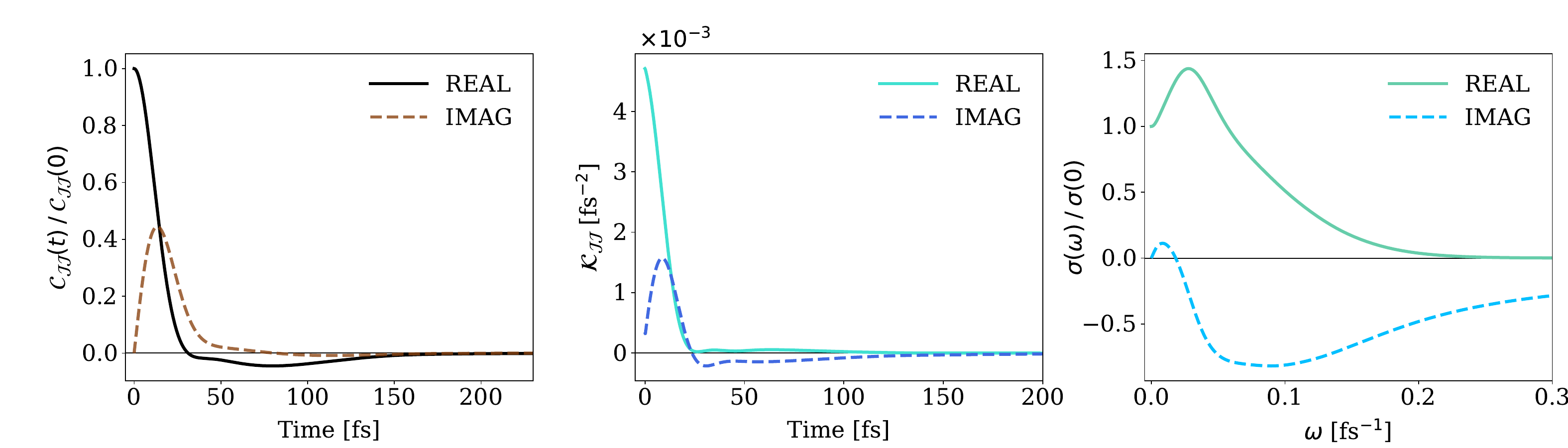}}
\vspace{-0pt}
\caption{\label{fig:set9} Parameters: $\eta / v = 6.0$, $\omega_c / v = 0.5$. \textbf{Left}: Current autocorrelation function, $C_{JJ}(t)$. \textbf{Middle}: Memory kernel, $\mathcal{K}(t)$ with lifetime $\tau_K = 305$~fs. \textbf{Right}: Real and imaginary part of the conductance, $\sigma(\omega)$.} 
\end{center}
\vspace{-14pt}
\end{figure*}


\begin{figure*}[!th]
\vspace{-8pt}
\begin{center} 
    \resizebox{.91\textwidth}{!}{\includegraphics[trim={15pt 5pt 2pt 0pt},clip]{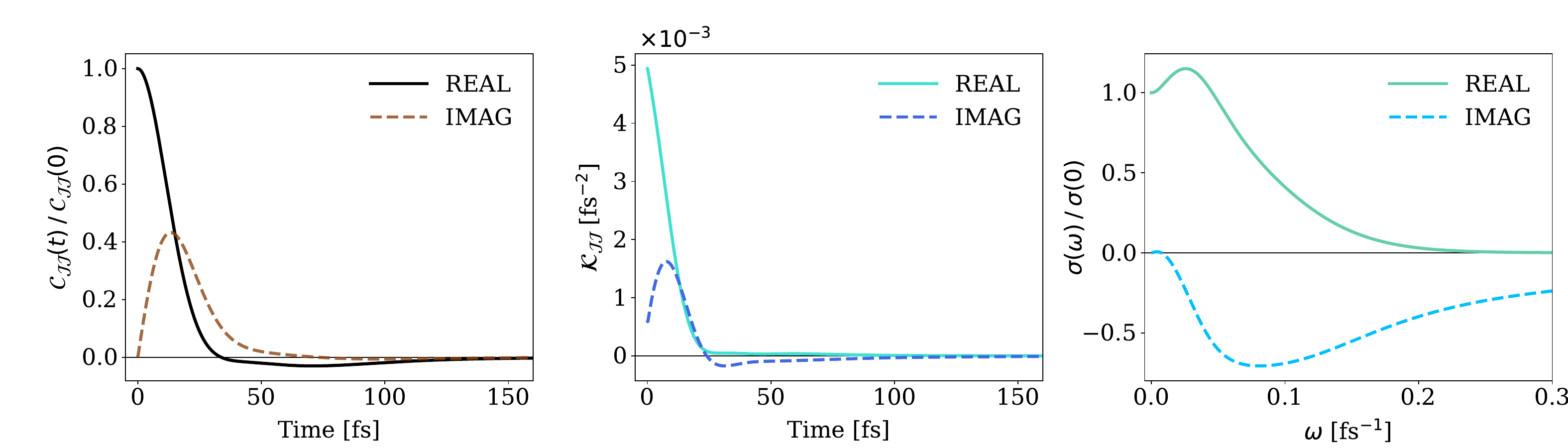}}
\vspace{-0pt}
\caption{\label{fig:set10} Parameters: $\eta / v = 6.0$, $\omega_c / v = 1.0$. \textbf{Left}: Current autocorrelation function, $C_{JJ}(t)$. \textbf{Middle}: Memory kernel, $\mathcal{K}(t)$ with lifetime $\tau_K = 189$~fs. \textbf{Right}: Real and imaginary part of the conductance, $\sigma(\omega)$.} 
\end{center}
\vspace{-14pt}
\end{figure*}


\begin{figure*}[!th]
\vspace{-8pt}
\begin{center} 
    \resizebox{.91\textwidth}{!}{\includegraphics[trim={15pt 5pt 2pt 0pt},clip]{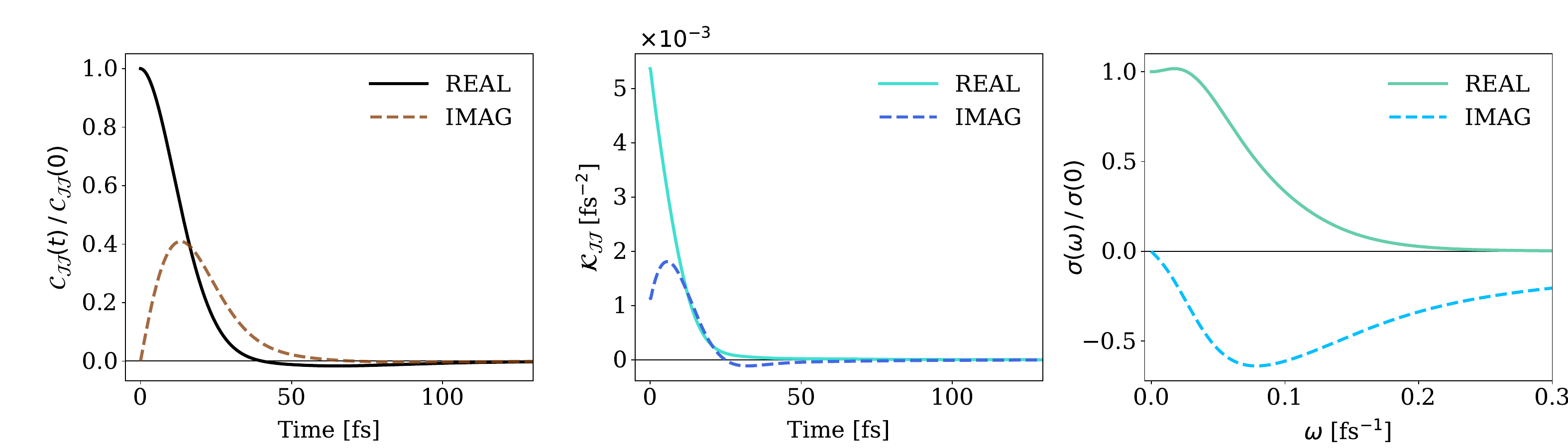}}
\vspace{-0pt}
\caption{\label{fig:set11} Parameters: $\eta / v = 6.0$, $\omega_c / v = 2.0$. \textbf{Left}: Current autocorrelation function, $C_{JJ}(t)$. \textbf{Middle}: Memory kernel, $\mathcal{K}(t)$ with lifetime $\tau_K = 96$~fs. \textbf{Right}: Real and imaginary part of the conductance, $\sigma(\omega)$.} 
\end{center}
\vspace{-14pt}
\end{figure*}


\begin{figure*}[!th]
\vspace{-8pt}
\begin{center} 
    \resizebox{.91\textwidth}{!}{\includegraphics[trim={15pt 5pt 2pt 0pt},clip]{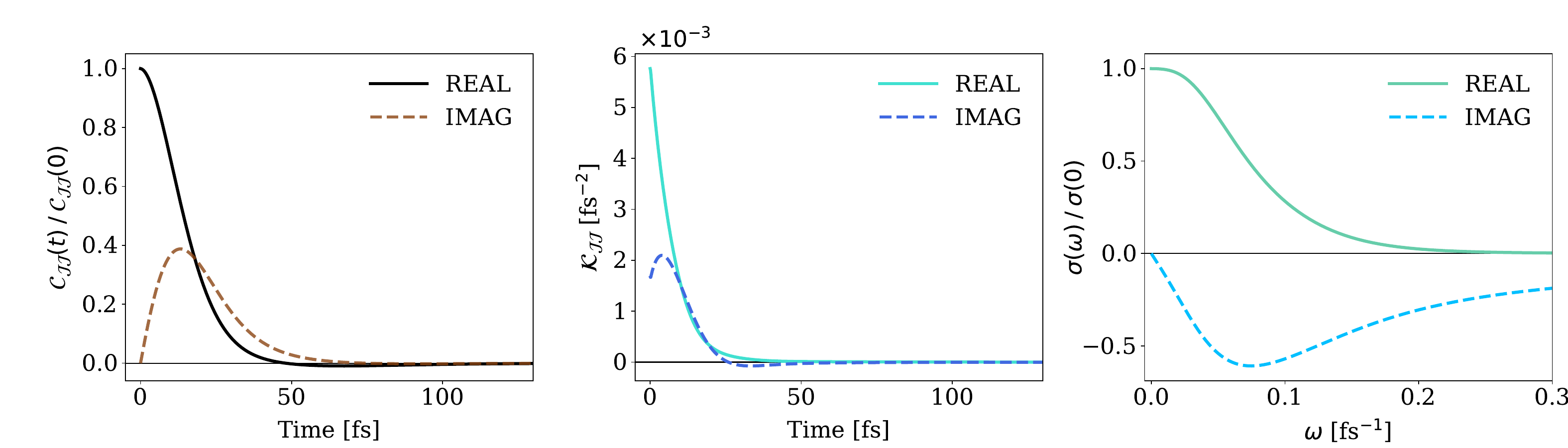}}
\vspace{-0pt}
\caption{\label{fig:set12} Parameters: $\eta / v = 6.0$, $\omega_c / v = 3.0$. \textbf{Left}: Current autocorrelation function, $C_{JJ}(t)$. \textbf{Middle}: Memory kernel, $\mathcal{K}(t)$ with lifetime $\tau_K = 61$~fs. \textbf{Right}: Real and imaginary part of the conductance, $\sigma(\omega)$.} 
\end{center}
\vspace{-14pt}
\end{figure*}


\begin{figure*}[!th]
\vspace{-8pt}
\begin{center} 
    \resizebox{.91\textwidth}{!}{\includegraphics[trim={15pt 5pt 2pt 0pt},clip]{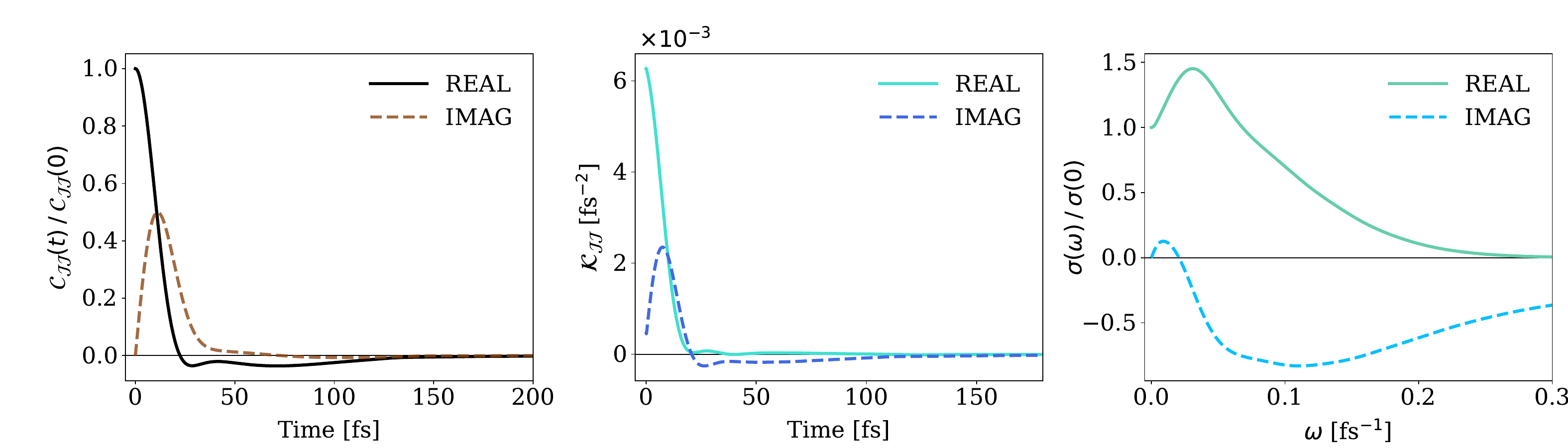}}
\vspace{-0pt}
\caption{\label{fig:set13} Parameters: $\eta / v = 8.0$, $\omega_c / v = 0.5$. \textbf{Left}: Current autocorrelation function, $C_{JJ}(t)$. \textbf{Middle}: Memory kernel, $\mathcal{K}(t)$ with lifetime $\tau_K = 378$~fs. \textbf{Right}: Real and imaginary part of the conductance, $\sigma(\omega)$.} 
\end{center}
\vspace{-14pt}
\end{figure*}


\begin{figure*}[!th]
\vspace{-8pt}
\begin{center} 
    \resizebox{.91\textwidth}{!}{\includegraphics[trim={15pt 5pt 2pt 0pt},clip]{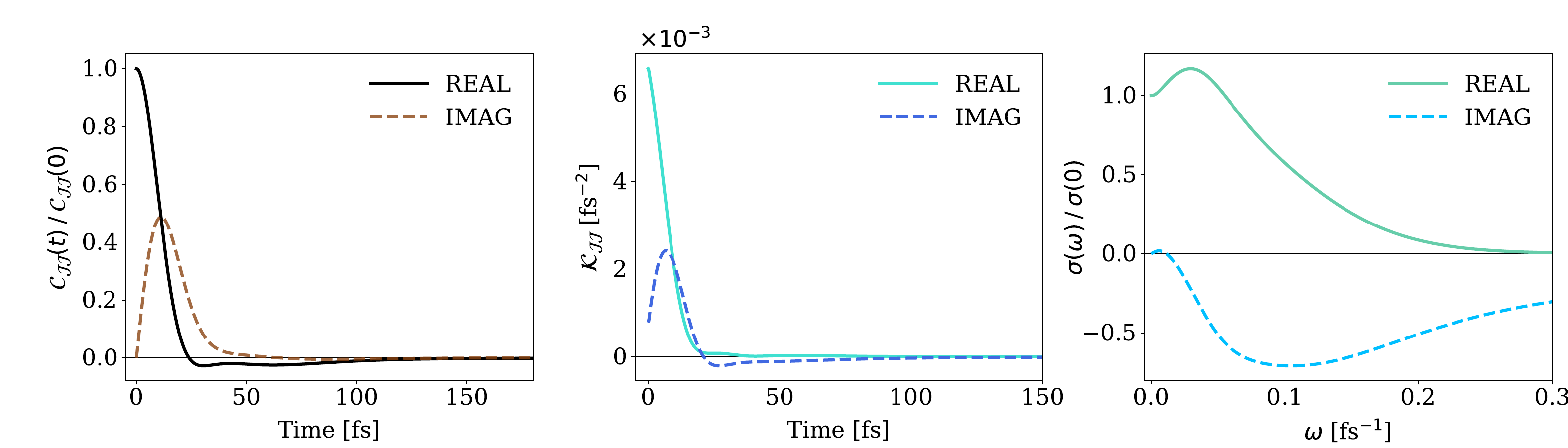}}
\vspace{-0pt}
\caption{\label{fig:set14} Parameters: $\eta / v = 8.0$, $\omega_c / v = 1.0$. \textbf{Left}: Current autocorrelation function, $C_{JJ}(t)$. \textbf{Middle}: Memory kernel, $\mathcal{K}(t)$ with lifetime $\tau_K = 195$~fs. \textbf{Right}: Real and imaginary part of the conductance, $\sigma(\omega)$.} 
\end{center}
\vspace{-14pt}
\end{figure*}


\begin{figure*}[!th]
\vspace{-8pt}
\begin{center} 
    \resizebox{.91\textwidth}{!}{\includegraphics[trim={15pt 5pt 2pt 0pt},clip]{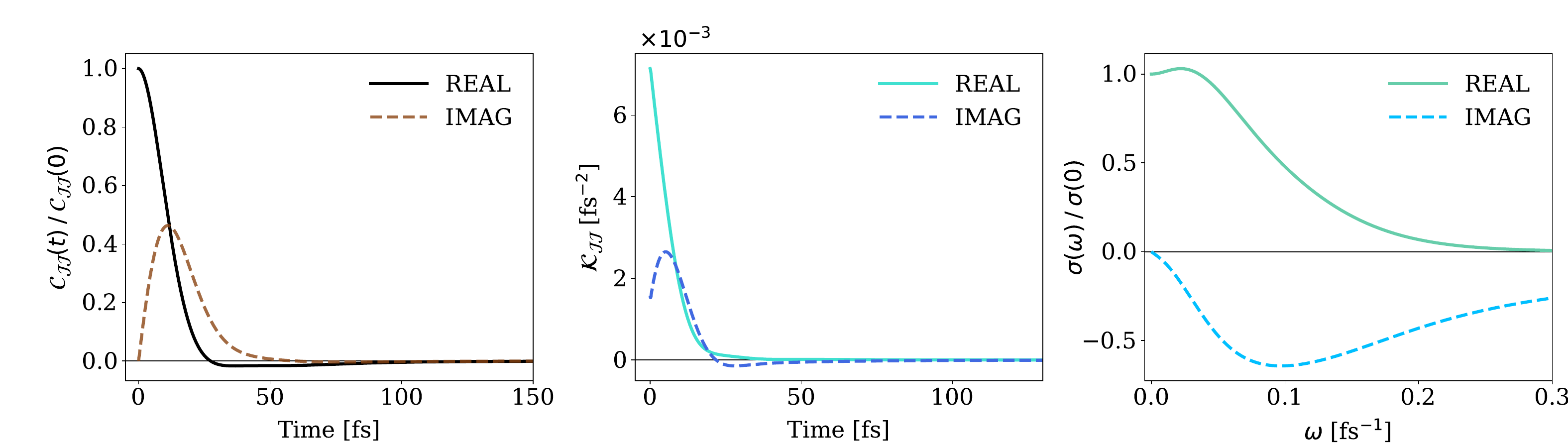}}
\vspace{-0pt}
\caption{\label{fig:set15} Parameters: $\eta / v = 8.0$, $\omega_c / v = 2.0$. \textbf{Left}: Current autocorrelation function, $C_{JJ}(t)$. \textbf{Middle}: Memory kernel, $\mathcal{K}(t)$ with lifetime $\tau_K = 96$~fs. \textbf{Right}: Real and imaginary part of the conductance, $\sigma(\omega)$.} 
\end{center}
\vspace{-14pt}
\end{figure*}


\begin{figure*}[!th]
\vspace{-8pt}
\begin{center} 
    \resizebox{.91\textwidth}{!}{\includegraphics[trim={15pt 5pt 2pt 0pt},clip]{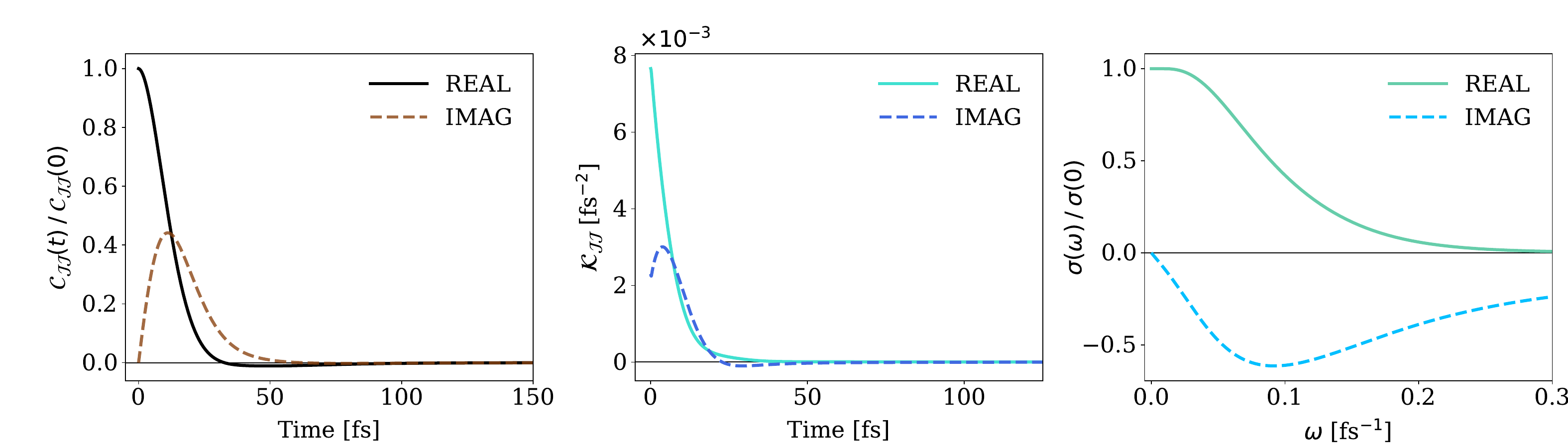}}
\vspace{-0pt}
\caption{\label{fig:set16} Parameters: $\eta / v = 8.0$, $\omega_c / v = 3.0$. \textbf{Left}: Current autocorrelation function, $C_{JJ}(t)$. \textbf{Middle}: Memory kernel, $\mathcal{K}(t)$ with lifetime $\tau_K = 62$~fs. \textbf{Right}: Real and imaginary part of the conductance, $\sigma(\omega)$.} 
\end{center}
\vspace{-14pt}
\end{figure*}


\begin{figure*}[!th]
\vspace{-8pt}
\begin{center} 
    \resizebox{.91\textwidth}{!}{\includegraphics[trim={15pt 5pt 2pt 0pt},clip]{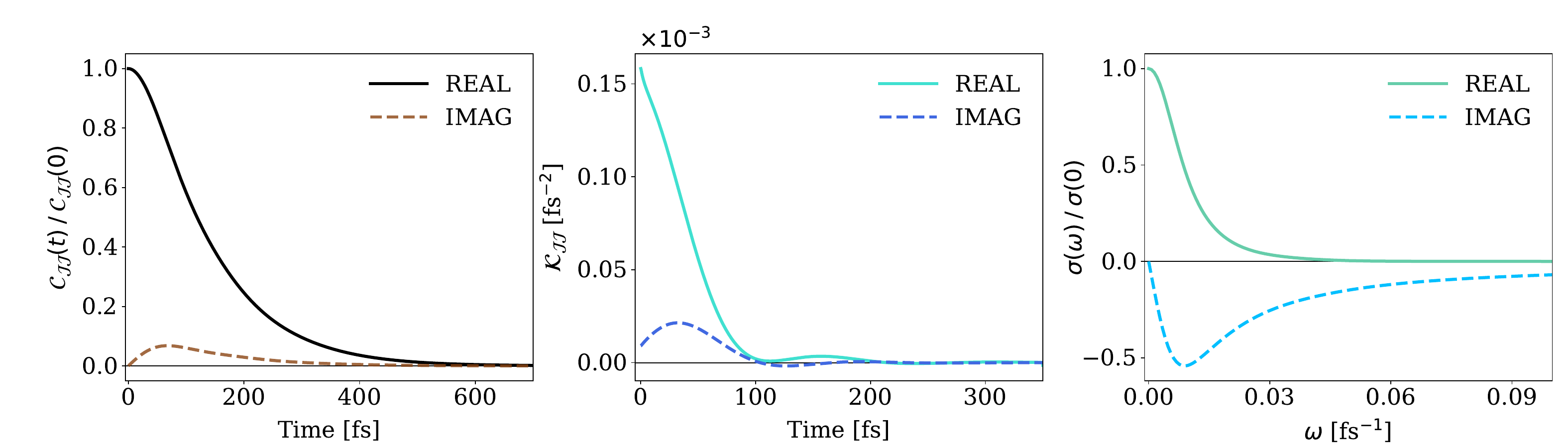}}
\vspace{-0pt}
\caption{\label{fig:set21} Parameters: $\eta / v = 0.2$, $\omega_c / v = 0.5$. \textbf{Left}: Current autocorrelation function, $C_{JJ}(t)$. \textbf{Middle}: Memory kernel, $\mathcal{K}(t)$ with lifetime $\tau_K = 173$~fs. \textbf{Right}: Real and imaginary part of the conductance, $\sigma(\omega)$.} 
\end{center}
\vspace{-14pt}
\end{figure*}


\begin{figure*}[!th]
\vspace{-8pt}
\begin{center} 
    \resizebox{.91\textwidth}{!}{\includegraphics[trim={15pt 5pt 2pt 0pt},clip]{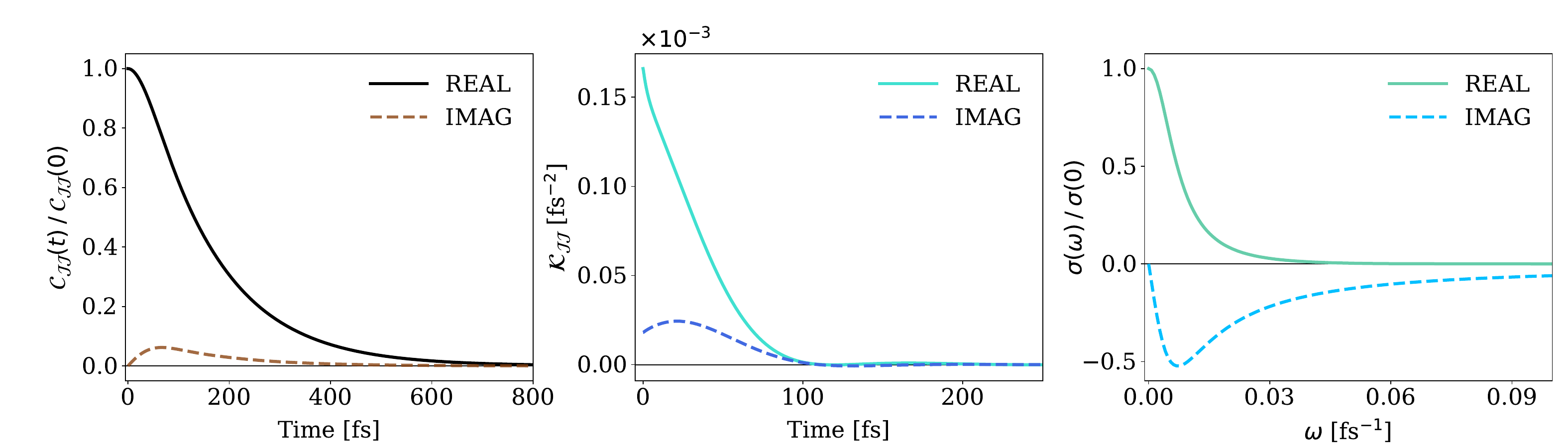}}
\vspace{-0pt}
\caption{\label{fig:set22} Parameters: $\eta / v = 0.2$, $\omega_c / v = 1.0$. \textbf{Left}: Current autocorrelation function, $C_{JJ}(t)$. \textbf{Middle}: Memory kernel, $\mathcal{K}(t)$ with lifetime $\tau_K = 105$~fs. \textbf{Right}: Real and imaginary part of the conductance, $\sigma(\omega)$.} 
\end{center}
\vspace{-14pt}
\end{figure*}


\begin{figure*}[!th]
\vspace{-8pt}
\begin{center} 
    \resizebox{.91\textwidth}{!}{\includegraphics[trim={15pt 5pt 2pt 0pt},clip]{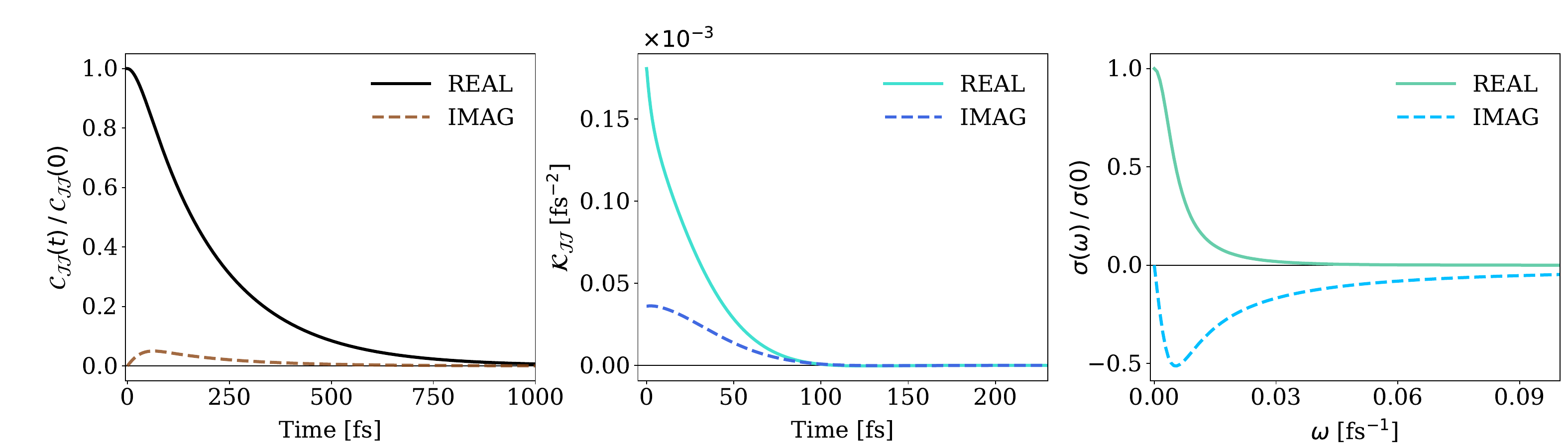}}
\vspace{-0pt}
\caption{\label{fig:set23} Parameters: $\eta / v = 0.2$, $\omega_c / v = 2.0$. \textbf{Left}: Current autocorrelation function, $C_{JJ}(t)$. \textbf{Middle}: Memory kernel, $\mathcal{K}(t)$ with lifetime $\tau_K = 80$~fs. \textbf{Right}: Real and imaginary part of the conductance, $\sigma(\omega)$.} 
\end{center}
\vspace{-14pt}
\end{figure*}


\begin{figure}[!th]
\vspace{-8pt}
\begin{center} 
    \resizebox{.91\textwidth}{!}{\includegraphics[trim={15pt 5pt 2pt 0pt},clip]{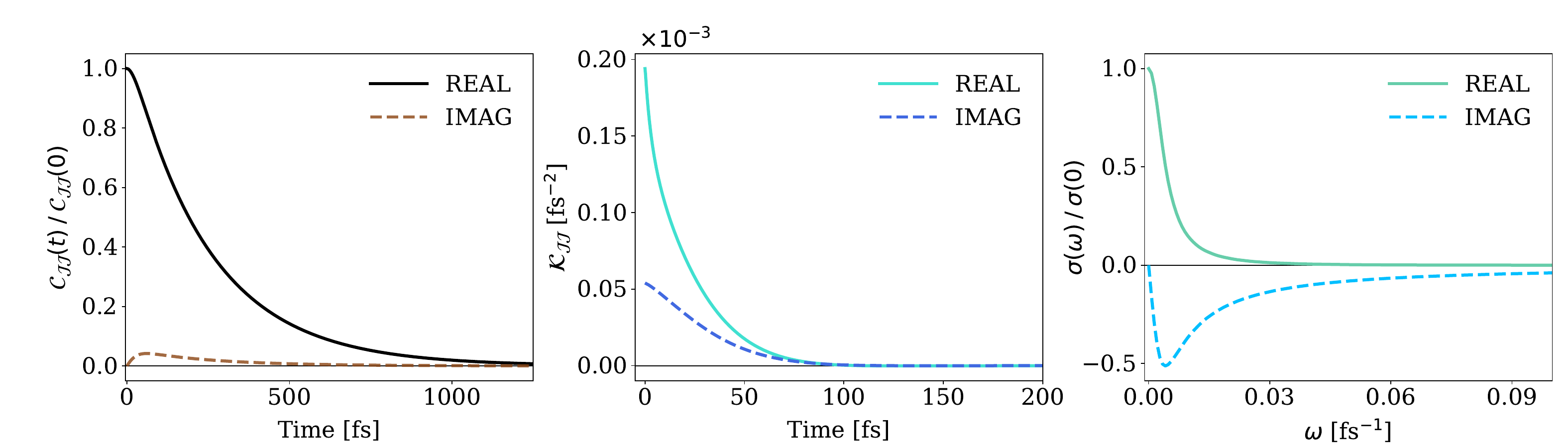}}
\vspace{-0pt}
\caption{\label{fig:set24} Parameters: $\eta / v = 0.2$, $\omega_c / v = 3.0$. \textbf{Left}: Current autocorrelation function, $C_{JJ}(t)$. \textbf{Middle}: Memory kernel, $\mathcal{K}(t)$ with lifetime $\tau_K = 74$~fs. \textbf{Right}: Real and imaginary part of the conductance, $\sigma(\omega)$.} 
\end{center}
\vspace{-14pt}
\end{figure}

\vfill
\pagebreak

\end{document}